\documentclass[
  journal=pasa,
  manuscript=research-paper, 
  year=2020,
  volume=37,
]{cup-journal}

\usepackage{microtype,siunitx,booktabs}

\usepackage{graphicx}
\usepackage{amsmath}	
\usepackage{amssymb}	
\usepackage{multicol}       
\usepackage{bm}		
\usepackage{pdflscape}	
\usepackage{xurl}
\usepackage{hyperref} 
\usepackage[T1]{fontenc}
\usepackage{ae,aecompl}
\usepackage{xcolor} 
\usepackage{booktabs}
\usepackage{pifont}
\usepackage{nicematrix}
\usepackage{longtable}
\usepackage{threeparttable}

\title[Odd Radio Circles in the First Year of the EMU Survey]{Discovery of Odd Radio Circles and Other Peculiars in the First Year of the EMU Survey using Object Detection}

\author{Nikhel Gupta$^{1}$} 
\author{Ray P. Norris$^{2,3}$}
\author{Zeeshan Hayder$^{4}$}
\author{Minh Huynh$^{1,5}$}
\author{Heinz Andernach$^{6}$}
\author{Andrew M. Hopkins$^{7}$}
\author{Stanislav Shabala$^{8}$}
\author{Lawrence Rudnick$^{9}$}
\author{Miroslav D. Filipovi\'c$^{2}$}
\author{B\"arbel S. Koribalski$^{3,2}$}
\author{Lars Petersson$^{4}$}
\author{X. Rosalind Wang$^{2}$}
\email[Nikhel Gupta]{Nikhel.Gupta@csiro.au}
\affiliation{
$^1$ CSIRO Space \& Astronomy, PO Box 1130, Bentley WA 6102, Australia \\
$^2$ Western Sydney University, Locked Bag 1797, Penrith, NSW 2751, Australia \\
$^3$ CSIRO Space \& Astronomy, P.O. Box 76, Epping, NSW 1710, Australia \\
$^4$ CSIRO Data61, Black Mountain ACT 2601, Australia \\
$^5$ International Centre for Radio Astronomy Research (ICRAR), M468, The University of Western Australia, 35 Stirling Highway, Crawley, WA 6009, Australia \\ 
$^6$ Depto.\ de Astronom\'{i}a, DCNE, Universidad de Guanajuato, Cj\'on.\ de Jalisco s/n, Guanajuato, CP 36023, Mexico \\
$^7$ School of Mathematical and Physical Sciences, 12 Wally’s Walk, Macquarie University, NSW 2109, Australia \\
$^{8}$ School of Natural Sciences, University of Tasmania, Private Bag 37, Hobart 7001, Australia \\
$^{9}$ University of Minnesota, 100 Church St SE, Minneapolis, MN 55455, USA \\
}

\received {dd Mmm YYYY}
\revised  {dd Mmm YYYY}
\accepted {dd Mmm YYYY}
\published{dd Mmm YYYY}


\usepackage{amsmath,amsfonts,amssymb}
\usepackage{color}
\usepackage{mathrsfs}
\usepackage{tabularx}
\usepackage{floatrow}
\usepackage{float}

\definecolor{ored}{rgb}{1.00,0.27,0.00}
\definecolor{mygreen}{rgb}{0.2,0.7,0.2}

\definecolor{Gray}{gray}{0.5}
\definecolor{LightCyan}{rgb}{0.88,1,1}

\def \BE{\begin{equation}}
\def \EE{\end{equation}}	
\def \BC{\begin{center}}
\def \EC{\end{center}}
\def \BEA{\begin{eqnarray}}
\def \EEA{\end{eqnarray}}

\def \SIGMA8{\sigma_{8}}

\usepackage{cellspace}
\begin{document}\sloppy\sloppypar\raggedbottom\frenchspacing


\begin{abstract}
We present a systematic search for Odd Radio Circles (ORCs) and other unusual radio morphologies using data from the first year of the EMU (Evolutionary Map of the Universe) survey. 
ORCs are rare, enigmatic objects characterized by edge-brightened rings of radio emission, often found in association with distant galaxies. 
To identify these objects, we employ a hybrid methodology combining supervised object detection techniques and visual inspection of radio source candidates. 
This approach leads to the discovery of five new ORCs and two additional candidate ORCs, expanding the known population of these objects. 
In addition to ORCs, we also identify 55 Galaxies with Large-scale Ambient Radio Emission (GLAREs), which feature irregular, rectangular, or circular shapes of diffuse radio emission mostly surrounding central host galaxies. 
These GLAREs may represent different evolutionary stages of ORCs, and studying them could offer valuable insights into their evolutionary processes. 
We also highlight a subset of Starburst Radio Ring Galaxies (SRRGs), which are star-forming galaxies exhibiting edge-brightened radio rings surrounding their central star-forming regions. 
We emphasize the importance of multi-wavelength follow-up observations to better understand the physical properties, host galaxy characteristics, and evolutionary pathways of these radio sources. 
\end{abstract}


\section{Introduction}
\label{SEC:Intro}
The advent of new technologies in radio astronomy has enabled faster and deeper scans across vast regions of the sky, producing highly sensitive continuum images of the Universe in shorter timescales. 
However, these advancements have also introduced the significant challenge of managing and using unprecedented volumes of Big Data. 
Telescopes such as the Australian Square Kilometre Array Pathfinder \citep[ASKAP;][]{johnston07ASKAP,DeBoer09,hotan21}, the Low-Frequency Array \citep[LOFAR;][]{vanharleem13}, the Murchison Widefield Array \citep[MWA;][]{wayth18}, MeerKAT \citep{jonas16}, and the Karl G. Jansky Very Large Array \citep[JVLA;][]{perley11} collectively produce more than 100 petabytes of data each year. 
ASKAP alone generates data at the staggering rate of 100 trillion bits per second, surpassing Australia’s entire internet traffic.
Efficient management of such datasets opens new avenues for detecting millions of galaxies at radio wavelengths. 
For instance, the ongoing Evolutionary Map of the Universe \citep[EMU;][]{norris21} survey, conducted with the ASKAP, is expected to uncover more than 20 million compact and extended radio galaxies during the five years of its planned operation \citep{norris21}.
Similarly, the LOFAR Two-metre Sky Survey \citep[LoTSS][]{shimwell22} is anticipated to detect over 10 million radio galaxies as it surveys the entire northern sky.
With the upcoming Square Kilometre Array (SKA\footnote{https://www.skatelescope.org/the-ska-project/}) set to become operational in the coming years, the number of radio galaxy detections could grow to several times
the current number, potentially reaching hundreds of millions. This vast data influx will profoundly impact our understanding of galaxy evolution and the broader history of the Universe. In addition, such extensive datasets are likely to lead to the discovery of new phenomena and deepen our understanding of the origins of radio emissions. 

\begin{figure*}
\centering
\includegraphics[scale=0.8]{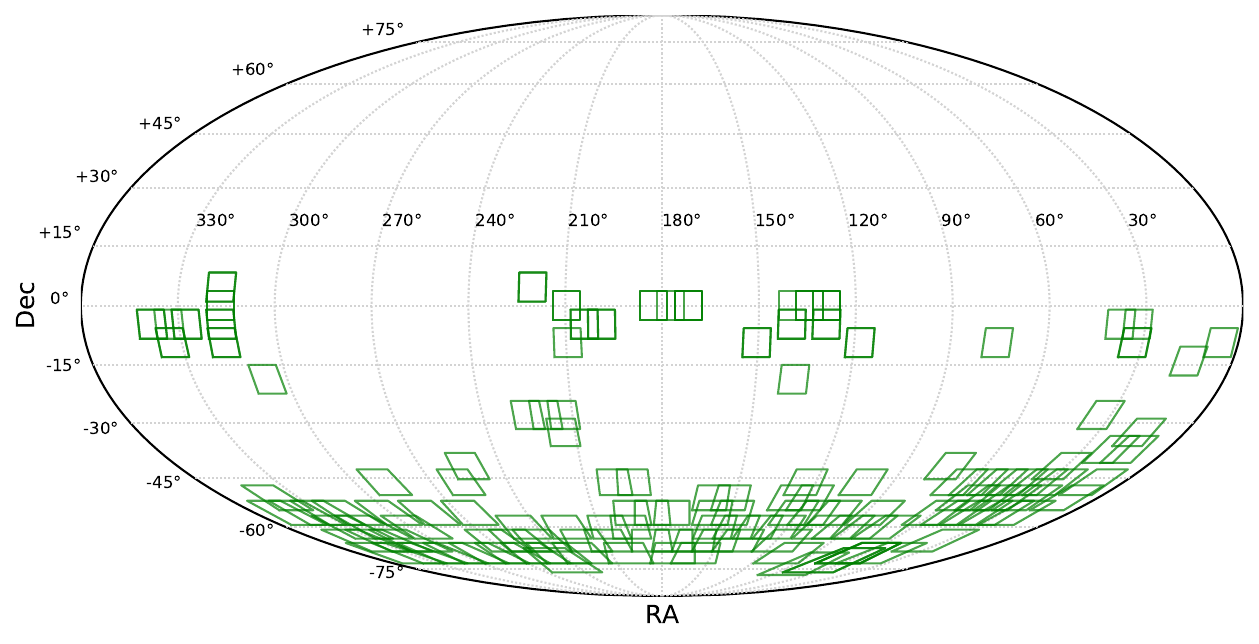}
\caption{The footprint of the first year of the EMU main survey, conducted between February 2023 and March 2024 (tiles SBID 45638 to 59612 in CASDA), includes 160 tiles covering an area of approximately 4,500 square degrees. The green squares represent the coverage of individual tiles, including their overlaps.} 
\label{FIG:EMUsurvey}
\end{figure*}

Historically, many of the most significant scientific discoveries made with major telescopes have been serendipitous, occurring beyond the original goals of the experiments \citep[][]{norris15}. \cite{ekers09} notes that only seven out of 18 major astronomical discoveries in the past 60 years were planned.
A prime example of this is the recent discovery of Odd Radio Circles \citep[ORCs;][]{norris21b}, which were found serendipitously during visual inspections of 270 square degrees of maps in the first pilot EMU survey \citep[EMU PS1;][]{norris21}.
However, with the main EMU survey having already scanned over 4,000 square degrees of the Southern Sky in its first year, relying on visual inspections to discover more ORCs or other rare phenomena is challenging.
To fully harness the potential of the EMU survey spanning entire Southern Sky in coming years and to discover more ORCs and other rare radio phenomena, it is essential to innovate and develop new technologies capable of handling and analysing the massive volumes of data. Unless we reconsider and overhaul our search methodologies, many unknown radio phenomena may go unnoticed for years or possibly never be discovered.

In recent years, machine learning has emerged as a powerful tool to accelerate the discovery of unknown and rare phenomena in the images of next-generation radio surveys \citep[e.g.][]{mostert21, gupta22, walmsley22, segal22, slijepcevic23, gupta23a, lochner23, Mohale24, Lastufka24, gupta24b, riggi24, lochner24}. 
For instance, \citet{gupta22} rediscovered previously known ORC J2103-6200 alongside an additional ORC J2223-4834 that was not recorded during the serendipitous inspections of the EMU PS1.
This success has bolstered confidence in machine learning as a key approach for discovering rare radio sources in large surveys, where visual inspections are increasingly impractical. 
Therefore, it is imperative to develop comprehensive methods to automate the detection of rare radio sources. 
Such advancements will enhance their statistics and provide crucial insights into their formation mechanisms.

With the advent of large-scale radio surveys generating unprecedented volumes of data, most machine learning methods produce a significant number of anomalies, many of which are not scientifically interesting. 
As a result, extensive visual inspections are often required to filter these anomalies further and identify rare or previously unknown morphologies.
In this work, we implement a supervised learning approach for object detection \citep[][]{gupta23b, gupta24a} first used to catalogue radio galaxies \citep[][]{gupta24b}, to identify rare and unusual radio morphologies and reduce the need for extensive visual inspections to confirm their rarity.
We use the object detection model Gal-DINO\footnote{\url{http://github.com/Nikhel1/Gal-DINO}}, trained on the RadioGalaxyNET dataset \citep[][]{gupta24a} and a dataset of atypical radio morphologies \citep[][]{gupta24b} from the EMU PS1. 
This model is applied to the first 160 observation tiles from the EMU main survey’s first-year observations \citep[][]{hopkins25}.
The model functions as a classifier to differentiate typical radio galaxies from rare and unusual morphologies. 
The selected sample of rare and unusual morphologies is then visually inspected to uncover new ORCs and other diffuse emission systems, which may represent different evolutionary stages of ORCs.

The paper is organized as follows. In Section~\ref{SEC:dataset}, we provide details on the EMU survey and infrared mosaics. 
Section~\ref{SEC:method} is dedicated to explaining the object detection method and detailed search methodology. 
Section~\ref{SEC:ORCs} and Section~\ref{SEC:OtherPecs} provide details about the new ORCs and other peculiar objects, respectively. 
Our findings are summarized in Section~\ref{SEC:conclusions}, where we also outline directions for future research.
Throughout this paper, we use cosmological parameters from \citet{planck18-1}.

\section{Data}
\label{SEC:dataset}
This section provides details about the EMU survey, the first year of EMU data, and the infrared observations used in this study.

\subsection{EMU Observations}
The Evolutionary Map of the Universe \citep[EMU\footnote{\url{https://emu-survey.org/}};][]{norris11,hopkins25} is a groundbreaking survey designed to create a comprehensive radio atlas of the southern sky using the Australian Square Kilometre Array Pathfinder (ASKAP) radio telescope \citep{hotan21}.
ASKAP, situated at Inyarrimahnha Ilgari Bundara, the Murchison Radio-astronomy Observatory (MRO), is a cutting-edge radio telescope equipped with phased array feed (PAF) technology, enabling rapid surveying through its expansive instantaneous field of view. 
It comprises 36 antennas with diverse baselines, the majority clustered within a 2.3 km diameter, while six antennas extend the baselines up to 6.4 km. 
This will result in 853 tile footprints, obtained through 1,014 individual tile observations. Of these, 692 tiles have an integration time of 10 hours, while 161 tiles with a 5-hour integration time are observed twice. 
The resulting RMS noise ranges from $25$ to $55~\mu$Jy/beam, depending on the integration time (10 or 5 hours), with a beamwidth of $13^{\prime\prime} \times 11^{\prime\prime}$ FWHM. 
Over its five-year duration, the EMU survey will cover the southern declination range from $-11^{\circ}.4$ to the south celestial pole, as well as selected equatorial regions extending up to $\delta = +7^{\circ}.0$.
Observations are conducted in the 800–1088 MHz frequency range, centred at 944 MHz. 
At these depths, the EMU survey is expected to catalogue up to 20 million radio sources across $2\pi$ sr of the sky, with completion anticipated by 2028.

In this work, we search for ORCs and other unusual radio morphologies using data from the first year of EMU observations. 
The dataset includes 160 tiles, covering approximately 4,500 square degrees of the sky. 
Data collection began in late 2022, with validated data arriving between February 2023 and March 2024. These include tiles with Scheduling Block unique ID (SBID) numbers ranging from 45638 to 59612, delivered through the CSIRO Data Access Portal (CASDA\footnote{https://research.csiro.au/casda/}).
We use restored images with a common $15^{\prime\prime}$ resolution for each beam in the mosaic of a tile (with ``conv'' as a suffix in the filename in CASDA).
Figure~\ref{FIG:EMUsurvey} illustrates the footprint of the first year of the EMU main survey, highlighting the overlap between the tiles. These tiles are processed using the ASKAPsoft pipeline \citep[][]{whiting17}, which operates on the raw telescope data. Following this, source catalogues of islands and components \citep[see][for details about the catalogues]{hopkins25} are generated using the \textit{Selavy} source finder \citep{whiting12}.
The sources are detected at a S/N threshold of 5, which corresponds to 0.17 mJy/beam for the 10-hour observations and 0.31 mJy/beam for the 5-hour observations.
Observations with a 10-hour integration time achieve a median RMS noise of $30~\mu$Jy/beam and typically detect approximately $750$ sources per square degree, while 5-hour integrations yield a median RMS noise of $46~\mu$Jy/beam, detecting around $460$ sources per square degree.
For the 160 tiles surveyed during the first year of the EMU survey, this corresponds to approximately 3 million detected radio sources.
In this study, we analyse each tile independently rather than creating super mosaics of overlapping tiles. While this method may result in the duplication of radio sources in overlapping regions, these duplications are addressed during the subsequent analysis.

\subsection{Infrared Mosaics}
In addition to the EMU observations, we create matching tiles for the AllWISE dataset from the Wide-field Infrared Survey Explorer \citep[WISE][]{wright10, cutri13} using the Montage image mosaic software\footnote{Implementation available at: \url{https://github.com/Nikhel1/wise_mosaics}}. 
WISE conducted an infrared survey of the entire sky across four bands: W1, W2, W3, and W4, corresponding to wavelengths of 3.4, 4.6, 12, and 22 $\mu$m. For this study, we focus on the W1 band from AllWISE, which offers a 5$\sigma$ point source detection limit of 28 $\mu$Jy and an angular resolution of $6.1^{\prime \prime}$.

\section{Search Methodology}
\label{SEC:method}
The section provides an overview of the machine learning model and catalogue pipeline, the dataset used for training the model, its application to the first year of EMU survey data, and the criteria established for visual inspections to identify ORCs and other unusual radio sources.

\begin{table*}[!ht]
\centering
\begin{threeparttable}
\resizebox{\textwidth}{!}{
\begin{tabular}{cccccc}
\hline
Property & ORC J0210-5710 & ORC J0402-5321 & ORC J0452-6231 & ORC J1313-4709 & ORC J2304-7129 \\
\hline
Host Name & $\text{WISEA J021009.39$-$571038.3}$ & $\text{WISEA J040214.70$-$532128.6}$ & $\text{WISEA J045230.76$-$623123.8}$ & $\text{WISEA J131335.54$-$470915.2}$ & $\text{WISEA J230420.71$-$712907.8}$ \\
RA (deg) & $32.5391$ & $60.5613$ & $73.1281$ & $198.3981$ & $346.0863$ \\
Dec (deg) & $-57.1774$ & $-53.3580$ & $-62.5232$ & $-47.1542$ & $-71.4855$ \\
$l$ (deg) & $283.4218$ & $263.1923$ & $272.6182$ & $306.8384$ & $314.2504$ \\
$b$ (deg) & $-56.8686$ & $-46.3305$ & $-37.4590$ & $15.5494$ & $-43.2118$ \\
$A_r$ & $0.06$ & $0.03$ & $0.06$ & $0.26$ & $0.06$ \\
$g$ & $21.04$ & $22.95$ & $20.85$ & $19.21$ & $17.53$ \\
$r$ & $19.25$ & $21.04$ & $19.06$ & $18.67$ & $17.02$ \\
$i$ & $18.61$ & $20.14$ & $18.40$ & $18.46$ & $16.66$ \\
$z$ & $18.26$ & $19.74$ & $18.06$ & $18.36$ & $16.55$ \\
W1 & $15.113\pm0.032$ & $16.240\pm0.028$ & $15.043\pm0.028$ & $15.935\pm0.042$ & $13.498\pm0.020$ \\
W2 & $14.582\pm0.106$ & $16.487\pm0.112$ & $14.921\pm0.043$ & $16.524\pm0.203$ & $12.756\pm0.022$ \\
W3 & $12.640\pm0.415$ & -- & -- & -- & $8.054\pm0.016$ \\
W1-W2 & $0.531\pm0.111$ & $-0.247\pm0.115$ & $0.122\pm0.051$ & $-0.589\pm0.207$ & $0.742\pm0.030$ \\
W2-W3 & $1.942\pm0.428$ & -- & -- & -- & $4.702\pm0.027$ \\
$z_{\rm sp}\tnote{a}$ & -- & -- & -- & -- & $0.1481$ \\
$z_{\rm ph}\tnote{b}$ & $0.410\pm0.018$ & $0.536\pm0.050$ & $0.430\pm0.014$ & -- & -- \\
$z_{\rm ph}\tnote{c}$ & $0.390\pm0.046$ & -- & $0.389\pm0.046$ & -- & $0.129\pm0.037$ \\
$z_{\rm ph}\tnote{d}$ & $0.399\pm0.035$ & -- & $0.412\pm0.041$ & -- & -- \\
$z_{\rm ph}\tnote{e}$ & $0.402\pm0.024$ & $0.560\pm0.017$ & $0.415\pm0.024$ & -- & -- \\
$z_{\rm ph}\tnote{f}$ & $0.442\pm0.024$ & $0.572\pm0.007$ & $0.453\pm0.024$ & -- & -- \\
$z_{\rm ph}\tnote{g}$ & $0.382\pm0.022$ & $0.548\pm0.044$ & $0.419\pm0.020$ & -- & -- \\
$\rm Flux_I$ (mJy) & $3.55\pm0.43$ & $1.55\pm0.11$ & $2.06\pm0.15$ & $3.01\pm0.18$ & $3.34\pm0.21$ \\
$\rm Lum_I$ ($\rm W~Hz^{-1}$) & $[1.3\pm0.2]\times10^{24}$ & $[9.0\pm1.7]\times10^{23}$ & $[8.0\pm0.8]\times10^{23}$ & -- & $[1.7\pm0.1]\times10^{23}$ \\
$\rm Flux_H$ (mJy) & $0.15\pm0.02$ & $0.08\pm0.02$ & $0.04\pm0.01$ & $0.22\pm0.02$ & $0.55\pm0.04$ \\
$\rm Lum_H$ ($\rm W~Hz^{-1}$) & $[5.3\pm0.8]\times10^{22}$ & $[4.6\pm1.4]\times10^{22}$ & $[1.6\pm0.4]\times10^{22}$ & -- & $[2.7\pm0.2]\times10^{22}$ \\
Size ($^{\prime\prime}$) & $78\pm3.9$ & $40\pm2.0$ & $52\pm2.6$ & -- & $60\pm3.0$ \\ 
Size (kpc) & $438\pm25$ & $260\pm18$ & $300\pm22$ & -- & $160\pm8$ \\
SFR ($M_{\odot}~\text{yr}^{-1}$) & $44.5\pm7.0$ & $38.9\pm11.8$ & $13.0\pm3.3$ & -- & $23.0\pm1.7$ \\
$\log M_*~(M_{\odot})$ & $11.44$ & -- & $11.41$ & -- & -- \\
EMU SBID & $46946$ & $59481$ & $50230$ & $51948$ & $53566$ \\
\hline
\end{tabular}
}
\begin{tablenotes}
\footnotesize
\item[a] \citet{jones09}
\item[b] \citet{zhou21}
\item[c] \citet{bilicki16}, where uncertainties are $0.033(1+z_{\rm ph})$
\item[d] \citet{duncan22}
\item[e] \citet{zou22}
\item[f] \citet{wen24}
\item[g] \citet{zhou25}
\end{tablenotes}
\end{threeparttable}
\caption{Characteristics of the ORCs and their potential host galaxies from the first year of the EMU survey. From left to right, the table lists the ORC names. From top to bottom, the rows provide details of the potential host galaxies, including their names, Right Ascension (RA), Declination (Dec), longitude (l), latitude (b), $r$-band foreground extinction ($A_r$, in mag), optical ($griz$ AB mag) and infrared (W1, W2, W3 Vega mag) photometry, spectroscopic ($z_{\rm sp}$) and photometric ($z_{\rm ph}$) redshifts where available. This is followed by the 944 MHz integrated radio flux densities ($\rm Flux_I$) estimated using ASKAP images, integrated radio luminosity ($\rm Lum_I$; assuming a spectral index of -0.7), 944 MHz radio flux density of the potential host ($\rm Flux_H$), its radio luminosity ($\rm Lum_H$), the largest angular and physical size of the ORC in arcseconds and kpc, the star formation rate (SFR) of the host (upper limits calculated using the relationship in \citet{murphy11} at 944 MHz) and stellar mass ($M_*$) of the host \citep[from][]{zou19}. The size, luminosity and SFR are calculated from the $z_{\rm sp}$ for ORC J2304-7129 and using $z_{\rm ph}$ from DESI LS DR9 \citep{zhou21} for the rest. Lastly, the table includes the ID of the EMU tile where the ORC is located. The $griz$ AB magnitudes are sourced from DESI LS DR10 and from \citet{tonry18} for ORC J1313-4709, while the W1, W2 and W3 Vega magnitudes for all comes from the WISE survey.}
\label{TAB:ORC-counterparts}
\end{table*}

\subsection{Object Detection Model}
\label{SEC:ObjectDetect}
In a first-of-its-kind attempt, this work employs a supervised machine-learning method to identify anomalous radio morphologies.
Specifically, we employ the RG-CAT catalogue construction pipeline, developed by \citet{gupta24b}, which leverages an object detection framework Gal-DINO \citep[][]{gupta24a} to construct a comprehensive catalogue of radio sources. For detailed descriptions of the Gal-DINO model and the RG-CAT pipeline, readers are referred to the respective papers; here, we provide a brief overview.
Gal-DINO is an advanced object detection model specifically designed to identify radio galaxies and locate their potential infrared hosts. 
The model is trained on a dataset of 5,000 radio galaxies, including 2,800 sources from the RadioGalaxyNET dataset \citep[][]{gupta24a}, which spans four classes of radio galaxies based on measurements of their total extent and the distance between peak positions. 
This classification follows the criteria outlined by \cite{fanaroff74}, which define the distinction between FR-I and FR-II galaxies based on the ratio of peak separation to total extent. Galaxies with a ratio below 0.45 are classified as FR-I, while those with a ratio above 0.55 are categorized as FR-II.

However, some galaxies cannot be classified as FR-I or FR-II due to image resolution limitations. These are assigned to the FR-x category, characterized by a peak-to-extent ratio between 0.45 and 0.55. Additionally, the R category includes radio galaxies with resolved emission but only a single visible central peak, resulting in a peak-to-extent ratio of zero.
This dataset is further augmented with 2,100 compact or unresolved radio galaxies and 100 sources exhibiting rare and peculiar morphologies.
The latter includes instances of uniquely shaped extended emissions, unusual bent-tailed radio galaxies, diffuse emissions from galaxy cluster halos, nearby resolved star-forming galaxies, and peculiar structures such as ORCs.
The training process focuses on optimizing the detection of both radio galaxies and their corresponding potential host galaxies by refining the accuracy of bounding box and keypoint predictions. 
Bounding box predictions encompass all components of a radio galaxy, while keypoint predictions pinpoint the locations of potential infrared hosts. The model's performance is evaluated using Average Precision (AP) at specified Intersection over Union (IoU) thresholds between the ground truth and predictions.
On test datasets, Gal-DINO achieves an AP$_{50}$ of 73.2\% for bounding box predictions and 71.7\% for keypoint detections. 
Moreover, when focusing on the central source in each image, 99\% of bounding box predictions achieve an IoU greater than 0.5, and 98\% of keypoint predictions fall within $<3^{\prime \prime}$ of the ground truth infrared host positions.

\subsection{Application to EMU main Survey}
We extend the RG-CAT pipeline, originally developed for the EMU PS1, to the first-year EMU main survey tiles. 
This involves generating $8^{\prime} \times 8^{\prime}$ cutouts at the locations of approximately 3 million \textit{Selavy}-based radio sources from the first year of the EMU survey. 
Each cutout is subsequently analysed using the trained Gal-DINO model to extract the bounding box, category, and prediction confidence score for the central source, along with its keypoint prediction.
The RG-CAT pipeline then uses Gal-DINO predictions to construct a catalogue of radio sources for each main survey tile.
Compact and extended radio galaxies are catalogued differently: compact sources are added individually, while multiple components of extended galaxies are grouped into a single entry.
The details of the main survey catalogue, including radio source classifications and their corresponding infrared/optical potential host galaxies, will be discussed in \citet{gupta25prep}. 
For this study, we focus solely on the predicted categories and confidence scores to identify ORCs and other rare radio morphologies.

\subsection{Visual Inspections}
We use the catalogues generated by the RG-CAT pipeline for each EMU tile and filter radio sources categorized as rare or peculiar radio sources. 
From the 160 tiles observed during the first year of the EMU survey, we identify 5,871 such sources where the prediction confidence score exceeds the model’s minimum estimated threshold.
For this analysis, we set a conservative confidence score threshold of 0.7, resulting in 1,794 radio sources in this category above the threshold. 
These are then visually inspected to identify ORCs and other unusual morphologies. 
This threshold was determined by randomly selecting 25 EMU tiles and visually inspecting all sources above the minimum confidence score. 
No interesting radio sources or ORCs were found below a confidence score of 0.8 in these tiles, leading us to adopt a slightly lower threshold of 0.7 to ensure no ORCs are missed during the visual inspection across all EMU tiles.
Note that while we use restored common $15^{\prime\prime}$ resolution images for training the model and constructing RG-CAT catalogues, we also use $\sim8^{\prime\prime}$ resolution images (with ``highres'' as a filename suffix in CASDA) during the visual inspection of 1,794 sources to identify any high-resolution features missed in the common-resolution images. 
However, all plots in this paper show only $15^{\prime\prime}$ resolution images.

Additionally, the confusion matrix for Gal-DINO predictions (see Figure~4 in \cite{gupta24b}) indicates that rare and peculiar sources could be misclassified as FR-I or FR-II radio galaxies. 
This may occur with unusual bent-tailed sources that resemble these categories. 
However, such misclassification is expected to be rare for circular sources like ORCs, and the likelihood is further reduced in the RG-CAT catalogue, which only includes predictions for the central source in each cutout (as discussed in Section\ref{SEC:ObjectDetect}). 
To confirm this, we reviewed all FR-I and FR-II predictions in the randomly selected 10 tiles and did not find any peculiar radio morphologies among them.

In summary, we classify the initial dataset of approximately 3 million radio sources using the RG-CAT pipeline. 
To focus on identifying ORCs and other rare morphologies, we filtered the catalogue for rare and peculiar predictions. 
Ultimately, we visually inspected just 1,794 radio sources from the initial set of $\sim$3 million for 160 EMU tiles.
This substantial reduction in manual effort for new discoveries highlights the transformative potential of advanced machine learning algorithms in minimizing the resources needed for Big Data analysis.

\begin{figure*}[htbp]
\centering
\vspace*{-0.2cm}
\includegraphics[trim=1.4cm 5.8cm 1.4cm 5.8cm, clip, width=18cm, scale=0.5]{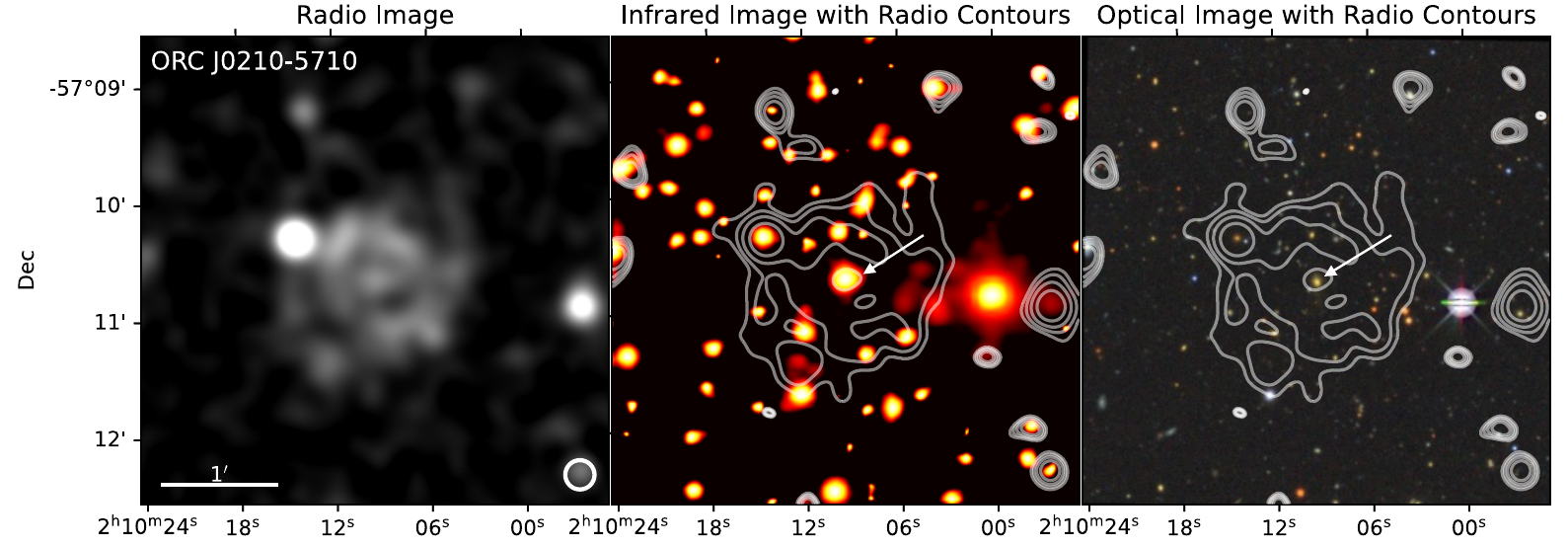}
\includegraphics[trim=1.4cm 5.8cm 1.4cm 6.3cm, clip, width=18cm, scale=0.5]{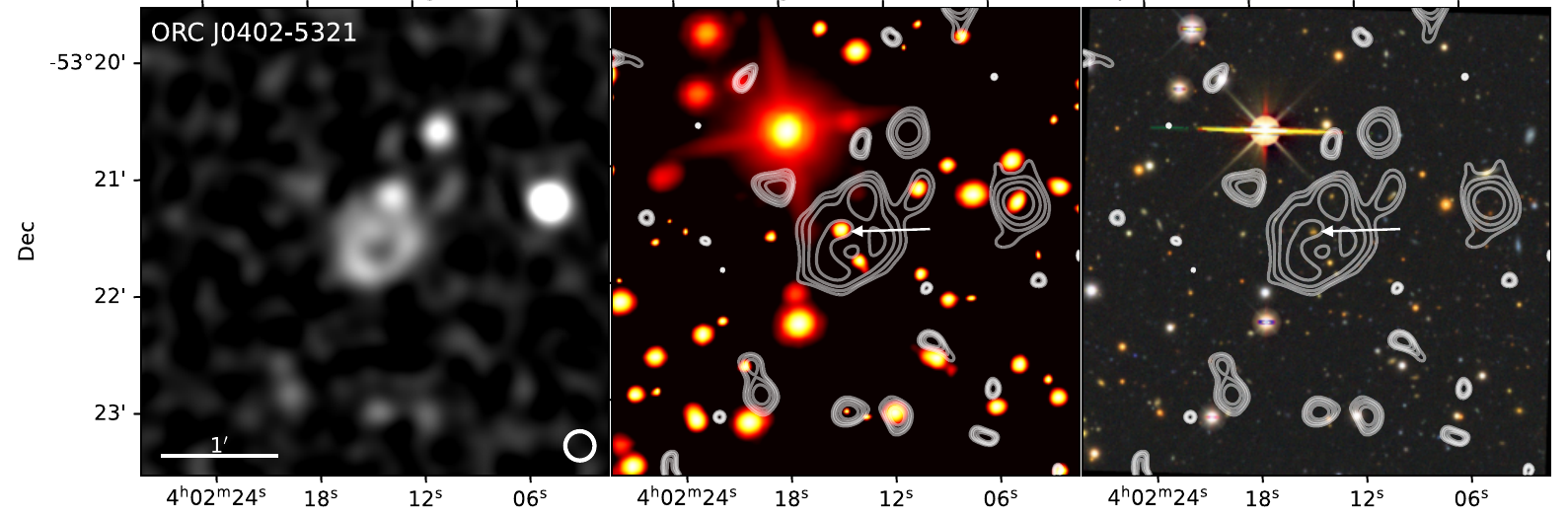}
\includegraphics[trim=1.4cm 5.8cm 1.4cm 6.3cm, clip, width=18cm, scale=0.5]{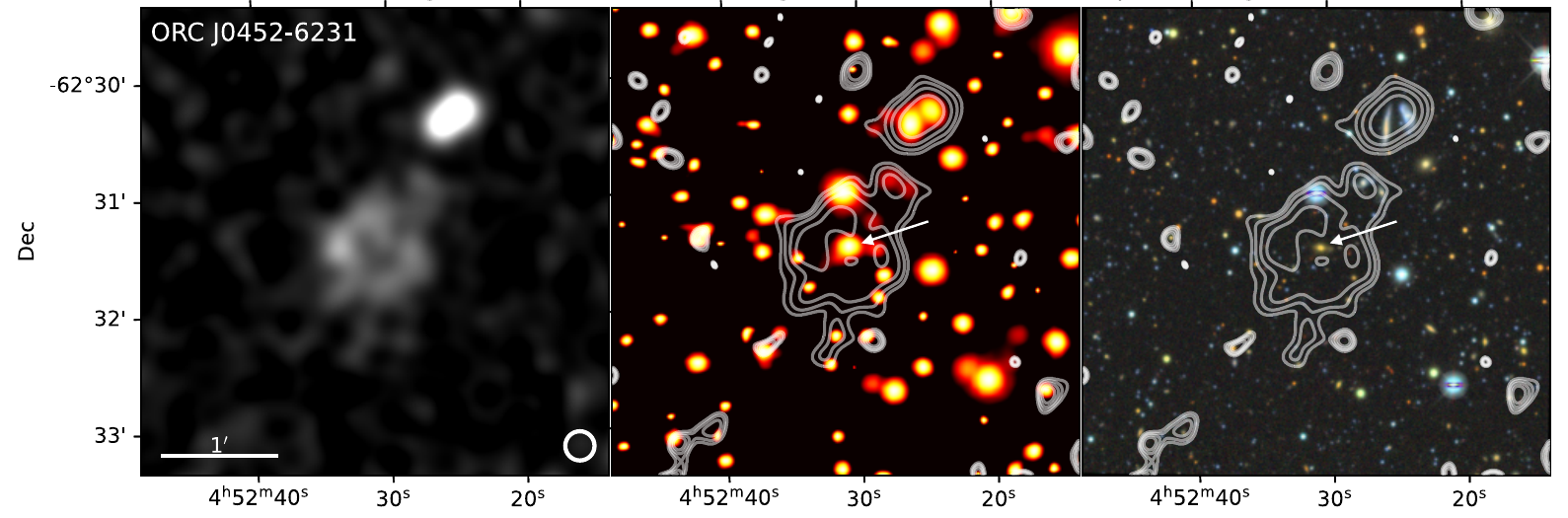}
\includegraphics[trim=1.4cm 5.8cm 1.4cm 6.3cm, clip, width=18cm, scale=0.5]{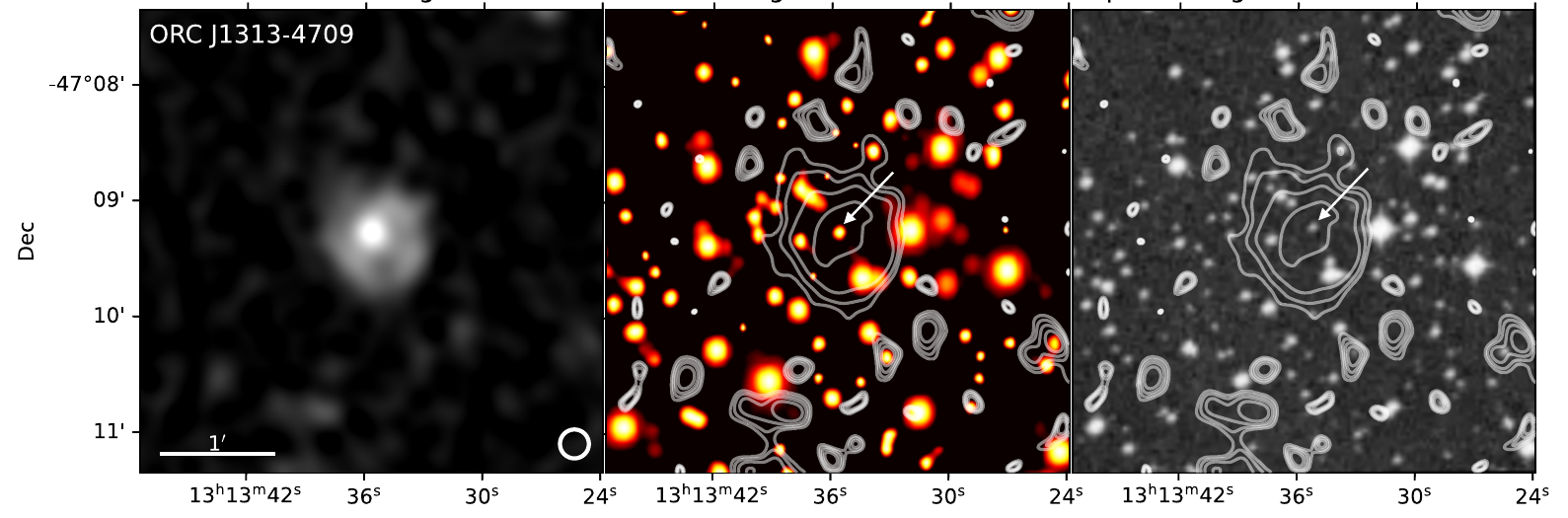}
\end{figure*}

\begin{figure*}[t!]
\centering
\vspace*{-0.2cm}
\includegraphics[trim=1.4cm 5.8cm 1.4cm 5.8cm, clip, width=18cm, scale=0.5]{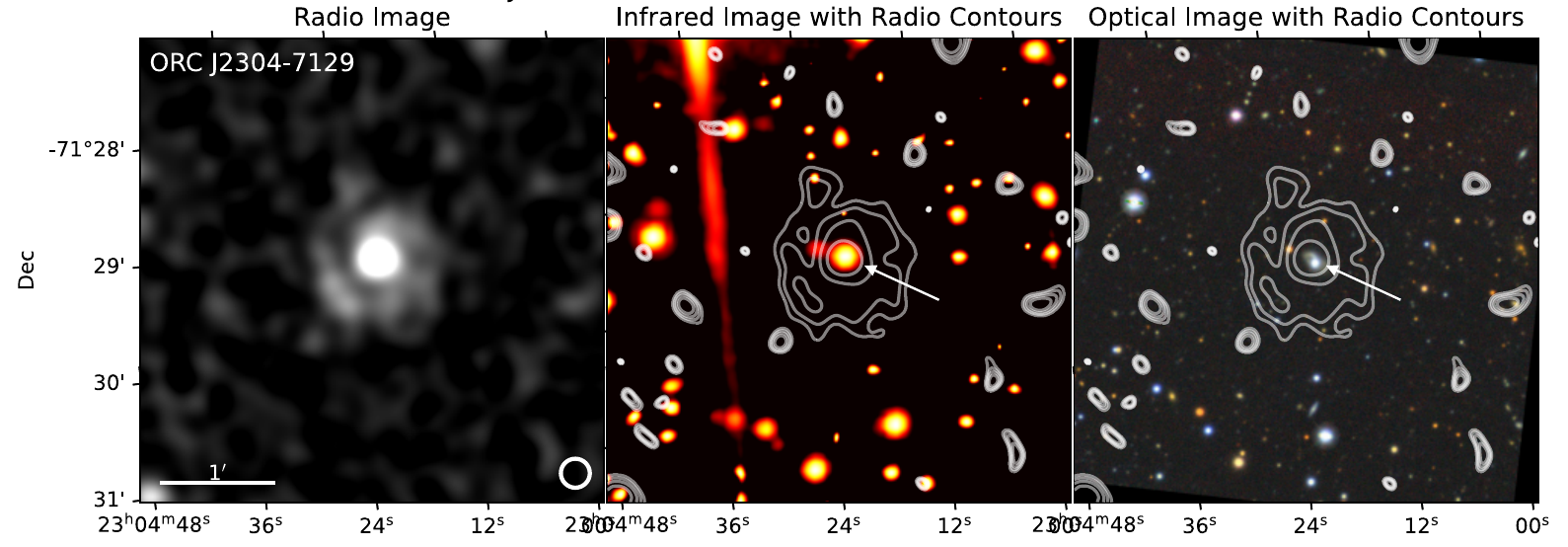}
\caption{ORCs in the first year of the EMU survey. The panels display radio images (left), corresponding infrared images (middle), and optical images (right), as indicated by the column titles. The infrared images are obtained from the AllWISE survey W1 band, while the optical images are primarily from DESI LS DR10. The optical image for ORC J1313-4709 is sourced from the DSS2 survey due to the lack of DESI LS DR10 coverage for this region. Each image has a frame size of $4^{\prime} \times 4^{\prime}$ on the sky, and the beam size is shown in the bottom right corner of the radio images. These ORCs exhibit edge-brightened radio emission surrounding distant host galaxies (marked with white arrows) and show no detectable emission at other wavelengths beyond their host galaxies. 
Table~\ref{TAB:ORC-counterparts} provides details about the host galaxies of these ORCs.}
\label{FIG:ORCs1}
\end{figure*}


\section{Odd Radio Circles}
\label{SEC:ORCs}
Odd Radio Circles (ORCs), first identified by \citet{norris21b} in ASKAP's EMU PS1 and later observed in ASKAP and other radio surveys \citep[][]{norris21b, koribalski21, gupta22, norris24}, are extragalactic, edge-brightened circles of radio emission surrounding distant host galaxies. 
These structures typically lack detectable emission at other wavelengths beyond their host galaxy but may exhibit diffuse radio emission within the bright ring structure \citep[][]{norris24}. 
This definition excludes related phenomena such as radio rings originating from the relic lobes of double-lobed radio galaxies \citep[][]{norris21b, omar22}, and diffuse radio emissions around galaxies that lack a well-defined ring structure \citep[][]{kumari24a, kumari24b}.
Following this definition, five ORCs are currently known: ORC J2103-6200 and ORC J1555+2726, presented in \citet{norris21b}; ORC J0102-2450, in \citet{koribalski21}; ORC J2223–4834, in \citet{gupta22}; and ORC J0219–0505, in \citet{norris24}.
Additionally, several ORC candidates and ORC-like radio shell systems have been discovered \citep[][]{gupta22, lochner23, dolag23, koribalski24a, koribalski24b}.

In this paper, we present five additional ORCs, characterized by edge-brightened circles of radio emission, each associated with a distant galaxy.
Table~\ref{TAB:ORC-counterparts} presents the properties of the potential host galaxies, including their infrared/optical names, positions, $r$-band foreground extinction magnitude, $griz$ AB magnitudes, W1, W2 and W3 Vega magnitudes from WISE survey, spectroscopic and photometric redshifts, integrated flux and luminosity, host flux and luminosity, ORC size, star formation rate (SFR), and the EMU tile SBID where the ORC is located.
The Galactic latitudes of these sources are consistent with an extragalactic origin. Furthermore, estimates of foreground extinction, derived from mean E(B–V) reddening values\footnote{S\&F values from: \url{https://irsa.ipac.caltech.edu/applications/DUST/}} based on Infrared Astronomical Satellite (IRAS) 100 micron data \citep{schlegel98} and adopting an extinction-to-reddening ratio of 2.176 \citep{schlafly11} for the Dark Energy Survey (DES) r-band, are likewise consistent with expectations for extragalactic sources. Additionally, we searched for the potential counterparts among the 3,709 planetary nebulae in the HASH catalogue\footnote{\url{http://202.189.117.101:8999/gpne/index.php}} \citep[][]{parker17} and the 310 Galactic supernova remnants in Green’s catalogue \citep[][]{green25}, but found no matches.
The $griz$ AB magnitudes in the 10th data release of the Dark Energy Spectroscopic Instrument's Legacy Imaging Surveys \citep[DESI LS DR10\footnote{\url{https://www.legacysurvey.org/dr10/}};][]{schlegel21}, are derived using the Tractor package \citep{lang2016tractor}.
The angular size was determined by visually inspecting each ORC in CARTA\footnote{\url{https://cartavis.org/}} and using its built-in ruler tool to measure the largest diameter, in order to account for the slightly asymmetric shapes of these systems. The uncertainty in angular size reflects a 5\% systematic manual error and is propagated in quadrature with redshift uncertainties when calculating the physical size.
All redshift dependent quantities are calculated using the $z_{\rm sp}$ from \citet{jones09} for ORC J2304-7129 and using $z_{\rm ph}$ from DESI LS DR9 \citep{zhou21} for the rest.
Additionally, we provide photometric measurements from \citet{zhou21}, \citet{bilicki16}, \citet{duncan22}, \citet{zou22}, \citet{wen24}, and \citet{zhou25}, which yield consistent estimates.
The flux density is measured in CARTA using the polygon tool to enclose the ORC boundaries. The associated uncertainty is estimated as $\sigma_{\rm rms} \sqrt{N_{\rm beam}}$, where $\sigma_{\rm rms}$ represents the RMS noise determined from a nearby source-free region, and $N_{\rm beam}$ is the number of synthesized beams covering the source region, both derived from CARTA measurements. An additional 5\% systematic uncertainty is incorporated in quadrature to account for potential manual measurement errors. The flux density and redshift uncertainties are then propagated in quadrature to estimate the errors in luminosity and star formation rate.
Figure~\ref{FIG:ORCs1} displays these ORCs.
Each row in these figures corresponds to an individual ORC, with the first column showing the radio image, the second column displaying radio contours overlaid on an infrared image, and the third column showing radio contours overlaid on an optical image.
This highlights the absence of corresponding extended circular emission in the infrared and optical wavelengths, which is clearly observed in the radio data.
Following the analysis of three ORCs presented in \citet{norris22}, we present colour–colour diagrams for the host galaxies of the five ORCs discussed in this work (Figure~\ref{FIG:color_color}). The top panel displays the WISE colours of the two ORC host galaxies with available W3 band measurements, plotted on a WISE colour–colour diagram adapted from Wright et al. (2010). Since the spectral energy distributions of these galaxies are not known, no k-corrections have been applied to these colours. The bottom panel shows the $gri$ colours of all five ORC host galaxies, plotted on a $gri$ colour–colour diagram adapted from Masters et al. (2011). These $gri$ colours, which are also not k-corrected, are taken from the DESI LS DR10 catalogues.

\subsection{ORC J0210-5710}
This ORC has a potential host galaxy at its centre, surrounded by edge-brightened circular emission in the radio band and has similar radio morphology as ORC1 in \citet{norris21b}.
The total integrated radio flux at 944 MHz is estimated to be $3.55\pm0.43$ mJy, and the size of this system is $438\pm25$ kpc. 
The central galaxy, WISEA J021009.39-571038.3, has counterparts in the DESI LS DR10.
The $griz$ AB magnitudes from DESI LS DR10 represent the brightness of the potential host galaxy in different optical bands. 
The galaxy is fainter in the $g$-band (21.04) compared to the $r$ (19.25), $i$ (18.61), and $z$ (18.26) bands, indicating that it is likely an elliptical or red-sequence galaxy. 
The brighter $z$-band magnitude may reflect rest-frame ultraviolet or optical light that has been redshifted into the infrared. 
This is confirmed by the photometric redshift estimate of $0.410 \pm 0.018$ from DESI LS DR9 \citep{zhou21}.
\citet{zou19} estimated stellar mass $\log M_*= 11.44~M_{\odot}$.
The integrated host flux is $0.15\pm0.02$ mJy, with a corresponding luminosity of $[5.3\pm0.8]\times10^{22}~\rm W~Hz^{-1}$ using DESI LS DR9 redshift and assuming a spectral index of -0.7.
Using the luminosity-SFR relation from Equation 17 in \citet{murphy11}, the star formation rate (SFR) at 944 MHz is estimated to be $44.5\pm7.0$ $M_{\odot} \rm yr^{-1}$. 
Note that this estimate assumes all radio emission is solely associated with recent star formation, and therefore represents an upper limit on the SFR. In contrast, SFR estimates derived from optical spectra \citep[e.g.,][]{rupke24} are more sensitive to unobscured, recent star formation and should be investigated in future studies.

The WISE colours shown in the top panel of Figure~\ref{FIG:color_color} indicate that the galaxy lies near the region occupied by spiral galaxies; however, its high redshift means that applying a k-correction could shift its position closer to that of LIRGs. The $gri$ colours in the bottom panel suggest that the galaxy is redder than typical star-forming galaxies. 
In addition to this central host galaxy, there is a bright radio source located over 1$^{\prime}$ to the northeast edge of the circular ring. This source has a much higher redshift of $0.745 \pm 0.029$ (DESI LS DR9) and is therefore unlikely to be related to the ORC system.
Additionally, a few faint galaxies with similar colours as the central galaxy are visible near the position of the radio ring, which may have contributed to the observed radio emission.

\subsection{ORC J0402-5321}
\begin{figure}[t!]
\centering
\includegraphics[trim=17cm 4cm 7cm 6.4cm, clip, width=12cm, scale=0.5]
{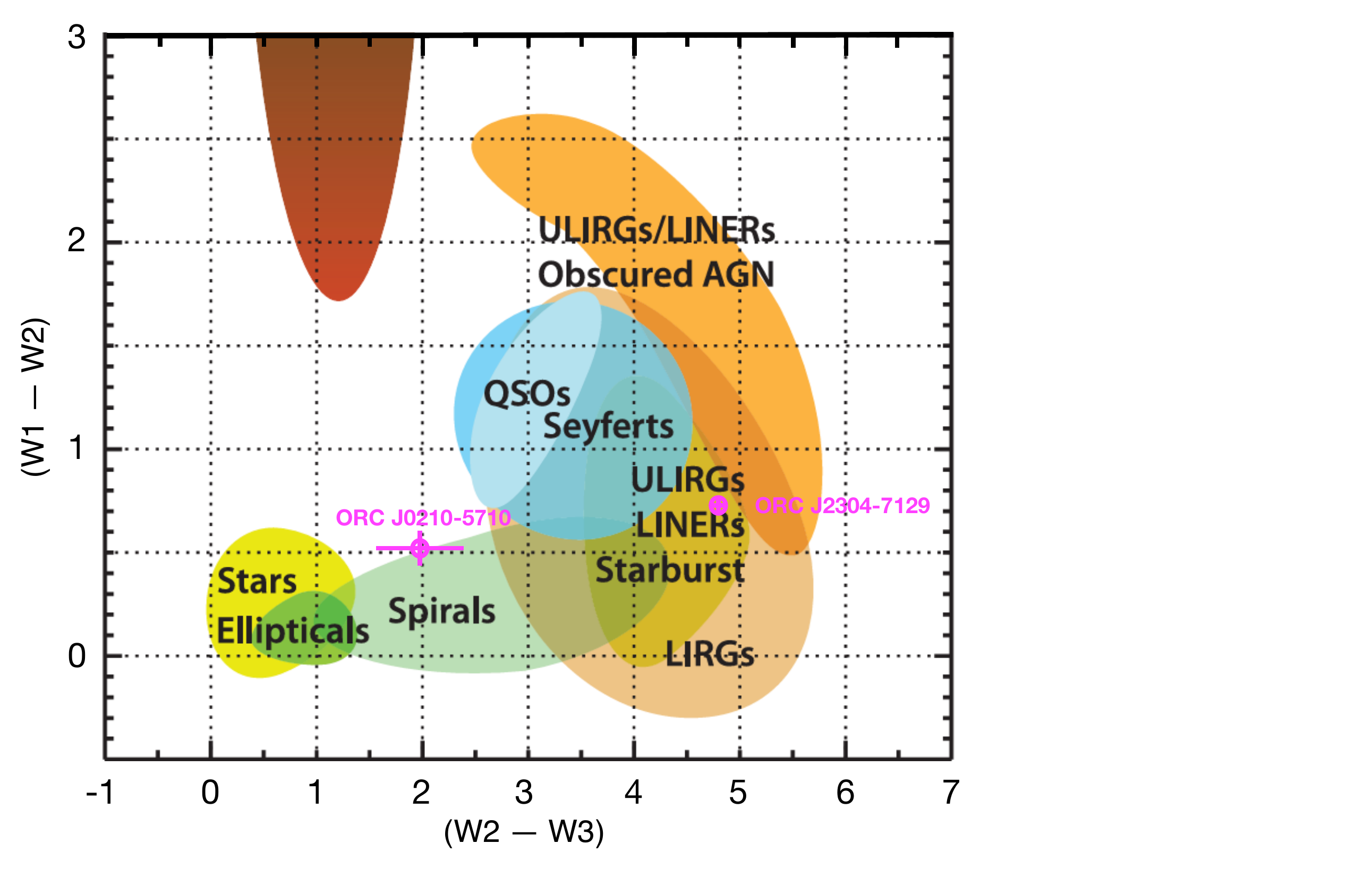}
\includegraphics[trim=8cm 0cm 9cm 0cm, clip, width=8.2cm, scale=0.5]
{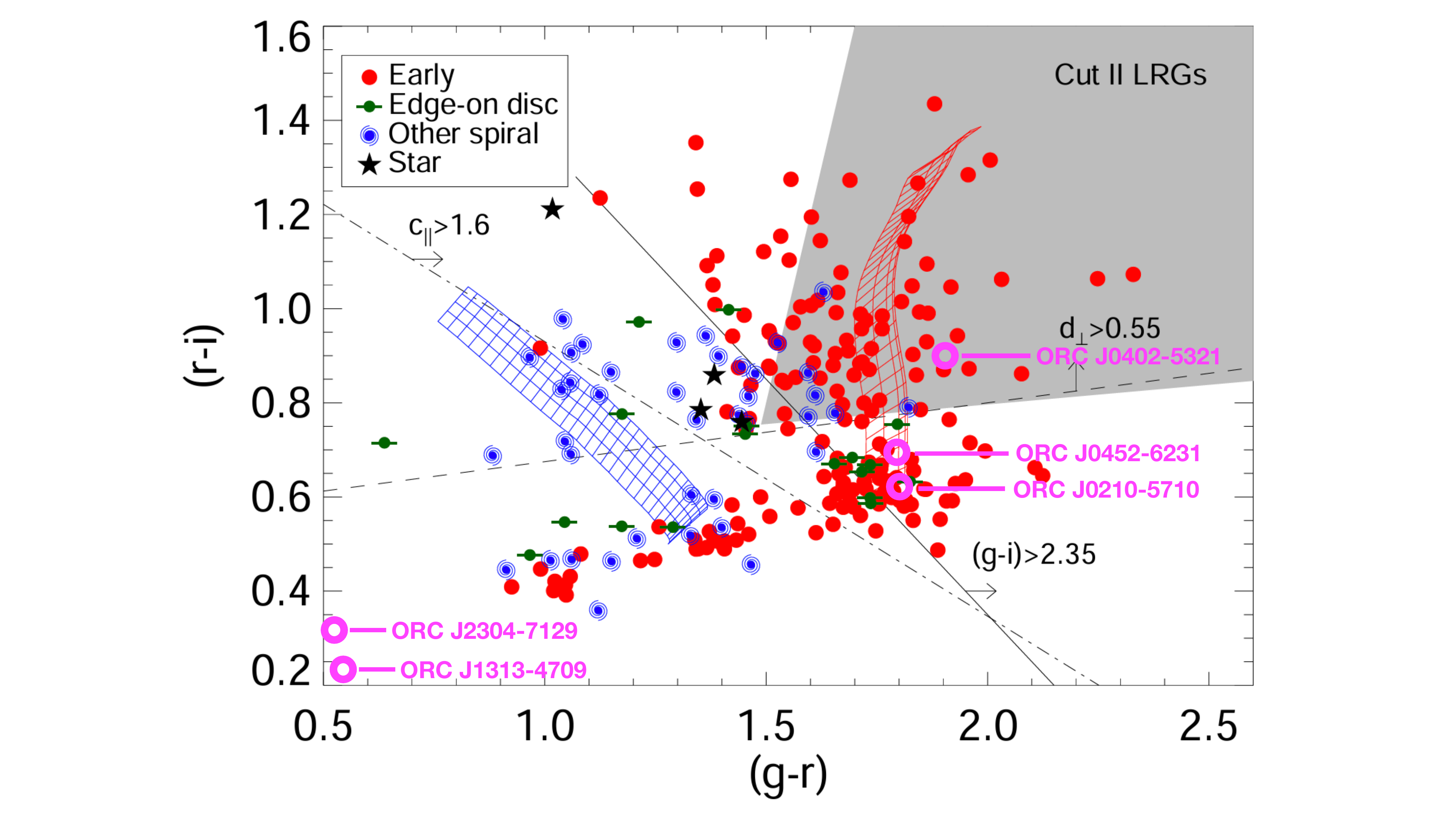}
\caption{The top panel shows the WISE colours of the two ORC host galaxies, for which W3 band measurements are available, plotted on the WISE colour–colour diagram adapted from \citet{wright10}. The bottom panel displays the $gri$ colours of all host galaxies of the five ORCs, plotted on the $gri$ colour–colour diagram adapted from \citet{masters11}. The $gri$ photometry is sourced from the DESI LS DR10 catalogues.}
\label{FIG:color_color}
\end{figure}

This ORC system features an edge-brightened circular ring with three galaxies within the region of radio emission, exhibiting a morphology similar to the ORC described in \citet{koribalski21}.
The total integrated radio flux at 944 MHz is estimated to be $1.55\pm0.11$ mJy, and the size of this system is $260\pm18$ kpc. 
The potential host galaxy, WISEA J040214.70-532128.6, is located near the centre of the radio emission.
The $g - r$ colour of 1.91 and $r - i$ colour of 0.90, as shown in the bottom panel of Figure~\ref{FIG:color_color}, underscore the red nature of the galaxy. 
These colours are typical of galaxies with older stellar populations, such as elliptical or lenticular galaxies.
The photometric redshift reported in DESI LS DR9 is $0.536 \pm 0.050$. The W1-W2 colour is consistent with that of an early-type galaxy.
The integrated host flux is $0.08\pm0.02$ mJy, with a corresponding luminosity of $[4.6\pm1.4]\times10^{22}~\rm W~Hz^{-1}$, and the SFR is estimated to be $38.9\pm11.8$ $M_{\odot} \rm yr^{-1}$.
In addition to the potential central host galaxy, two more galaxies are located near the circular boundary of the radio emission. 
These are WISEA J040213.71-532144.9 (toward the south-west) and WISEA J040213.69-532109.8 (toward the north-west). 
Both galaxies exhibit similar colours and consistent redshifts within the error bars as the central galaxy, with redshifts of $0.346 \pm 0.183$ and $0.574 \pm 0.034$, respectively.
This suggests that the radio emission from these galaxies may have contributed to the circular structure of this ORC.

\subsection{ORC J0452-6231}
This ORC exhibits nearly circular bright edges in the radio regime with a clear central galaxy.
The total combined radio flux at 944 MHz is estimated to be approximately $2.06\pm0.15$ mJy, and the system has a size of $300\pm22$ kpc.
The potential host galaxy at the centre, WISEA J045221.40-623422.7, has a counterpart in DESI LS DR10, as shown in Figure~\ref{FIG:ORCs1}.
The increase in brightness from $g$ to $z$, as presented in Table~\ref{TAB:ORC-counterparts}, suggests that the galaxy is likely a red-sequence galaxy, typically elliptical or lenticular, where star formation has largely ceased.
The estimated photometric redshift is 0.3891 (DESI LS DR9).
The integrated host flux is $0.04\pm0.01$ mJy, with a corresponding luminosity of $[1.6\pm0.4]\times10^{22}~\rm W~Hz^{-1}$, and the SFR is estimated to be $13.0\pm3.3$ $M_{\odot} \rm yr^{-1}$.
\citet{zou19} estimated stellar mass $\log M_*= 11.41~M_{\odot}$.
The WISE magnitudes (W1 = $15.043\pm0.028$, W2 = $14.921\pm0.043$) are relatively faint and the low W1–W2 colour ($0.122\pm0.051$) aligns with the galaxy’s low star formation activity.
The $gri$ colours of this galaxy place it in the redder region of the diagram shown in the bottom panel of Figure~\ref{FIG:color_color}.
Aside from the central galaxy, no other galaxies at a similar redshift are observed within the region of radio emission.

\subsection{ORC J1313-4709}
This system exhibits bright central radio emission along with edge-brightened circular emission. 
The total integrated flux at 944 MHz is measured as 3 mJy from the radio image. 
A central galaxy, WISEA J131335.54-470915.2, is visible in the infrared and optical images shown in Figure~\ref{FIG:ORCs1}. 
Due to the lack of coverage in the DESI LS DR10, we present the optical image from the Digitized Sky Survey \citep[DSS;][]{lasker90}. 
The $gri$ colours of this galaxy, shown in the bottom panel of Figure~\ref{FIG:color_color}, suggest it is moderately blue, indicating ongoing star formation.
An unusual negative W1-W2 value indicates that the galaxy's flux decreases in the mid-infrared bands from W1 to W2. 
This indicates minimal influence from warm dust and active star formation, while the radio emission points to significant AGN activity.
The faint infrared magnitudes further imply that the galaxy may either be at a moderate-to-high redshift or possess a lower stellar mass. 
Additional sources of emission seen in the infrared and optical images, particularly toward the northeast and southwest, may also be associated with the ORC. 
Improved optical coverage of this region would provide further insights into the nature of the optical sources within the area of the circular radio emission.

\begin{figure*}[t!]
\centering
\includegraphics[trim=1.4cm 5.8cm 1.4cm 5.8cm, clip, width=18cm, scale=0.5]
{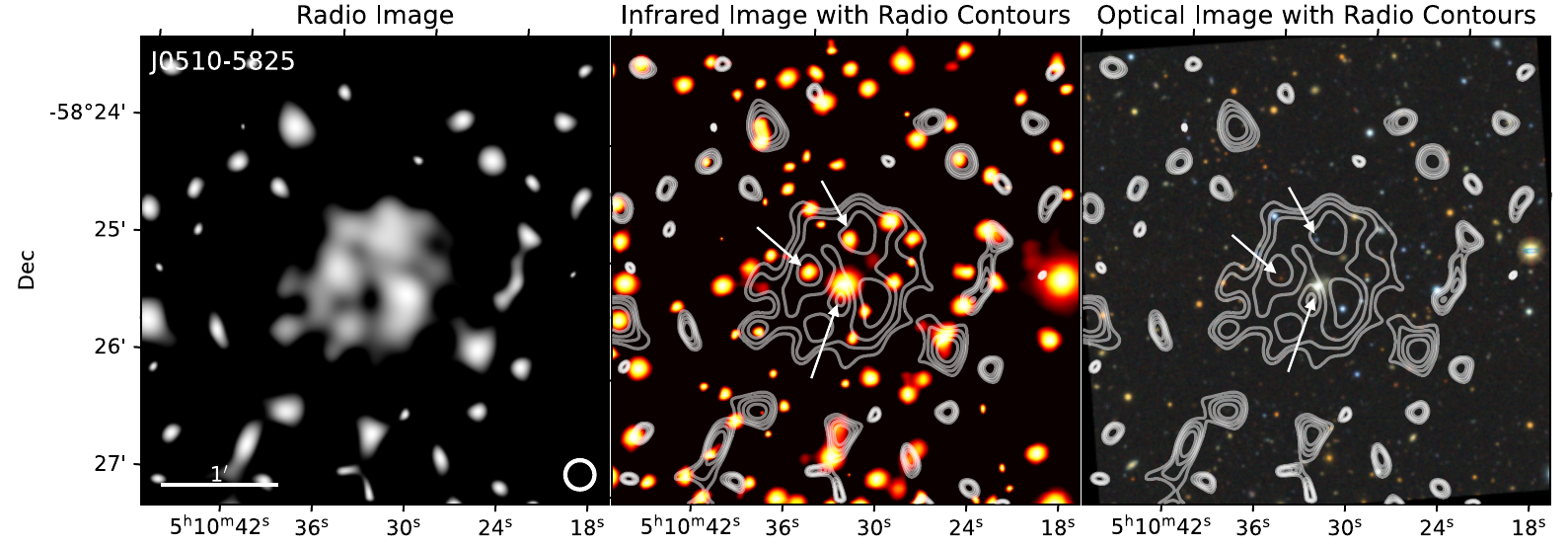}
\includegraphics[trim=0cm 0cm 0cm 1.2cm, clip, width=18cm, scale=0.5]{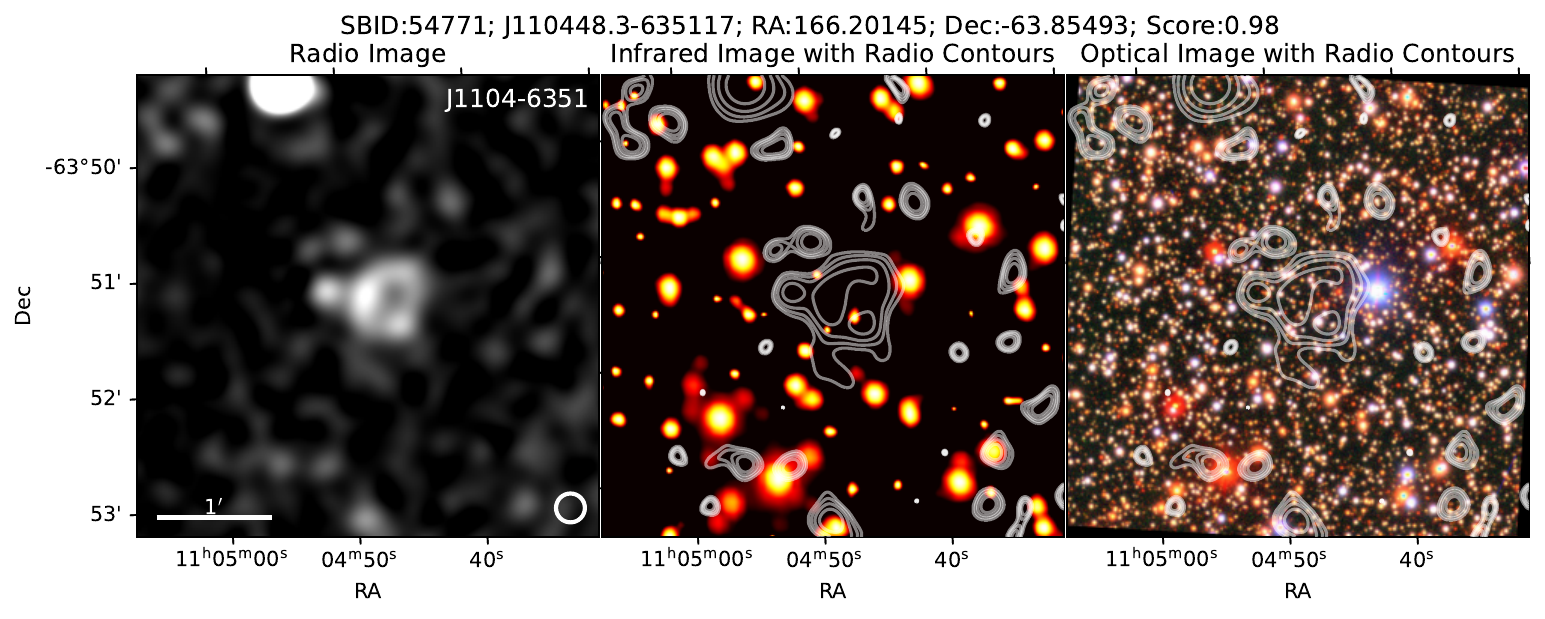}
\caption{ORC candidates with unconfirmed host galaxies are presented. Details about the panels are provided in Figure~\ref{FIG:ORCs1}. The corresponding optical images are sourced from the DECaPS2 survey due to the absence of DESI LS DR10 coverage. In the top panel, white arrows indicate one central galaxy and two additional galaxies at the locations of radio emission peaks, making it challenging to conclusively identify the origin of the circular emission with the current data. In the bottom panel, the high density of sources in the optical image complicates the confirmation of a host galaxy.} 
\label{FIG:ORC_Candidates}
\end{figure*}

\subsection{ORC J2304-7129}
This ORC system exhibits edge-brightened circular radio emission with a potential host galaxy at its centre.
The total integrated radio flux at 944 MHz is estimated to be approximately $3.34\pm0.21$ mJy, and the system has a size of $160\pm8$ kpc.
The central galaxy, WISEA J230420.71-712907.8, has counterparts 2MASS J23042070-7129078 and 6dF J2304207-712907.
The colours ($g - r = 0.51$ and $r - i = 0.36$) displayed in the bottom panel of Figure~\ref{FIG:color_color} suggest that the galaxy is moderately blue, likely indicating it is a star-forming galaxy.
The WISE colours suggest the potential presence of both ongoing star formation and AGN activity, as evidenced in the radio image.
The estimated spectroscopic redshift of this galaxy is 0.1481 \citep[][]{jones09}.
The integrated host flux is $0.55\pm0.04$ mJy, with a corresponding luminosity of $[2.7\pm0.2]\times10^{22}~\rm W~Hz^{-1}$, and the SFR is estimated to be $23.0\pm1.7$ $M_{\odot} \rm yr^{-1}$.
Additionally, two more galaxies are visible to the northeast (WISEA J230423.51-712902.5) and northwest (WISEA J230419.53-712853.2) of the central galaxy.
However, these galaxies have redder colours and do not appear to be at the same redshift as the central source, suggesting they may not contribute to the observed circular radio emission.

\subsection{Other ORC Candidates}
In addition to the ORCs presented in the above subsections, we identify two more unconfirmed ORC candidates through visual inspections of the same 1,794 sources.
These ORC candidates are shown in Figure~\ref{FIG:ORC_Candidates}.
Here, we present radio, infrared, and optical images at the locations of these candidate ORCs.

The ORC candidate, J0510-5825, has edge-brightened circular radio emission with no counterpart in HASH and Greens's catalogue.
This candidate shares morphological similarities in radio continuum with SAURON \citep[][]{lochner23}, a complex, ring-like radio structure centred around a luminous red galaxy.
The total integrated flux at 944 MHz is measured as $2.5\pm0.2$ mJy from the radio image.
The central galaxy, WISEA J051032.87-582518.8 (longitude, latitude $ = 267.1155^{\circ}, -35.9488^{\circ}$ and $A_r=0.04$) is relatively bright across all four bands ($g$, $r$, $i$, and $z$), with the brightness increasing toward the redder bands.
The $g$-band magnitude (17.72) is slightly fainter than the $r$-band (17.08), $i$-band (16.87), and $z$-band (16.70).
The colours ($g - r = 0.64$ and $r - i = 0.21$) suggest the galaxy might be elliptical or lenticular, or it could belong to the "red sequence," indicating an older stellar population.
The photometric redshift reported in DESI LS DR9 is $0.120\pm0.059$.
The integrated host flux is $0.08\pm0.02$ mJy, with a corresponding luminosity of $[2.6\pm3.2]\times10^{21}~\rm W~Hz^{-1}$, and the SFR is estimated to be $2.2\pm2.7$ $M_{\odot} \rm yr^{-1}$.
Additionally, two more nearby galaxies are located at the peaks of the radio emission of the ORC.
These galaxies are approximately 0.5$^{\prime}$ towards the east (WISEA J051035.28-582513.7), and north (WISEA J051032.75-582455.6) of the central galaxy. 
Their redshifts ($1.059\pm0.099$ and $0.782\pm0.416$, respectively from DESI LS DR9) suggest that these galaxies may not be associated with the circular radio emission.
However, the coincidence of these galaxies with the peaks of the radio emission means we cannot conclusively determine that the circular emission originates solely from the central galaxy. 
Future high-resolution radio observations will be necessary to further investigate this system.

Another ORC candidate, J1104-6351, exhibits a morphology similar to that of ORC J0402-5321, shown in Figure~\ref{FIG:ORCs1}, as well as the ORC discovered by \citet{koribalski21}.
However, the identification of potential hosts for this system is challenging using infrared and optical images.
Due to the absence of DESI LS DR10 coverage for these sources, we include optical images from the Dark Energy Camera Plane Survey 2 \citep[DECaPS2;][]{saydjari23}.
The DECaPS2 optical image reveals several potential host galaxies near the centre of the ORC candidate. 
However, due to the high galaxy density and the absence of redshift information in this region, it is not possible to confirm whether the radio emission is associated with one or more of these galaxies.
Additionally, in contrast to all other ORCs listed in Table~\ref{TAB:ORC-counterparts} and the ORC candidate J0510–5825, this system lies much closer to the Galactic plane, at longitude of $291.51^{\circ}$, latitude of $-3.38^{\circ}$, and experiences significant foreground extinction with $A_r = 2.13$. 
However, similar to the ORCs and the candidate J0510–5825 presented in this work, it lacks a counterpart in both the HASH and Green’s catalogues, leaving its origin uncertain.
Future studies should further investigate this system to confirm its multi-wavelength counterpart.

\begin{figure*}[t!]
\centering
\includegraphics[trim=1.5cm 3.7cm 1.5cm 3.8cm, clip, width=9cm, scale=0.5]
{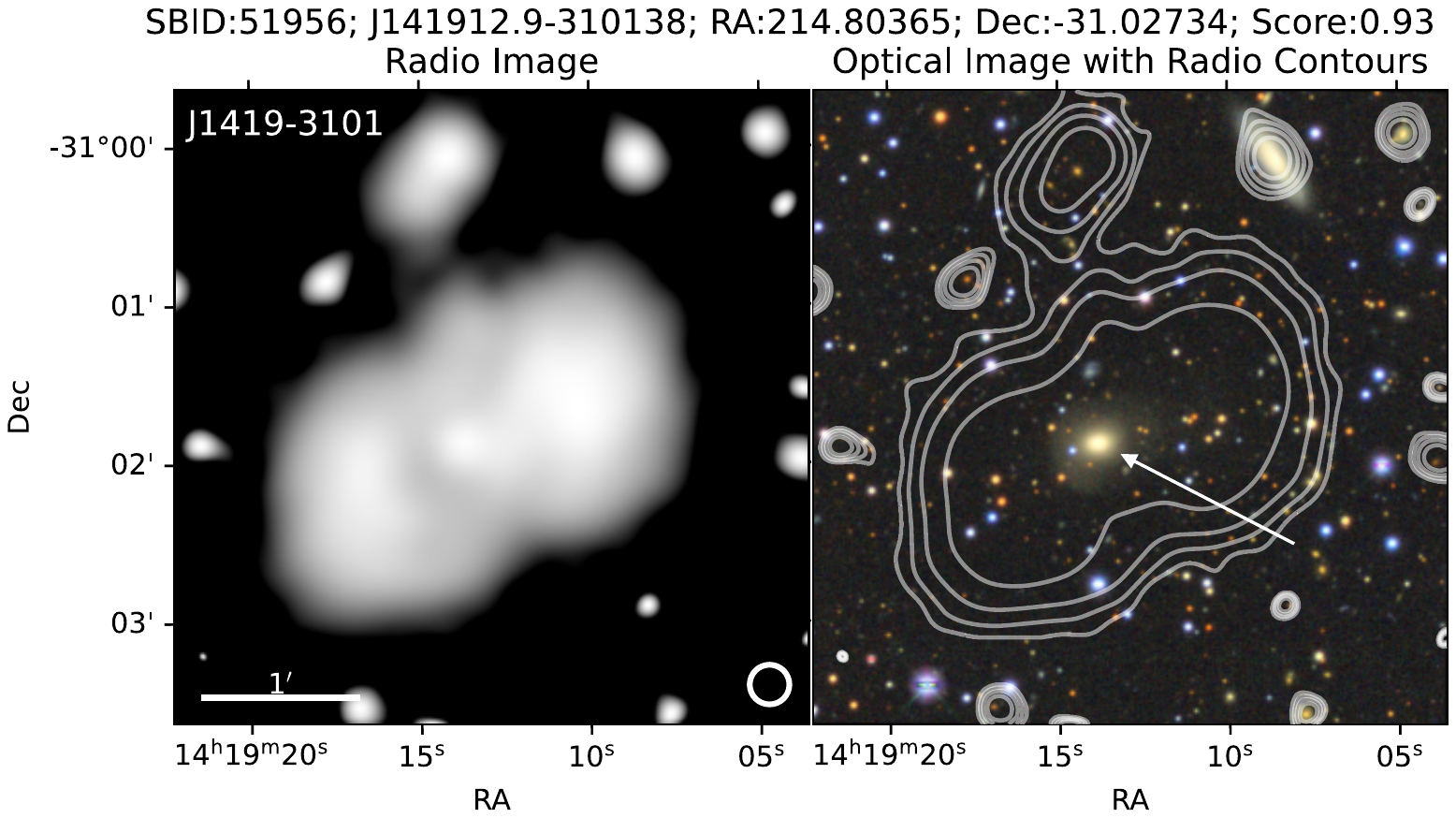}
\includegraphics[trim=1.5cm 3.7cm 1.5cm 3.8cm, clip, width=9cm, scale=0.5]{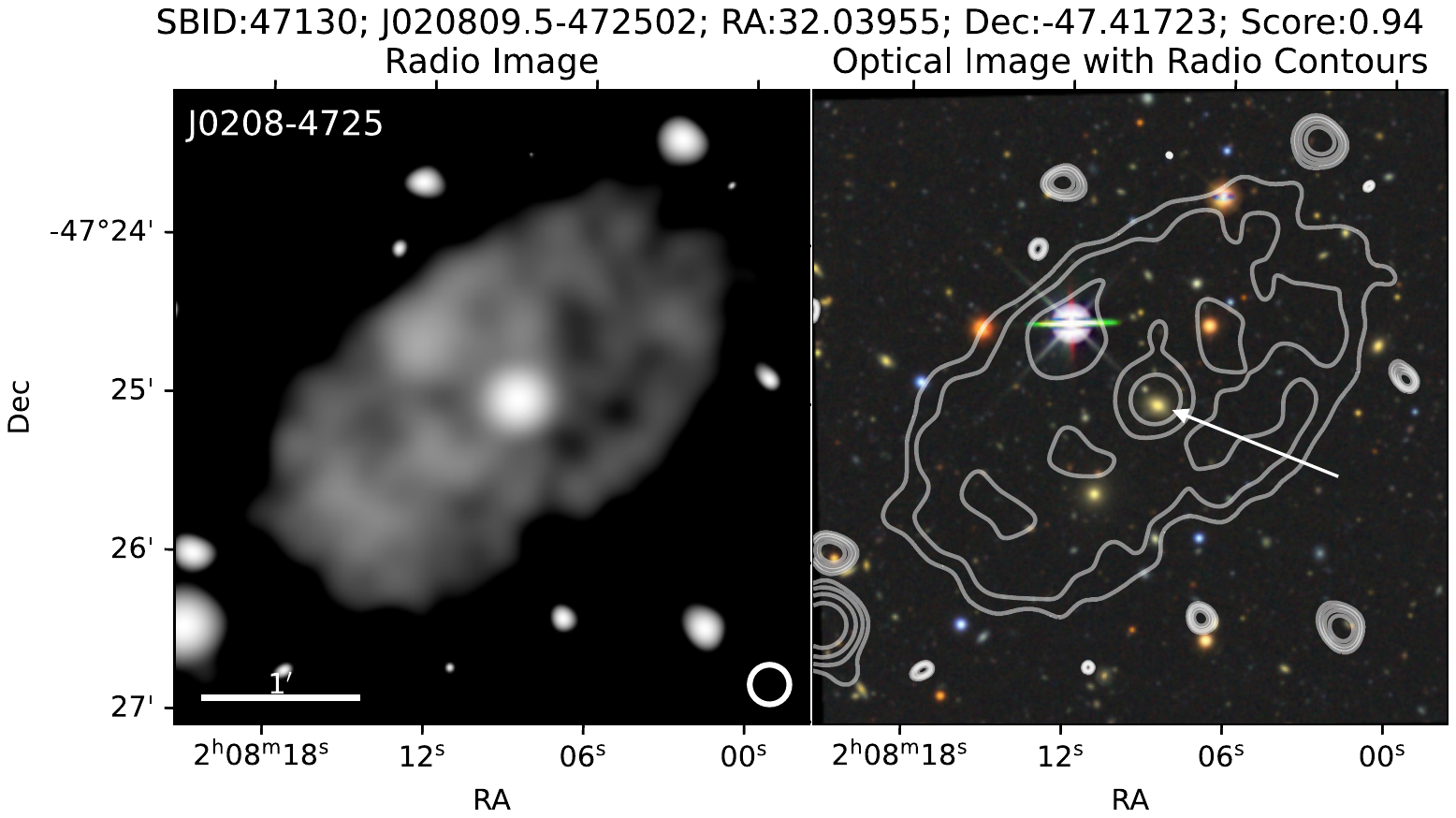}
\includegraphics[trim=1.5cm 3.8cm 1.5cm 4.5cm, clip, width=9cm, scale=0.5]{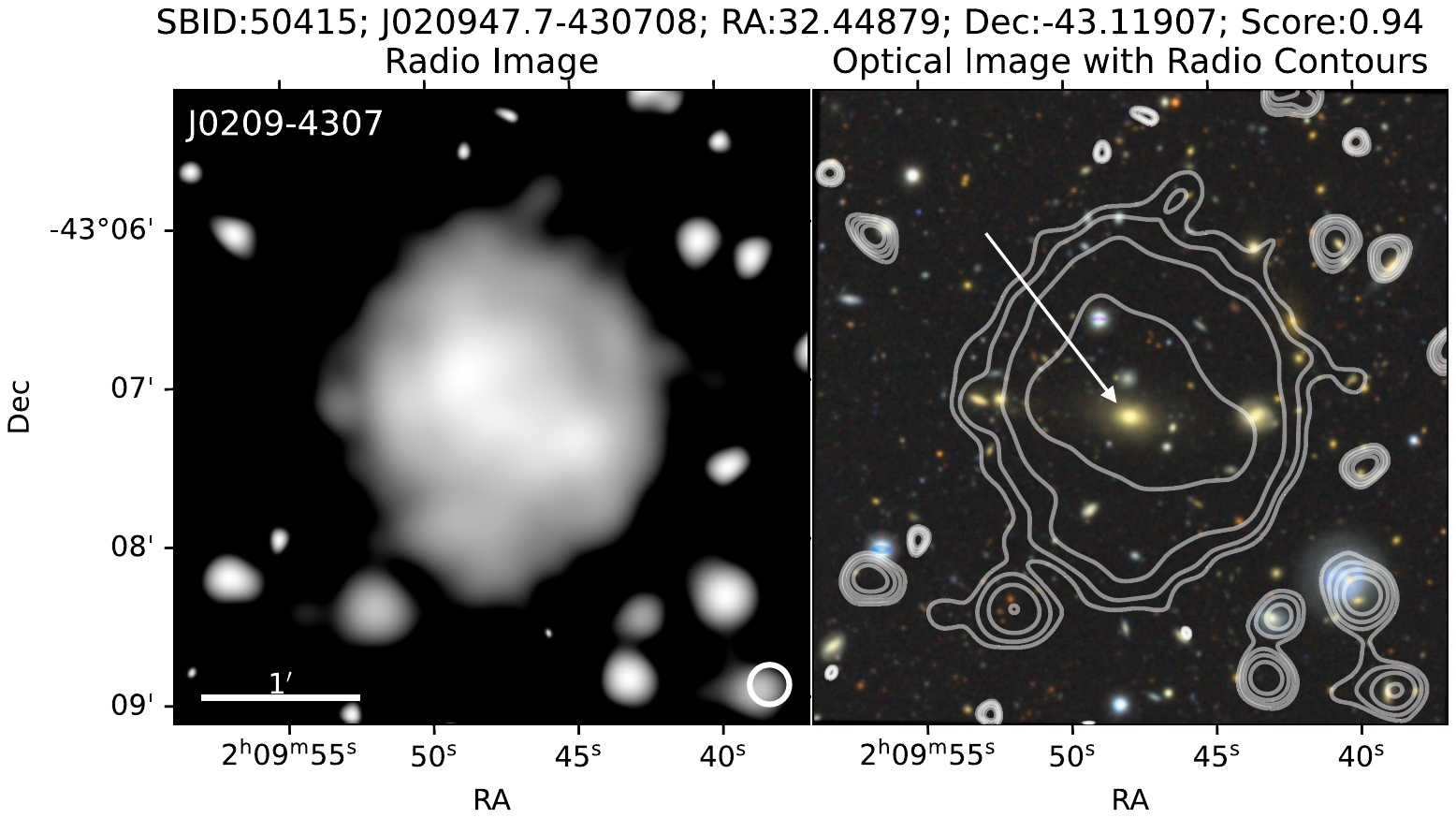}
\includegraphics[trim=1.5cm 3.8cm 1.9cm 4.6cm, clip, width=9cm, scale=0.5]{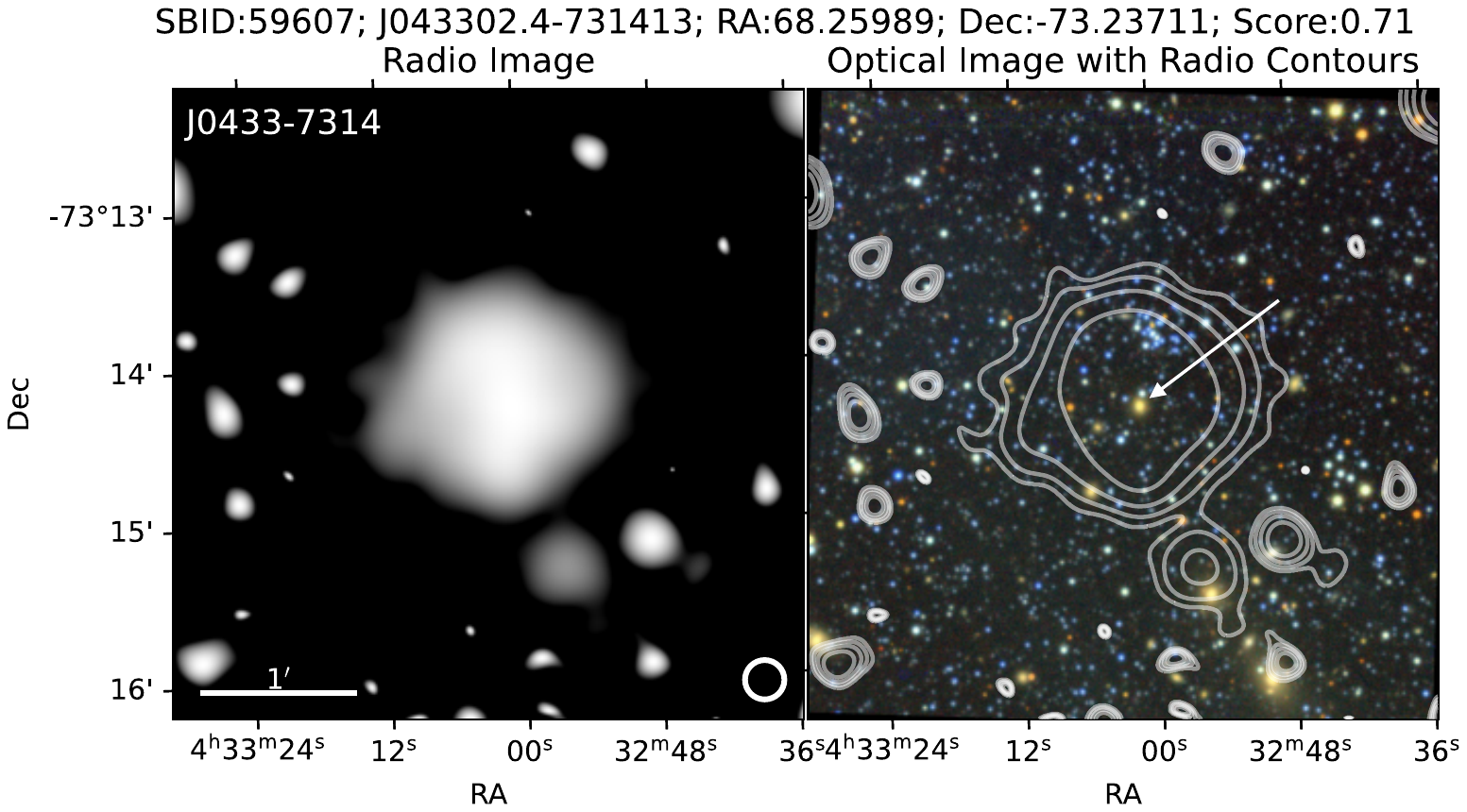}
\includegraphics[trim=1.5cm 3.8cm 1.5cm 4.5cm, clip, width=9cm, scale=0.5]{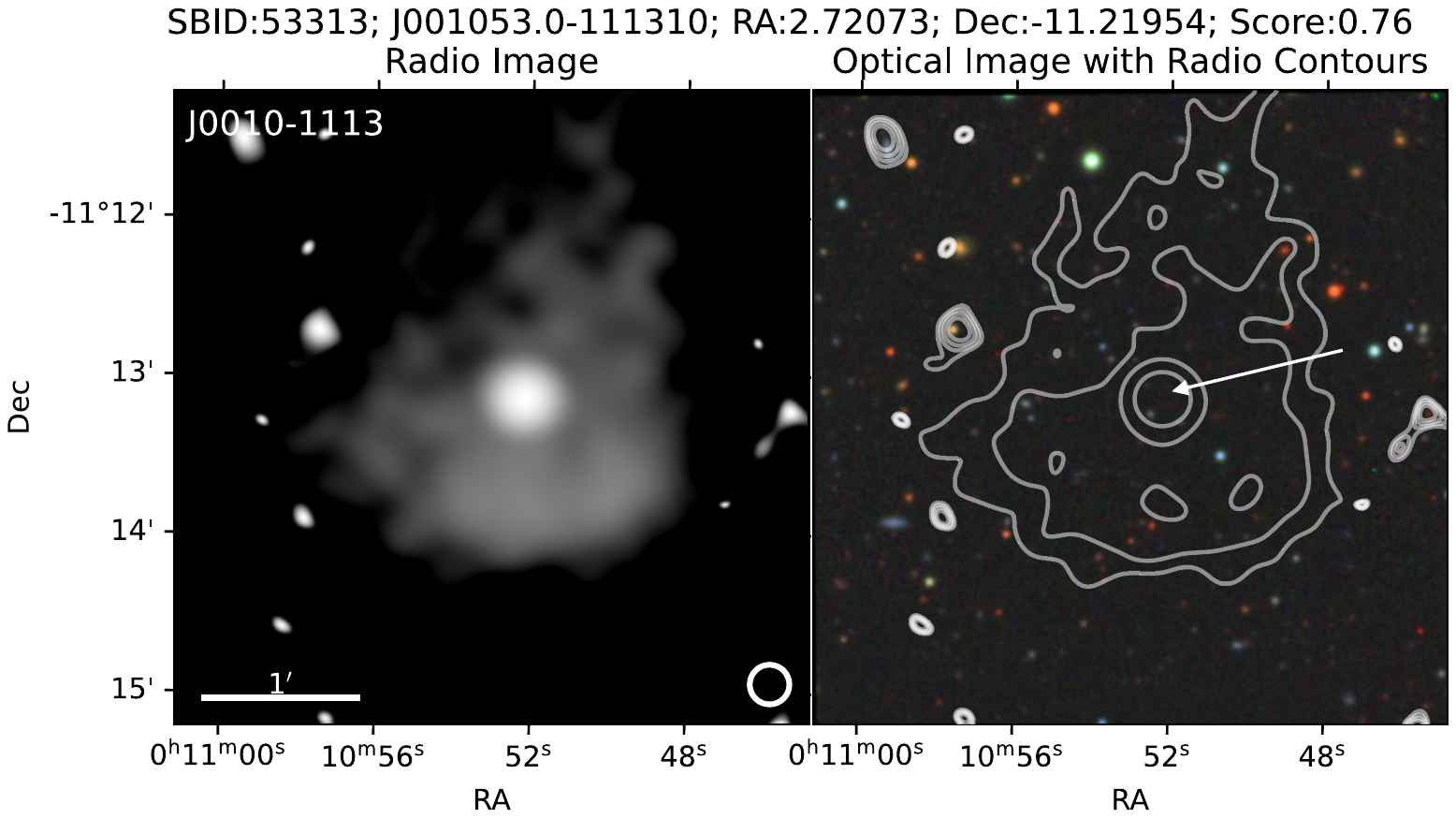}
\includegraphics[trim=1.5cm 3.8cm 1.5cm 4.5cm, clip, width=9cm, scale=0.5]
{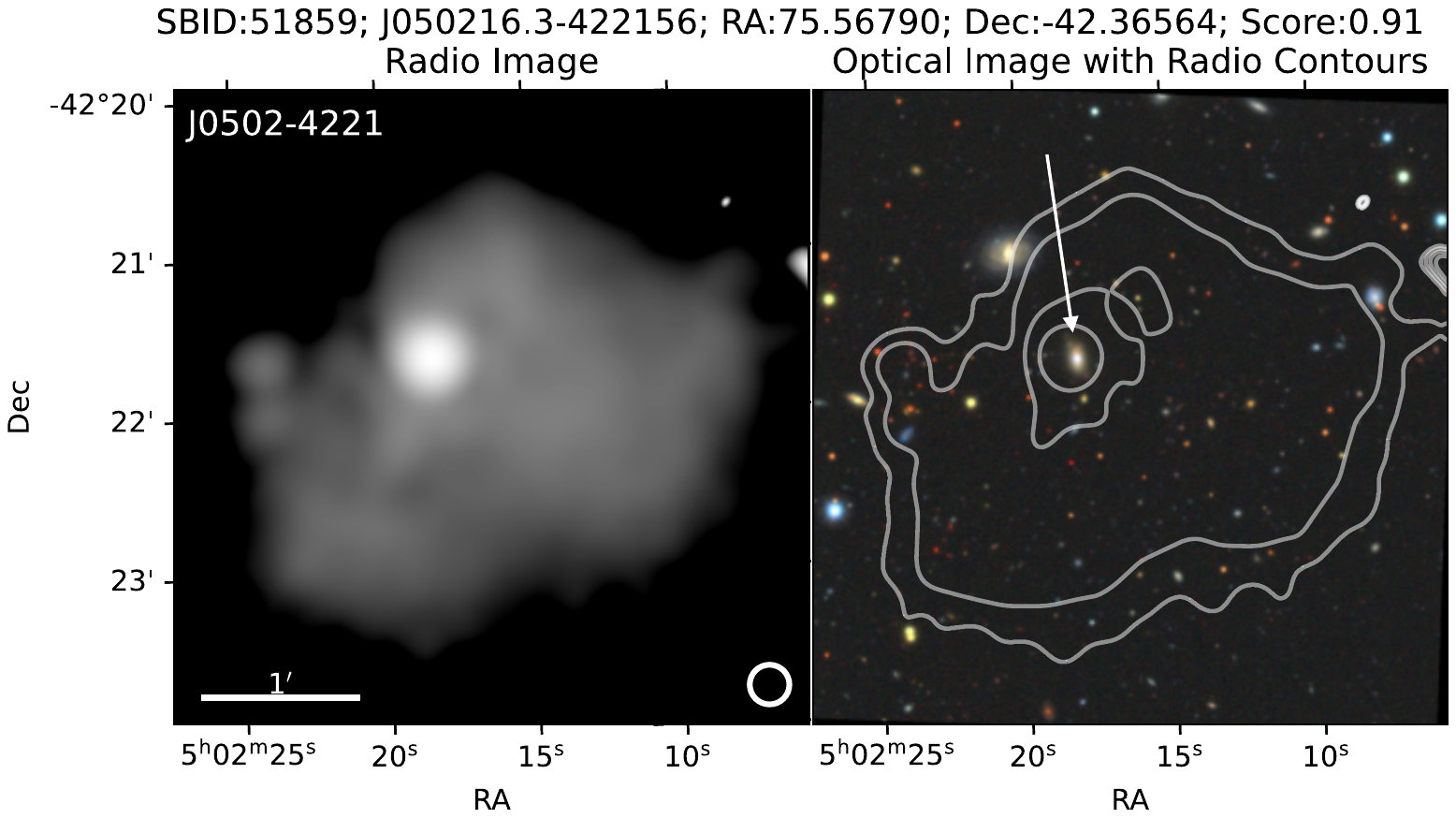}
\includegraphics[trim=1.5cm 2.5cm 1.5cm 4.5cm, clip, width=9cm, scale=0.5]{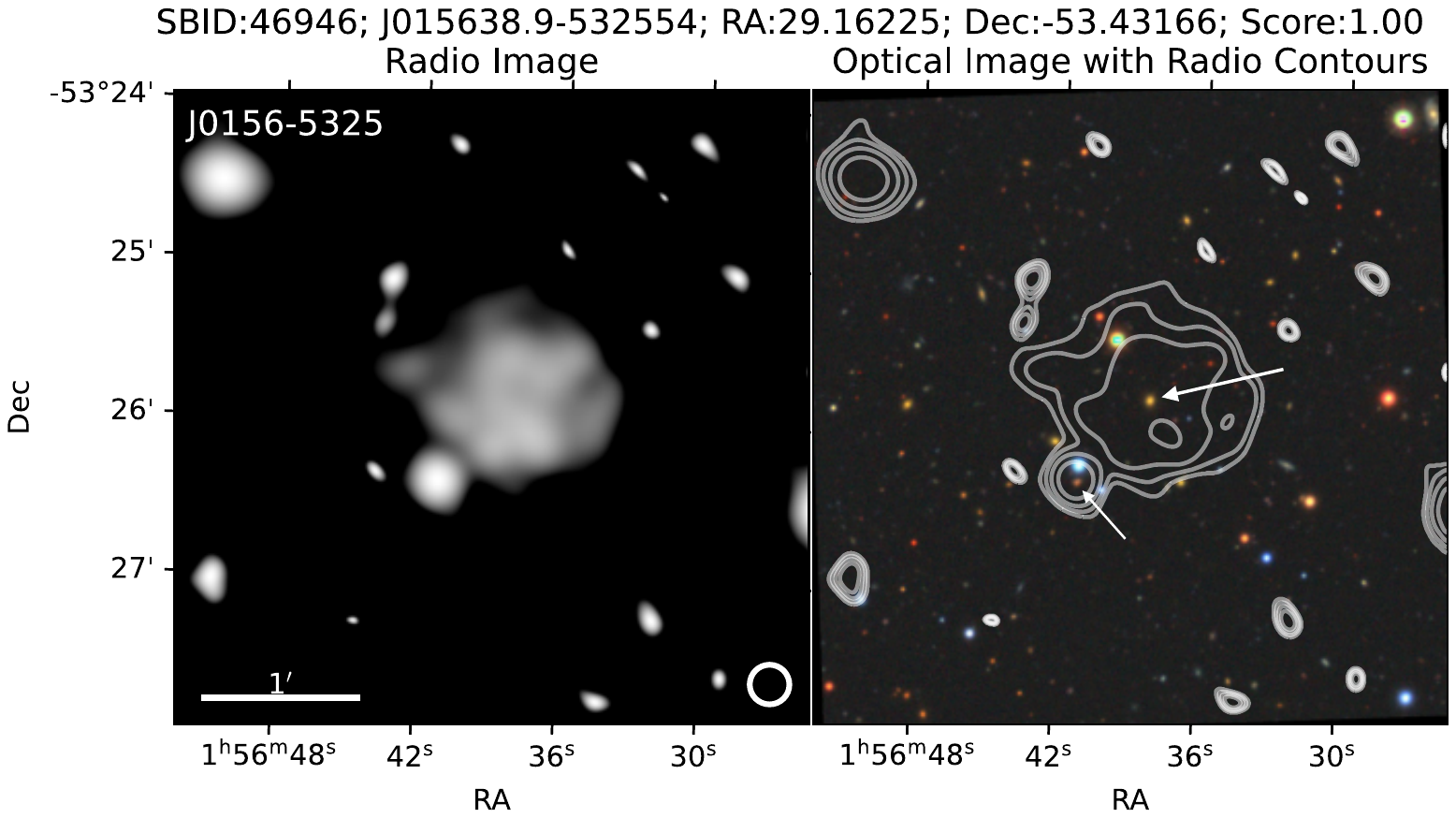}
\includegraphics[trim=1.5cm 2.5cm 1.5cm 4.5cm, clip, width=9cm, scale=0.5]{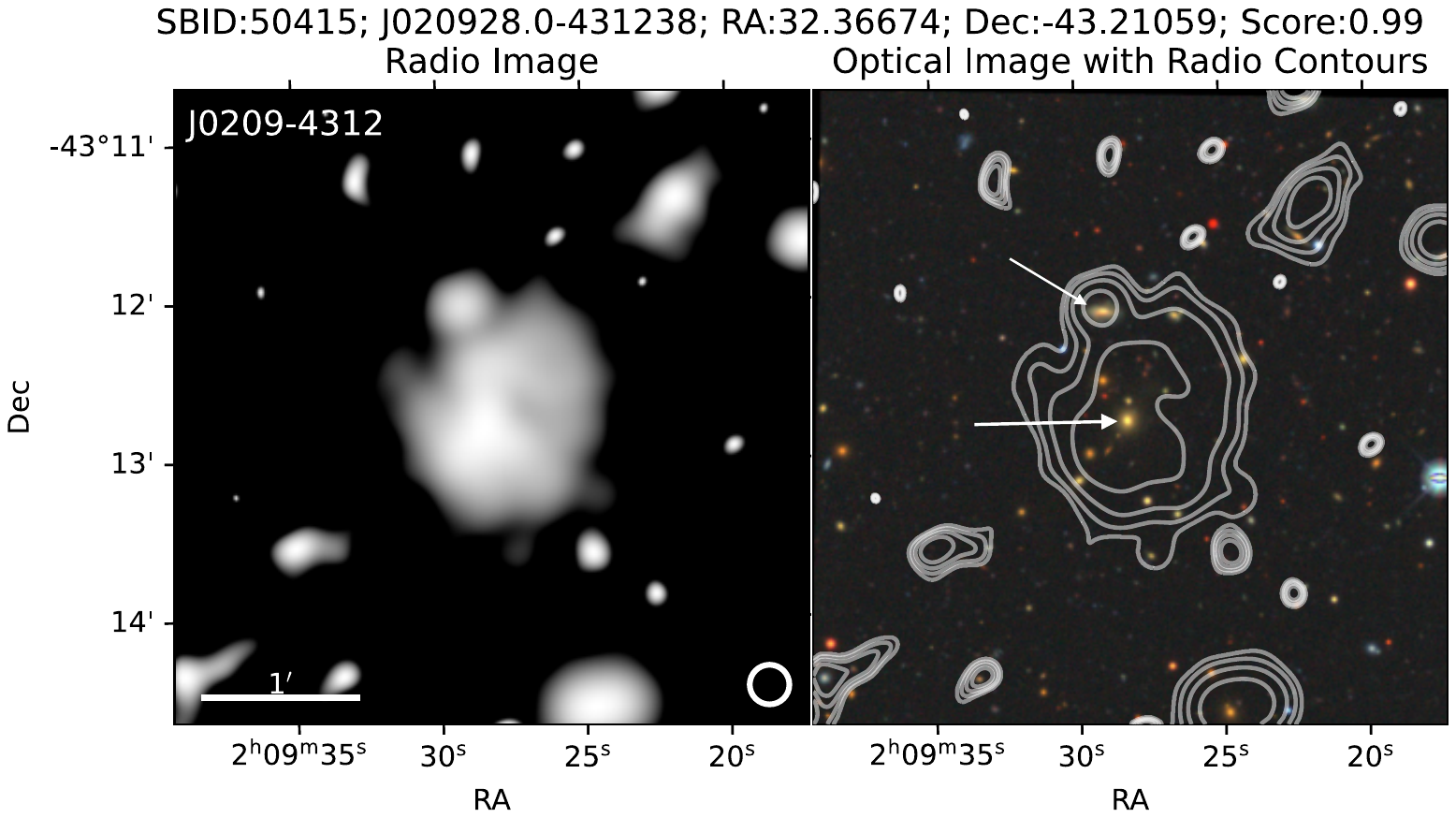}
\caption{Examples of Galaxies with Large-scale Ambient Radio Emission (GLAREs). The left panels of each column display radio images from the EMU survey, while the right panels present DESI LS DR10 images overlaid with radio emission contours in white. From top to bottom, each row displays two GLAREs featuring rectangular, circular, or irregularly shaped diffuse radio emissions surrounding their potential host galaxies, marked by white arrows. In the bottom panel, a smaller white arrow indicates a secondary galaxy. The final row features two GLAREs where the potential host galaxies are centrally located, and another likely unrelated galaxy near the edge of the radio emission.
While only a few examples are presented here, Table~\ref{TAB:GLAREs} provides a comprehensive list of all GLAREs identified in the first year of the EMU survey. These can also be viewed at \url{https://doi.org/10.25919/cvz8-4d27}. Diffuse radio emission around distant galaxies could offer insights into the origins of ORCs. Additionally, the intriguing morphologies of these systems merit further investigation to explore the physics behind their shape and potential connections with ORCs.} 
\label{FIG:GLAREs}
\end{figure*}

%

\section{Other Peculiars}
\label{SEC:OtherPecs}
While searching for ORCs using the criteria described in Section~\ref{SEC:method}, we identified several radio sources with peculiar radio morphologies. In this section, we present these radio sources, which include Galaxies with Large-scale Ambient Radio Emission (GLAREs), Starburst Radio Ring Galaxies (SRRGs), and some unusual morphologies.

\subsection{Galaxies with Large-scale Ambient Radio Emission}
\label{SEC:GLAREs}
We identify several systems with diffuse radio emission around distant galaxies, which may serve as potential precursors to ORCs or may represent different evolutionary stages of ORCs \citep[see examples in][]{gupta22,kumari24a,kumari24b}. 
We call these sources GLAREs (Galaxies with Large-scale Ambient Radio Emission). 
Figure~\ref{FIG:GLAREs} provides examples, with the left panels showing radio images and the right panels displaying radio contours overlaid on optical images from the DESI LS DR10.
In addition to these examples, several other such GLAREs were identified in the first year of the EMU survey. 
All GLAREs are listed in Table~\ref{TAB:GLAREs}, along with details of their potential host galaxies, including names and redshifts where available. 
While we do not display all these sources here for brevity, they can be viewed at \url{https://doi.org/10.25919/cvz8-4d27}, where we also provide infrared images from the AllWISE survey and optical images from DESI LS DR10, where available.
All GLAREs are located at Galactic latitudes $|b| > 5^{\circ}$, and only 10 have $|b| < 10^{\circ}$ (see Table~\ref{TAB:GLARE_latitude}). The $r$-band foreground extinction magnitudes for these 10 GLAREs are moderate, except for J1053-5317, where $A_r = 0.93$. This modest extinction is consistent with -- though not definitive evidence for -- an extragalactic origin for most GLAREs. Furthermore, no cross-matches were found for any GLAREs in the HASH or Green’s catalogues, reinforcing the likelihood that, like ORCs and ORC candidates, these sources are not associated with planetary nebulae or supernova remnants.
The diffuse radio emission in these systems exhibits various shapes, such as rectangular, circular, and irregular. 
A few GLAREs in the table also feature a secondary source near the radio emission's edge.
Spectroscopic redshifts (subscript sp') are from \citet{jones09} and from Sloan Digital Sky Survey\citep[SDSS DR16][]{ahumada20} for J0202-0218, while photometric redshifts are primarily from DESI LS DR9, with superscripts `a' from \citet{wen24}, `b' from \citet{bilicki16}, and `c' from \citet{bilicki14}. Additional photometric redshifts from \citet{duncan22}, \citet{zou22}, \citet{wen24}, and \citet{zhou25} are consistent with those reported here and are omitted for brevity.

\floatsetup[table]{font=scriptsize}
\begin{table*}[htbp]
\centering
\begin{center}
\begin{tabular}{lccccccccc}
\hline
\multicolumn{1}{c}{Name} & \multicolumn{1}{c}{Host Name} & \multicolumn{1}{c}{RA} & \multicolumn{1}{c}{Dec} & \multicolumn{1}{c}{$z$} & \multicolumn{1}{c}{S} & \multicolumn{1}{c}{Secondary Source Name} & \multicolumn{1}{c}{RA} & \multicolumn{1}{c}{Dec} & \multicolumn{1}{c}{$z$} \\
\hline
J0010-1113 & WISEA J001052.00-111307.5 & 2.7167 & -11.2188 & $0.530\pm 0.147$ & I & - & - & - & - \\ 
J0047-4555 & WISEA J004756.03-455550.7 & 11.9835 & -45.9308 & $0.541\pm 0.046$ & R & - & - & - & - \\ 
J0146-4752 & WISEA J014637.65-475205.9 & 26.6569 & -47.8683 & $0.162\pm 0.010$ & R & - & - & - & - \\ 
J0156-5325 & WISEA J015638.10-532551.2 & 29.1588 & -53.4309 & $0.379\pm0.025$ & C & WISEA J015641.21-532615.2 & 29.1717 & -53.4376 & $0.676\pm0.061$ \\ 
J0201-6227 & WISEA J020117.35-622754.3 & 30.3223 & -62.4651 & $0.293\pm0.019$ & I & - & - & - & - \\
J0201-5139 & WISEA J020147.44-513907.4 & 30.4477 & -51.6521 & $0.408\pm 0.017$ & I & - & - & - & - \\ 
J0202-0218 & WISEA J020213.37-021846.8 & 30.5557 & -2.3130 & 0.177$_{\rm sp}$ & C & - & - & - & - \\
J0208-4725 & WISEA J020808.61-472500.0 & 32.0359 & -47.4167 & $0.210\pm0.012$ & R & - & - & - & - \\ 
J0209-4312 & WISEA J020928.22-431241.3 & 32.3676 & -43.2115 & $0.260\pm0.009$ & C & WISEA J020929.14-431200.7 & 32.3715 & -43.2002 & $0.231\pm0.021$\\ 
J0209-4307 & WISEA J020947.80-430712.3 & 32.4492 & -43.1201 & $0.199\pm0.005$ & C & - & - & - & - \\  
J0214-1121 & WISEA J021455.10-112131.4 & 33.7296 & -11.3587 & $0.194\pm0.004$ & C & - & - & - & - \\ 
J0219-6433 & WISEA J021946.05-643306.9 & 34.9419 & -64.5519 & $0.747\pm 0.059$ & C & - & - & - & - \\ 
J0224-4436 & WISEA J022430.58-443614.7 & 36.1274 & -44.6041 & $0.462\pm0.051$ & I & - & - & - & - \\  
J0317-5236 & WISEA J031706.96-523641.7 & 49.2790 & -52.6116 & $0.394\pm 0.034$ & C & - & - & - & - \\ 
J0334-5434 & WISEA J033409.47-543458.1 & 53.5395 & -54.5828 & $0.055\pm0.012$ & I & - & - & - & - \\
J0349-5149 & WISEA J034918.21-514904.3 & 57.3259 & -51.8179 & $0.192\pm0.024$ & C & - & - & - & - \\  
J0351-7251 & WISEA J035124.82-725128.2 & 57.8535 & -72.8578 & $0.811\pm 0.054^a$ & I & - & - & - & - \\ 
J0433-7314 & WISEA J043301.23-731412.8 & 68.2552 & -73.2369 & $0.304\pm0.043^b$ & C & - & - & - & - \\ 
J0502-4221 & WISEA J050218.16-422137.0 & 75.5757 & -42.3603 & $0.108\pm0.043$ & I & - & - & - & - \\  
J0701-7003 & WISEA J070156.30-700340.0 & 105.4849 & -70.0611 & - & C & - & - & - & - \\ 
J0739-5503 & WISEA J073953.80-550310.8 & 114.9742 & -55.0530 & $0.139\pm0.038^b$ & R & - & - & - & - \\ 
J0741-5355 & WISEA J074156.77-535551.6 & 115.4866 & -53.9310 & - & C & - & - & - & - \\ 
J0858-6637 & WISEA J085829.77-663728.6 & 134.6241 & -66.6246 & 0.063$_{\rm sp}$ & C & - & - & - & - \\ 
J0911-1756 & WISEA J091119.05-175643.6 & 137.8294 & -17.9455 & $0.839\pm 0.024^a$ & R & - & - & - & - \\ 
J0918-6227 & WISEA J091841.91-622755.4 & 139.6746 & -62.4654 & - & I & - & - & - & - \\ 
J0918-1949 & WISEA J091851.75-194948.5 & 139.7156 & -19.8301 & - & C & - & - & - & - \\ 
J0929-6508 & WISEA J092919.29-650853.5 & 142.3304 & -65.1482 & - & R & - & - & - & - \\ 
J0934-6116 & WISEA J093414.13-611655.7 & 143.5589 & -61.2821 & - & I & - & - & - & - \\ 
J0953-6131 & WISEA J095300.69-613135.7 & 148.2529 & -61.5266 & $0.043\pm0.015^c$ & I & - & - & - & - \\ 
J0955-0738 & WISEA J095514.69-073804.1 & 148.8112 & -7.6345 & $0.107\pm 0.085$ & I & WISEA J095514.47-073715.8 & 148.8103 & -7.6211 & $1.058\pm 0.105$\\ 
J1014-6351 & WISEA J101415.55-635150.1 & 153.5648 & -63.8639 & $0.202\pm0.015^c$ & R & - & - & - & - \\ 
J1053-5318 & WISEA J105322.98-531800.4 & 163.3458 & -53.3001 & - & I & - & - & - & - \\ 
J1054-4847 & WISEA J105440.47-484702.5 & 163.6686 & -48.7840 & - & I & - & - & - & - \\ 
J1149+0044 & WISEA J114900.49+004459.7 & 177.2521& 0.7499 & $0.235\pm0.013$ & I & - & - & - & - \\
J1208-5531 & WISEA J120857.69-553104.7 & 182.2404 & -55.5179 & - & C & - & - & - & - \\ 
J1214-0233 & WISEA J121415.55-023318.3 & 183.5648 & -2.5551 & $0.801\pm 0.043$ & C & - & - & - & - \\ 
J1309-4444 & WISEA J130923.34-444404.6 & 197.3473 & -44.7346 & $0.088\pm0.036^b$ & I & - & - & - & - \\ 
J1322-7121 & WISEA J132229.55-712126.5 & 200.6232 & -71.3574 & - & I & - & - & - & - \\ 
J1419-3101 & WISEA J141913.79-310152.0 & 214.8075 & -31.0311 & 0.096$_{\rm sp}$ & R & - & - & - & - \\ 
J1438-2736 & WISEA J143848.83-273639.9 & 219.7035 & -27.6111 & 0.048$_{\rm sp}$ & R & - & - & - & - \\ 
J1446-2743 & WISEA J144645.68-274310.0 & 221.6903 & -27.7195 & $0.113\pm0.037^b$ & I & - & - & - & - \\ 
J1452-6632 & WISEA J145210.97-663233.7 & 223.0457 & -66.5427 & - & R & - & - & - & - \\ 
J1552-6606 & WISEA J155209.52-660657.2 & 238.0397 & -66.1159 & - & I & - & - & - & - \\ 
J1648-6908 & WISEA J164815.60-690821.7 & 252.0650 & -69.1394 & $0.051\pm0.035^b$ & I & WISEA J164821.75-690847.4 & 252.0906 & -69.1465	& -	\\
J1723-6614 & WISEA J172343.89-661457.3 & 260.9329 & -66.2493 & - & I & - & - & - & - \\ 
J1733-6528 & WISEA J173316.13-652819.8 & 263.3172 & -65.4722 & 0.030$_{\rm sp}$ & R & - & - & - & - \\ 
J1841-5623 & WISEA J184156.65-562312.9 & 280.4861 & -56.3869 & - & I & - & - & - & - \\ 
J1912-7120 & WISEA J191205.59-712051.2 & 288.0233 & -71.3476 & $0.222\pm0.040^b$ & I & WISEA J191146.27-712023.5 & 287.9428 & -71.3399	& -\\ 
J2012-5848 & WISEA J201201.18-584826.7 & 303.0052 & -58.8092 & $0.601\pm 0.066$ & C & WISEA J201203.16-584804.3 & 303.0132 & -58.8012	& $0.155\pm 0.006$\\ 
J2018-1718 & WISEA J201823.92-171817.6 & 304.5997 & -17.3049 & $0.112\pm0.037^b$ & I & - & - & - & - \\ 
J2058-0255 & WISEA J205823.13-025544.2 & 314.5964 & -2.9289 & $0.862\pm 0.034$ & I & - & - & - & - \\ 
J2102+0356 & WISEA J210249.68+035645.5 & 315.7070 & 3.9459 & $0.057\pm0.007$ & I & WISEA J210253.08+035636.6 & 315.7212 & 3.9435	& $0.886\pm0.311$\\ 
J2116-0718 & WISEA J211630.09-071831.7 & 319.1280 & -7.3068 & $0.800\pm 0.167$ & I & WISEA J211629.26-071731.6 & 319.1219 & -7.2921 & $0.275\pm 0.018$\\ 
J2120-6900 & WISEA J212022.85-690041.4 & 320.0949 & -69.0114 & - & R & - & - & - & - \\ 
J2133-5409 & WISEA J213339.17-540931.8 & 323.4132 & -54.1588 & $0.280\pm0.011$ & I & - & - & - & - \\ 
\hline
\end{tabular}
\end{center} 
\caption{Galaxies with Large-scale Ambient Radio Emission (GLAREs) identified in the first year of the EMU survey. The columns, from left to right, list the source name, potential host galaxy, host's RA and Dec (in degrees), host redshift, and the shape (S) of the diffuse radio emission (R: rectangular, C: circular, I: irregular; see Figure~\ref{FIG:GLAREs} for examples). For GLAREs with another source near the edge of the radio emission, the source’s name, RA, Dec, and redshift are also provided. Spectroscopic redshifts (subscript sp') are from \citet{jones09} and from \citet{ahumada20} for J0202-0218, while photometric redshifts are primarily from DESI LS DR9, with superscripts `a' from \citet{wen24}, `b' from \citet{bilicki16} with error $0.033(1+z)$, `c' from \citet{bilicki14} with error $0.015$. All GLAREs can be viewed at \url{https://doi.org/10.25919/cvz8-4d27}.}
\label{TAB:GLAREs}
\end{table*}

\begin{table}[!ht]
    \centering
    \begin{tabular}{cccc}
    \hline
        \multicolumn{1}{c}{Name} & \multicolumn{1}{c}{l (deg)} & \multicolumn{1}{c}{b (deg)} & \multicolumn{1}{c}{$A_r$} \\
    \hline
         J0918–6227 & 280.9207 & -9.0789 & 0.36\\
         J0934–6117 & 281.4023 & -6.9520 & 0.65\\
         J0953–6131 & 283.2721 & -5.6670 & 0.84\\
         J1014-6352 & 286.6386 & -6.0823 & 0.60\\
         J1053-5317 & 285.7258 &  5.5816 & 0.93\\
         J1054-4847 & 283.9030 &  9.7258 & 0.43\\
         J1208-5531 & 296.9035 &  6.8528 & 0.63\\
         J1322-7121 & 305.4353 & -8.6368 & 0.50\\
         J1452-6632 & 314.5520 & -6.4345 & 0.61\\
         J1552-6607 & 319.8550 & -9.3774 & 0.19\\
    \hline
    \end{tabular}
    \caption{Galactic coordinates of GLAREs with source name, longitude (l), latitude (b) and $r$-band foreground extinction magnitude ($A_r$) for those with $\rm |b| < 10^{\circ}$.}
    \label{TAB:GLARE_latitude}
\end{table}

The exact mechanism behind the formation of ORCs remains uncertain, but two primary models are proposed. 
A recent plausible model suggests that the rings result from the collision of a fading relic radio lobe with an external shock front, re-energizing the aged electrons \citep{shabala24}.
Another widely supported hypothesis suggests that the rings are projections of a spherical shell created by a shockwave originating from the central host galaxy. 
Potential triggers for the shock include a merger of supermassive black holes \citep{norris22}, a galaxy merger \citep{dolag23}, or a starburst-driven shockwave \citep{norris22, coil24}. 

The first row of Figure~\ref{FIG:GLAREs} shows two GLAREs with rectangular shapes, each with a potential host galaxy near the geometric centre of the diffuse radio emission. 
The radio source J1419-3101 is linked to the central galaxy WISEA J141913.79-310152.0, with a spectroscopic redshift of 0.096, while J0208-4725 is associated with WISEA J020808.61-472500.0, at a photometric redshift of $0.210\pm0.012$.
A total of 12 such GLAREs were identified in the first year of the EMU survey and are listed in Table~\ref{TAB:GLAREs}, denoted by `R' in the shape column.
The second row of Figure~\ref{FIG:GLAREs} highlights circular diffuse radio emission surrounding potential central host galaxies. 
The central galaxy for J0209-4307 is WISEA J020947.80-430712.3, with a photometric redshift of $0.199\pm0.005$, while J0433-7314 is associated with WISEA J043301.23-731412.8, with a photometric redshift of 0.304. 
These systems may represent different evolutionary stages of ORCs identified by \citet{norris21b} and \citet{gupta22}, as well as the ORCs J0452-6231, J2304-7129, J1313-4709, and ORC candidate J0510-5825 discussed in this work. 
Table~\ref{TAB:GLAREs} lists 16 such systems, denoted by `C' in the shape column.
Both rectangular and circular GLAREs may represent precursors or evolutionary stages of known ORCs, potentially forming from shockwaves originating in the central galaxy and later developing into edge-brightened rings. Alternatively, rectangular GLAREs e.g. J1419-3101 could be classical double-lobed or remnant radio galaxies viewed near their major radio axis, appearing foreshortened.

\begin{table}[!ht]
    \centering
    \begin{tabular}{cc}
    \hline
        \multicolumn{1}{c}{Shape} & \multicolumn{1}{c}{Count} \\
    \hline
         Rectangular (R) & 12 \\
         Circular (C) & 16 \\
         Irregular (I) & 27 \\
    \hline
    \end{tabular}
    \caption{Count of Galaxies with Large-scale Ambient Radio Emission (GLAREs) categorized by the shape of their radio continuum emission. See Table~\ref{TAB:GLAREs} for details.}
    \label{TAB:GLARE_Nrs}
\end{table}

The third row of Figure~\ref{FIG:GLAREs} presents GLAREs with irregularly shaped diffuse radio emission. 
Their potential host galaxy is either near the centre of the diffuse emission (e.g. left panel), or, it is located closer to the edge (e.g. right panel).
The radio source J0010-1113 is likely hosted by WISEA J001052.00-111307.5, which has a photometric redshift from DESI LS DR9 of $0.530\pm0.147$.
Although the host galaxy is too faint to be easily seen in the optical image, it can be recognised upon close inspection and is catalogued in DESI LS DR10, with its redshift recorded in the DR9 photometric redshift catalogues. The galaxy also shows bright emission in the infrared, as illustrated in the online image repository of GLAREs.
One possibility behind the higher infrared brightness could be enhanced dust absorption of optical light, followed by re-emission at infrared wavelengths, a scenario that warrants further investigation in future studies of this system.
The source J0502-4221 is hosted by WISEA J050218.16-422137.0 at a redshift of $0.108\pm0.043$. 
A total of 27 such irregularly shaped GLAREs are listed in Table~\ref{TAB:GLAREs}, denoted by `I' in the shape column.
The irregular diffuse emission could result from anisotropic shockwaves originating from the host galaxies or galactic winds. While these systems may not necessarily evolve into ORCs, they could potentially develop edge-brightened circles over time, similar to the ORC described in \citet{koribalski21} and ORC J0402-5321 presented in this work.

The fourth row of Figure~\ref{FIG:GLAREs} shows GLAREs J0156-5325 and J0209-4312, each with potential central host galaxies, WISEA J015638.10-532551.2 and WISEA J020928.22-431241.3, respectively. 
Both systems feature a prominent secondary galaxy near the edge of the diffuse radio emission.
These secondary galaxies might be at similar redshifts to the central host. 
For example, in the case of J0209-4312, WISEA J020928.22-431241.3 and WISEA J020929.14-431200.7 have redshifts of $0.260\pm0.009$ and $0.231\pm0.021$, respectively. 
Such systems could potentially evolve into ORCs, similar to the one described by \citet{koribalski21} and ORC J0402-5321 presented in this work.
Conversely, systems like J0156-5325, where the central galaxy WISEA J015638.10-532551.2 is at a redshift of $0.379\pm0.025$ and the secondary galaxy WISEA J015641.21-532615.2 is at $0.676\pm0.061$, suggest that the galaxy near the edge may not contribute to the diffuse emission. 
Such systems might evolve into structures like ORC J0210-5710, shown in the top panel of Figure~\ref{FIG:ORCs1}.
Table~\ref{TAB:GLAREs} lists 9 such systems, providing details about both the central and secondary galaxies.

The GLAREs discussed here appear to have potential host galaxies that could be responsible for the observed diffuse radio emission. This emission could evolve into edge-brightened rings through merger-induced or starburst-driven shockwaves originating from their host galaxy \citep[e.g.,][]{norris22,dolag23,coil24}, or through an external shock front during the course of their evolution \citep[e.g.,][]{shabala24}.
Diffuse radio emission in galaxies is often associated with galactic winds, which are large-scale outflows of gas and cosmic rays driven by processes such as star formation and AGN activity \citep[e.g.,][and references therein]{nims15}, and these winds could also influence the evolution of these systems.
This suggests that some of the GLARE systems might be precursors or a later evolutionary stage of ORCs.
Given that two ORCs were identified in the 270~$\deg^2$ EMU-PS1 \citep[ORC J2103-6200 and ORC J2223–4834;][]{norris21b,gupta22}, we would expect approximately 33 ORCs in the $\sim4,500~\deg^2$ covered by the first year of the EMU survey. 
With five ORCs, two ORC candidates, and the possibility that even half of the GLAREs represent an evolutionary stage of ORCs, the observed numbers align well with these expectations.
Additionally, in our search, we visually inspected sources with a confidence score threshold of 0.7. While we tested this threshold using 10 tiles for ORCs (see Section~\ref{SEC:method}), some GLAREs may still be missed with scores below 0.7.
The hypothesized connection between ORCs and GLAREs remains unconfirmed until detailed studies of these systems and their host galaxies are conducted.
Such investigations are necessary to determine whether the emission results from galaxy mergers, or starburst-driven shockwaves. 
Future research should focus on examining the properties of the GLAREs' potential hosts as well as those of known ORCs to uncover the underlying physics driving their morphological evolution.

While the connection between GLAREs and ORCs is an intriguing hypothesis, the diffuse radio emission around these galaxies itself warrants further investigation into the mechanisms producing it. 
Notably, although the emission may originate from a single host galaxy in most cases, there are systems where multiple galaxies within the diffuse radio emission region are located at similar redshifts as the central galaxy.
Examples of such GLAREs include J0209-4307, J0214-1121, J0219-6433, J0318-5708, J1214-0233, and J2133-5409, all of which have nearby galaxies with redshifts consistent with the central galaxy. 
There are additional examples where optical images from DESI LS DR10 are not available, but nearby sources are visible in the infrared images, such as J0351-7251, J0701-7003, J0739-5503, J0741-5355, J0911-1756, J0929-6508, J1014-6352, J1208-5531, J1309-4444, J1332-7121, J1452-6632, J1723-6614 and J1733-6528.
Understanding the differences in emission mechanisms between single-host systems and those involving galaxy groups or clusters is a compelling area for future research.
Additionally, some GLAREs may share underlying physical mechanisms with classical double-lobed radio galaxies or with remnant and restarted radio galaxies that are poorly resolved at EMU resolution, though they may exhibit slight variations in their jet emission processes. This could be the case for some rectangular GLAREs, e.g. top-left panel of Figure~\ref{FIG:GLAREs} and some GLAREs with a secondary source e.g. bottom-left panel of Figure~\ref{FIG:GLAREs}.
Detailed future studies of these systems, such as spectral index analysis and investigations into the correlation between radio luminosity and infrared emission, will be crucial for advancing our understanding of radio jet emission physics.

\begin{table}[!ht]
\centering
\caption{Coordinates of radio sources without plausible hosts in available infrared and optical data, listed to aid their localisation in Equatorial (RA and Dec) and Galactic (l, b) reference frames. The Galactic latitudes with $\rm |b|>9$ for all sources indicate that none are likely located in the Galactic plane. The images of these sources can be accessed at \url{https://doi.org/10.25919/cvz8-4d27}}
\begin{tabular}{cccccc}
\hline
Name & RA (deg) & Dec (deg) & l (deg) & b (deg) & $A_r$\\
\hline
J0048-4713 & 12.2404  & -47.2300 & 304.1548 & -69.8924 & 0.02\\
J0107-3624 & 16.8013  & -36.4042 & 284.0816 & -80.1400 & 0.02\\
J0318-5708 & 49.5083  & -57.1444 & 272.4037 & -50.3927 & 0.06\\
J0903-6121 & 135.7693 & -61.3616 & 278.8400 & -9.6975  & 0.40\\
J1009-1034 & 152.4989 & -10.5701 & 251.3455 & 35.6621  & 0.11\\
J1407-0917 & 211.7863 & -9.2841  & 332.2821 & 49.2248  & 0.08\\
J1615-7011 & 243.9400 & -70.1878 & 318.6919 & -13.8563 & 0.21\\
J1637-7520 & 249.2728 & -75.3426 & 315.7716 & -18.4552 & 0.14\\
J1750-7156 & 267.5040 & -71.9335 & 321.6310 & -21.1257 & 0.12\\
\hline
\end{tabular}
\label{TAB:noHost}
\end{table}

\begin{figure*}[t!]
\centering
\includegraphics[trim=0cm 0cm 0cm 0.6cm, clip, width=18cm, scale=0.5]
{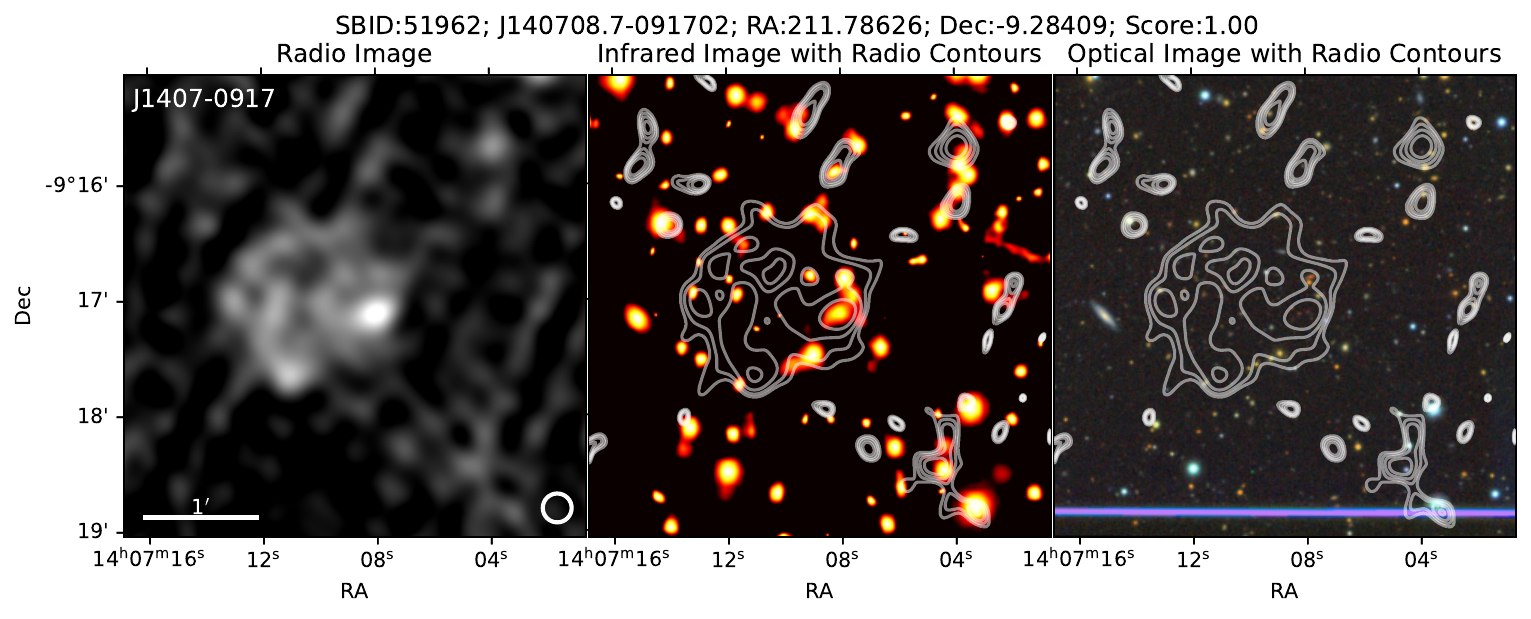}
\caption{An example of a diffuse radio source without a plausible host galaxy (left panel), as seen in the infrared W1 band from the AllWISE survey (middle panel) and in the optical image from DESI LS DR10 (right panel). All similar diffuse sources without a plausible host galaxy can be viewed at \url{https://doi.org/10.25919/cvz8-4d27}.} 
\label{FIG:DiffuseOrphans}
\end{figure*}

In addition to the GLAREs, we have also identified a few diffuse radio sources that lack a plausible host galaxy.
These sources were identified through visual inspection of the $\sim1,800$ sources selected under the search criteria in Section~\ref{SEC:method}.
Table~\ref{TAB:noHost} provides the approximate positions of these sources, which are intended to aid in locating them within the survey maps. These coordinates do not correspond to any identified infrared or optical counterparts.
Figure~\ref{FIG:DiffuseOrphans} presents one such diffuse source, and eight more can be accessed at \url{https://doi.org/10.25919/cvz8-4d27}.
All of these sources are located at Galactic latitudes $\rm |b| > 9^{\circ}$ and have $r$-band extinction $A_r < 0.4$. Additionally, none have cross-matches in the HASH catalogue of planetary nebulae and Green's catalogue of supernova remnants, supporting the likelihood of an extragalactic origin.
The source J1407-0197, shown in Figure~\ref{FIG:DiffuseOrphans}, does not have a plausible host galaxy that could be responsible for the diffuse radio emission.
The overdensity of radio emission toward the southwest edge corresponds to two faint sources in the optical image with unknown redshifts, but these are unlikely to be the hosts leading to the observed $\sim1^{\prime}$ diffuse radio emission.
The potential relationship between these diffuse radio sources and ORCs is unclear. 
ORCs are defined as requiring a host galaxy at the centre of their edge-brightened ring structure \citep[][]{norris24}.
While all GLAREs listed in Table~\ref{TAB:GLAREs} have potential host galaxies and may represent an evolutionary stage of ORCs, these diffuse sources without hosts are unlikely to serve as such.
It is worth noting that some GLAREs in Table~\ref{TAB:GLAREs} and displayed in the online image repository might also fall into this category, as confirming the host galaxies based solely on infrared images remains challenging.
Notable examples include J1322-7122, J1912-7120, and J2018-1718. In the case of J0334-5435, although a nearby spiral galaxy is identified as a potential host in the optical image, if it is not the true host, identifying an alternative plausible host for the diffuse radio emission would be difficult.
The same is true for sources listed without plausible hosts, such as J0318-5708, J1615-7011, and J1637-7520, where bright infrared sources located at the edge or within the diffuse emission could be the hosts, potentially classifying these as GLAREs.
Future focused studies of these systems will be necessary to address these ambiguities.

Additionally, due to the absence of corresponding diffuse emission around these sources, it is improbable that they represent the resolved radio lobes of typical double-lobed radio galaxies. 
However, more sensitive radio images of these diffuse systems and their surroundings will be required to confirm this. 
Independent of this, one possibility is that they are fading relic radio lobes with re-energized, aged electrons due to an external shock \citep[][]{shabala24}, a mechanism that has been suggested to explain ORC2 and ORC3\footnote{Note that these systems have now been removed from the list of ORCs \citep[see][]{norris24}.} \citep[][]{norris21b, shabala24}.
Future studies of these systems will be crucial in testing the relic lobe hypothesis and exploring potential connections with ORCs.

\begin{table}[t]
    \centering
    \begin{tabular}{c c c c c}
        \hline
        Name & Host Name & RA (deg) & Dec (deg) & $z$ \\
        \hline
        J0132-3840 & ESO 013015-3856.2 & 23.1144 & -38.6796 & 0.0122$_{\rm sp}$ \\
        J0208-5644 & ESO 020716-5658.4 & 32.2314 & -56.7369 & 0.0214$_{\rm sp}$ \\
        J0250-5458 & ESO 024925-5510.8 & 42.7239 & -54.9759 & 0.0187$_{\rm sp}$ \\
        J0252-5756 & ESO 025100-5809.1 & 43.0891 & -57.9479 & 0.0295$_{\rm sp}$ \\
        J0450-6120 & ESO 044950-6125.7 & 72.6116 & -61.3454 & 0.0197$_{\rm sp}$ \\
        J0831-0111 & UGC 04455 & 127.8871 & -1.1979 & 0.0309 \\
        J0938-6149 & ESO 093708-6136.2 & 144.6216 & -61.8297 & 0.0097$_{\rm sp}$ \\
        J0938-6356 & ESO 093717-6342.4 & 144.6317 & -63.9334 & 0.0153$_{\rm sp}$ \\
        J1123-0106 & 6dF J1123465-010618 & 170.9436 & -1.1049 & 0.0184$_{\rm sp}$ \\
        J1255-4554 & ESO 125211-4538.1 & 193.7529 & -45.9057 & 0.0366$_{\rm sp}$ \\
        J1406-3418 & ESO 140338-3404.5 & 211.6479 & -34.3116 & 0.0161$_{\rm sp}$ \\
        J1440+0618 & UGC 09454 & 220.0994 & 6.3072 & 0.0238$_{\rm sp}$ \\
        J1508-2546 & ESO 150508-2535.0 & 227.0217 & -25.7737 & 0.0977$_{\rm ph}$ \\
        J1654-7235 & ESO 164902-7230.3 & 253.7403 & -72.5866 & 0.0195$_{\rm sp}$ \\
        J1737-5531 & ESO 173344-5529.3 & 264.4771 & -55.5181 & 0.0233$_{\rm sp}$ \\
        J1954-5842 & ESO 195012-5850.7 & 298.5984 & -58.7137 & 0.0071$_{\rm sp}$ \\
        J2044-6844 & ESO 203948-6855.7 & 311.1408 & -68.7476 & 0.0104$_{\rm sp}$ \\
        J2101-0011 & UGC 11663 & 315.2823 & -0.1952 & 0.0237$_{\rm sp}$ \\
        \hline
    \end{tabular}
    \caption{Starburst Radio Rings (SRRGs) detected during the first year of the EMU survey are listed with their corresponding details. The columns, arranged from left to right, include the source name, host galaxy, host's right ascension (RA) and declination (Dec) in degrees, and the host's redshift ($z$). Spectroscopic redshifts (subscript `sp’) are available for all galaxies except J1508-2546, for which the photometric redshift (subscript `ph’) is taken from \citet{bilicki16}.
    The spectroscopic redshift values for J0252-5756, J1123-0106, J1255-4554, and J1406-3418 are taken from \citet{jones09}, while the rest are from \citet{zaw19}. Examples of these galaxies are illustrated in Figure~\ref{FIG:SRRGs}, and all 18 SRRGs can be accessed at \url{https://doi.org/10.25919/cvz8-4d27}.}
    \label{TAB:SRRG}
\end{table}
\begin{figure*}[t!]
\centering
\includegraphics[trim=0cm 0.65cm 0cm 0.6cm, clip, width=9cm, scale=0.5]
{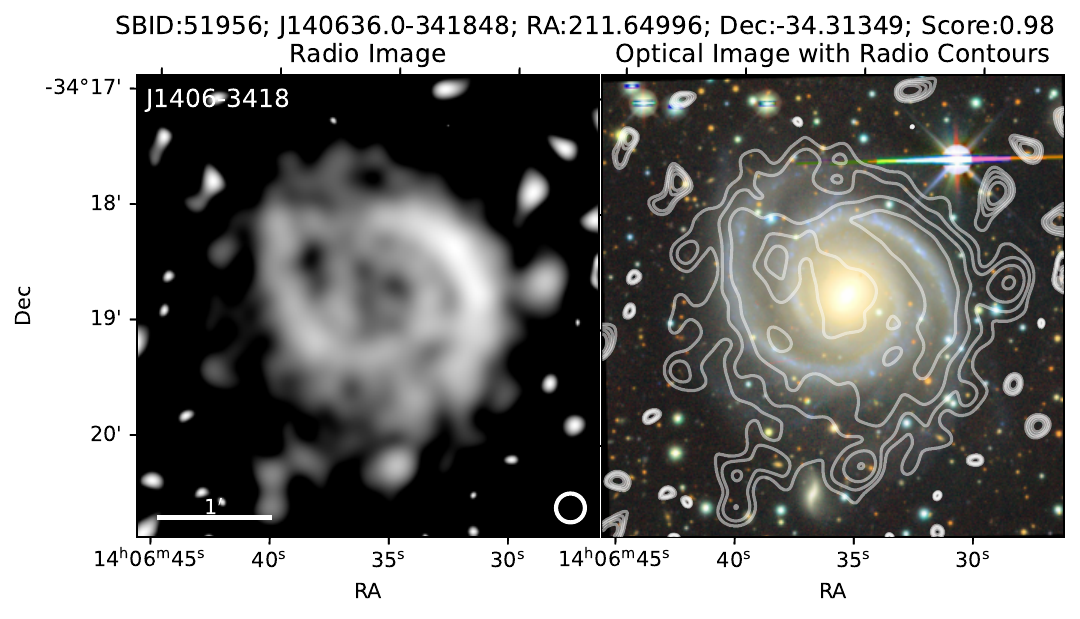}
\includegraphics[trim=0cm 0.65cm 0cm 0.6cm, clip, width=9cm, scale=0.5]{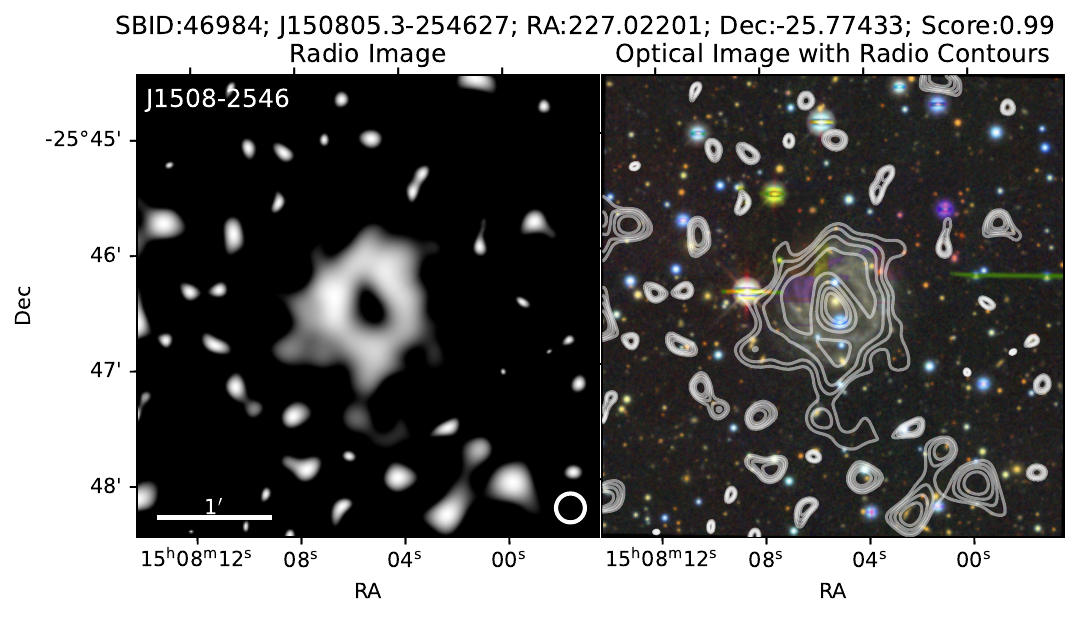}
\includegraphics[trim=0cm 0cm 0cm 1.2cm, clip, width=9cm, scale=0.5]{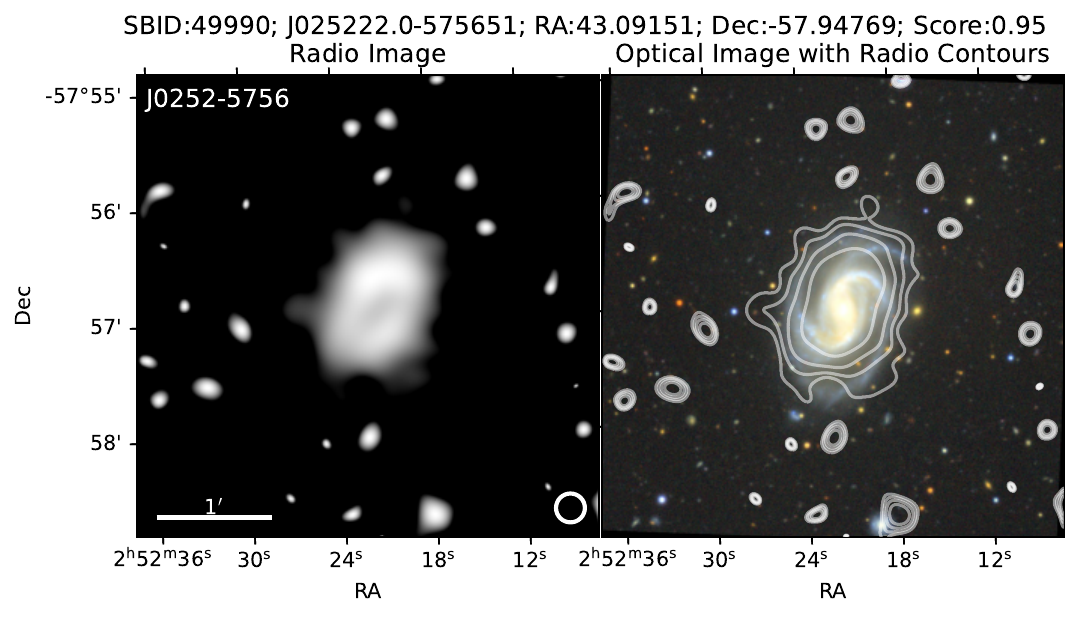}
\includegraphics[trim=0cm 0cm 0cm 1.2cm, clip, width=9cm, scale=0.5]{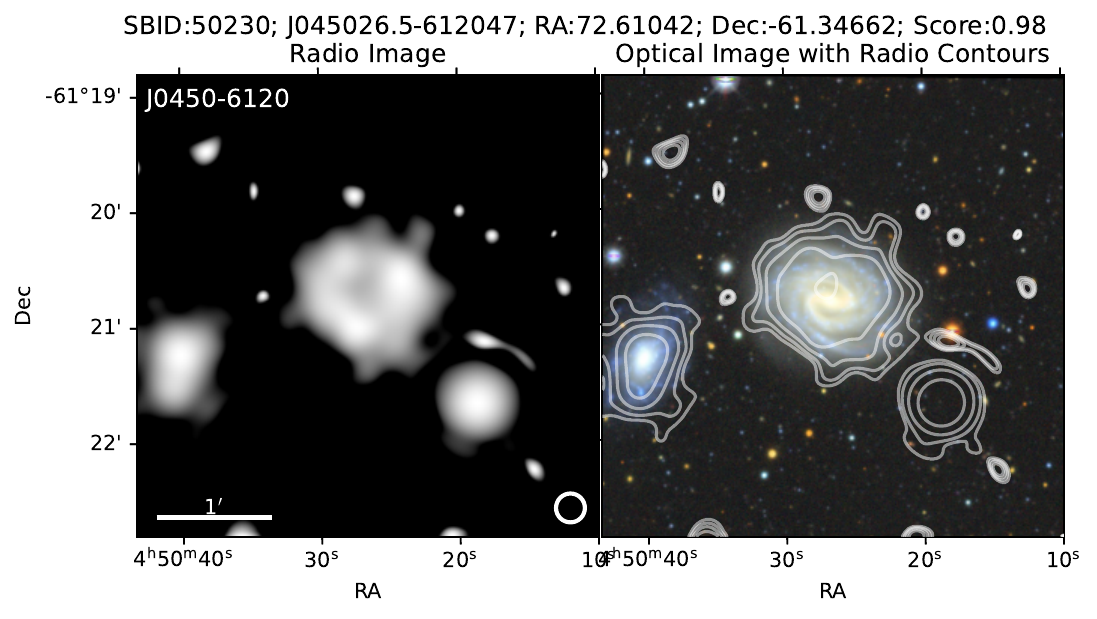}
\caption{Eaxamples of Starburst Radio Ring Galaxies (SRRGs) identified in the first year of the EMU survey. A notable characteristic of these systems is the minimal radio emission from the nuclear region. In each column, the left panels show radio images from the EMU survey, while the right panels display DESI LS DR10 images with white radio emission contours overlaid. All SRRGs listed in Table~\ref{TAB:SRRG} can be accessed at \url{https://doi.org/10.25919/cvz8-4d27}.} 
\label{FIG:SRRGs}
\end{figure*}
%


\subsection{Starburst Radio Ring Galaxies}
\label{SEC:SRRGs}
As discussed in previous sections, ORCs and GLAREs exhibit extended radio emissions that appear to originate from their potential host galaxies. 
However, while these systems display radio emission, they lack corresponding extended emission at infrared or optical wavelengths. 
In contrast, Starburst Radio Ring Galaxies \citep[hereafter SRRGs; see, e.g.,][and references therein]{forbes94} are bright, star-forming galaxies characterized by edge-brightened radio rings surrounding the resolved star-forming regions, with minimal radio emission from the central nuclear region. 
Examples of such SRRGs in the EMU-PS1 can be found in Figure 10 of \citet{norris21b} and Figure 12 of \citet{gupta22}.
Table~\ref{TAB:SRRG} lists 18 additional Starburst Radio Ring Galaxies identified in the first year of the EMU survey.
Figure~\ref{FIG:SRRGs} shows examples of a subset of these SRRGs.
All SRRGs can be viewed at \url{https://doi.org/10.25919/cvz8-4d27}.
In each column of the figure, the left panels display radio images from the EMU survey, while the right panels show W1 band AllWISE images overlaid with white radio emission contours.

The radio source J1406-3418 is associated with the host galaxy WISEA J140635.51-341841.9, located at RA $211.6479^\circ$ and Dec $-34.3116^\circ$, with counterpart ESO 140338-3404.5. It has a spectroscopic redshift of 0.0161.  
The host galaxy of J1508-2546 is ESO 150508-2535.0 is a ring galaxy \citep[see,][]{buta99}, situated at RA $227.0214^\circ$ and Dec $-25.7745^\circ$. 
The ring structure of the host can be seen in the DESI LS DR10.
Its photometric redshift is reported as 0.098, 0.099, and 0.073 by \citet{bilicki16}, \citet{beck21}, and \citet{wen24}, respectively.
J0252-5756 is linked to the spiral host galaxy WISEA J025221.37-575652.7, at RA $43.0891^\circ$ and Dec $-57.9479^\circ$. Its counterparts, ESO 025100-5809.1 and 2MASX J02522129-5756531, are located at a spectroscopic redshift of 0.0295. This source exhibits a well-defined spiral structure in DESI LS DR10.
The host of J0450-6120 is another spiral galaxy, WISEA J045026.79-612043.7, with counterparts ESO 044950-6125.7 and 2MASX J04502674-6120438. It is positioned at RA $72.6116^\circ$ and Dec $-61.3454^\circ$, with a spectroscopic redshift of 0.0197. 
This source exhibits a well-defined spiral structure along with a nearby resolved star-forming galaxy to the east, as seen in DESI LS DR10.

All these SRRGs are nearby, resolved, star-forming galaxies and do not appear to have any connection with ORCs or GLAREs.
Among the possible physical mechanisms leading to these rings, \citet{buta86} suggested that they could be explained by the accumulation of gas at Lindblad resonances, which are regions in a galaxy where gravitational forces from a non-axisymmetric component, such as a bar, cause stars and gas to experience periodic perturbations. The gravitational torques from a bar structure could drive gas toward these resonant regions, leading to enhanced star formation in the resulting rings.
Alternative explanations for the formation of starburst rings include gravitational instabilities within the ring itself. \citet{elmegreen94} proposed that gas accumulating at the Inner Lindblad Resonance (ILR) can become gravitationally unstable, leading to the fragmentation of the ring and subsequent starburst activity.
Recent studies have further investigated the role of Lindblad resonances in the formation of nuclear rings. For instance, \citet{sormani24} proposed that nuclear rings form as gas accumulates at the inner edge of a gap that develops around the ILR. They suggest that the bar potential excites trailing waves near the ILR, which remove angular momentum from the gas, causing it to move inward and accumulate at the inner edge of the gap, forming the nuclear ring.
While these mechanisms could also explain the ring structure observed in radio wavelengths (although it is not seen in optical except for J1508-2546), each of these galaxies is unique, and the absence of radio emission at their centres warrants further investigation in future studies.

\section{Conclusions}
\label{SEC:conclusions}
This paper presents a systematic search for Odd Radio Circles (ORCs) and other unusual radio morphologies using first-year data from the EMU survey, part of the ongoing Evolutionary Map of the Universe (EMU) project. 
The aim is to identify peculiar radio sources by combining advanced machine learning, specifically supervised object detection, with traditional visual inspections. 
This approach enables the efficient processing of vast radio datasets while reducing the likelihood of missing intriguing radio structures, such as ORCs with edge-brightened rings and Galaxies with Large-scale Ambient Radio Emissions (GLAREs).

The first step in our methodology involves training the object detection model Gal-DINO on a subset of known radio sources to detect specific morphological patterns like the distinctive edge-brightened rings characteristic of ORCs and other unusual radio morphologies in the EMU-PS1. 
The model is then applied to the broader dataset from the first year of the EMU survey to flag potential ORC candidates and other peculiar sources, which are further visually inspected to verify their true nature. 
Additionally, we focus on GLAREs, another interesting morphological class that may serve as precursors or an evolutionary stage of ORCs. 
These GLAREs typically show diffuse radio structures around galaxies, with the emission exhibiting various shapes such as rectangular, circular or irregular.
While all GLAREs have a potential host galaxy, we also identify a few diffuse radio sources without a plausible host galaxy visible in the infrared and optical images.
Another set of objects we present are Starburst Radio Ring Galaxies (SRRGs), which are bright, star-forming galaxies characterized by distinct edge-brightened radio rings that surround the resolved star-forming regions, with little to no central radio emission. 

Through this approach, we uncover 5 new ORCs, 2 candidate ORCs, 55 GLAREs, and 18 SRRGs, providing new insights into their potential formation mechanisms. 
The identified ORCs are particularly interesting, as they represent rare objects that challenge our understanding of radio source morphology. 
These ORCs are characterized by their edge-brightened rings of diffuse radio emission, which could be formed through mechanisms like the collision of fading relic radio lobes with external shock fronts, or the projection of spherical shells created by shockwaves originating from the central galaxy. 
In addition to ORCs, we identify a substantial number of GLAREs, which may either represent precursor systems that could evolve into ORCs over time or a later evolutionary stage of ORCs. 
These objects provide a unique opportunity to study the evolutionary pathways of such systems. 
The GLAREs exhibit various shapes and diffuse emissions, with some showing a clear connection to their host galaxies. 
These findings offer new clues into the mechanisms driving the formation of edge-brightened rings and the role of galactic interactions or shockwaves in shaping the morphology of these radio structures.

While our study focuses on radio data from the EMU survey, it is crucial to examine these objects across other wavelengths, such as infrared, optical, and X-rays, to uncover the full picture of their physical nature. 
This multi-wavelength approach will help us determine whether the radio emission is associated with starburst-driven shockwaves, galaxy mergers, or interactions with supermassive black holes. 
In particular, follow-up observations will allow us to test the hypothesis that some of the GLAREs identified in this work may evolve into ORCs in the future, offering an exciting avenue for future research.
Some GLAREs might share fundamental physical mechanisms with classical double-lobed radio sources, though they could display subtle differences in their jet emission processes. This may apply to rectangular GLAREs or those with a secondary galaxy located at the edge of the radio emission. Future detailed studies of these systems will be essential for advancing our understanding of the physics governing radio jet emissions.
Moreover, detailed studies of the host galaxies, including their stellar populations, gas content, and merger histories, will provide valuable insights into the broader context of galaxy evolution and the role of radio emission in these processes. 
Thus, this work lays the groundwork for future investigations that will deepen our understanding of these peculiar radio sources and their connection to galaxy evolution.
Since this work relies on catalogues generated by the RG-CAT pipeline, which in turn are derived from \textit{Selavy}-based catalogues, any undetected diffuse radio sources in \textit{Selavy}-based catalogues will also be absent from this analysis. Future studies should work towards expanding the search criteria using a catalogue-agnostic approach, using images directly to identify any previously overlooked diffuse sources.
Finally, with the ORCs, GLAREs, and SRRGs presented in this study, future efforts should focus on improving object detection and other machine-learning models. Incorporating these radio sources into training datasets will enhance detection probabilities and confidence scores.

\section*{Data Availability}
Network architectures for Gal-DINO can be cloned from \url{http://hdl.handle.net/102.100.100/602494?index=1} or \url{https://github.com/Nikhel1/Gal-DINO}.
ASKAP observations for the EMU survey are available at \url{https://research.csiro.au/casda/}.
AllWISE image mosaics for each corresponding EMU tile can be generated using \url{https://github.com/Nikhel1/wise_mosaics}.
Images of all GLAREs listed in Table~\ref{TAB:GLAREs} and all SRRGs listed in Table~\ref{TAB:SRRG} can be visualized at \url{https://doi.org/10.25919/cvz8-4d27}.

\section*{Acknowledgements}
NG acknowledges support from CSIRO’s Machine Learning and Artificial Intelligence Future Science Impossible Without You (MLAI FSP IWY) Platform.
H.A. has benefited from grant CIIC 211/2024 of Universidad de Guanajuato, Mexico.
\subsection*{ASKAP}
This scientific work uses data obtained from Inyarrimanha Ilgari Bundara / the Murchison Radio-astronomy Observatory. We acknowledge the Wajarri Yamaji People as the Traditional Owners and native title holders of the Observatory site. The Australian SKA Pathfinder is part of the Australia Telescope National Facility (https://ror.org/05qajvd42) which is managed by CSIRO. Operation of ASKAP is funded by the Australian Government with support from the National Collaborative Research Infrastructure Strategy. ASKAP uses the resources of the Pawsey Supercomputing Centre. The establishment of ASKAP, the Murchison Radio-astronomy Observatory and the Pawsey Supercomputing Centre are initiatives of the Australian Government, with support from the Government of Western Australia and the Science and Industry Endowment Fund.

This paper includes archived data obtained through the CSIRO ASKAP Science Data Archive, CASDA (\url{http://data.csiro.au}).

\subsection*{NED}
This research has made use of the NASA/IPAC Extragalactic Database (NED), which is operated by the Jet Propulsion Laboratory, California Institute of Technology, under contract with the National Aeronautics and Space Administration.

\subsection*{Legacy Surveys Sky Viewer}
The Legacy Surveys consist of three individual and complementary projects: the Dark Energy Camera Legacy Survey (DECaLS; Proposal ID \#2014B-0404; PIs: David Schlegel and Arjun Dey), the Beijing-Arizona Sky Survey (BASS; NOAO Prop. ID \#2015A-0801; PIs: Zhou Xu and Xiaohui Fan), and the Mayall $z$-band Legacy Survey (MzLS; Prop. ID \#2016A-0453; PI: Arjun Dey). DECaLS, BASS and MzLS together include data obtained, respectively, at the Blanco telescope, Cerro Tololo Inter-American Observatory, NSF’s NOIRLab; the Bok telescope, Steward Observatory, University of Arizona; and the Mayall telescope, Kitt Peak National Observatory, NOIRLab. Pipeline processing and analyses of the data were supported by NOIRLab and the Lawrence Berkeley National Laboratory (LBNL). The Legacy Surveys project is honoured to be permitted to conduct astronomical research on Iolkam Du’ag (Kitt Peak), a mountain with particular significance to the Tohono O’odham Nation.

NOIRLab is operated by the Association of Universities for Research in Astronomy (AURA) under a cooperative agreement with the National Science Foundation. LBNL is managed by the Regents of the University of California under contract to the U.S. Department of Energy.

This project used data obtained with the Dark Energy Camera (DECam), which was constructed by the Dark Energy Survey (DES) collaboration. Funding for the DES Projects has been provided by the U.S. Department of Energy, the U.S. National Science Foundation, the Ministry of Science and Education of Spain, the Science and Technology Facilities Council of the United Kingdom, the Higher Education Funding Council for England, the National Center for Supercomputing Applications at the University of Illinois at Urbana-Champaign, the Kavli Institute of Cosmological Physics at the University of Chicago, Center for Cosmology and Astro-Particle Physics at the Ohio State University, the Mitchell Institute for Fundamental Physics and Astronomy at Texas A\&M University, Financiadora de Estudos e Projetos, Fundacao Carlos Chagas Filho de Amparo, Financiadora de Estudos e Projetos, Fundacao Carlos Chagas Filho de Amparo a Pesquisa do Estado do Rio de Janeiro, Conselho Nacional de Desenvolvimento Cientifico e Tecnologico and the Ministerio da Ciencia, Tecnologia e Inovacao, the Deutsche Forschungsgemeinschaft and the Collaborating Institutions in the Dark Energy Survey. The Collaborating Institutions are Argonne National Laboratory, the University of California at Santa Cruz, the University of Cambridge, Centro de Investigaciones Energeticas, Medioambientales y Tecnologicas-Madrid, the University of Chicago, University College London, the DES-Brazil Consortium, the University of Edinburgh, the Eidgenossische Technische Hochschule (ETH) Zurich, Fermi National Accelerator Laboratory, the University of Illinois at Urbana-Champaign, the Institut de Ciencies de l’Espai (IEEC/CSIC), the Institut de Fisica d’Altes Energies, Lawrence Berkeley National Laboratory, the Ludwig Maximilians Universitat Munchen and the associated Excellence Cluster Universe, the University of Michigan, NSF’s NOIRLab, the University of Nottingham, the Ohio State University, the University of Pennsylvania, the University of Portsmouth, SLAC National Accelerator Laboratory, Stanford University, the University of Sussex, and Texas A\&M University.

BASS is a key project of the Telescope Access Program (TAP), which has been funded by the National Astronomical Observatories of China, the Chinese Academy of Sciences (the Strategic Priority Research Program “The Emergence of Cosmological Structures” Grant \# XDB09000000), and the Special Fund for Astronomy from the Ministry of Finance. The BASS is also supported by the External Cooperation Program of Chinese Academy of Sciences (Grant \# 114A11KYSB20160057), and Chinese National Natural Science Foundation (Grant \# 12120101003, \# 11433005).

The Legacy Survey team makes use of data products from the Near-Earth Object Wide-field Infrared Survey Explorer (NEOWISE), which is a project of the Jet Propulsion Laboratory/California Institute of Technology. NEOWISE is funded by the National Aeronautics and Space Administration.

The Legacy Surveys imaging of the DESI footprint is supported by the Director, Office of Science, Office of High Energy Physics of the U.S. Department of Energy under Contract No. DE-AC02-05CH1123, by the National Energy Research Scientific Computing Center, a DOE Office of Science User Facility under the same contract; and by the U.S. National Science Foundation, Division of Astronomical Sciences under Contract No. AST-0950945 to NOAO.

\subsection*{Legacy Survey Photometric Redshifts}
The Photometric Redshifts for the Legacy Surveys (PRLS) catalogue used in this paper was produced thanks to funding from the U.S. Department of Energy Office of Science, Office of High Energy Physics via grant DE-SC0007914.


\bibliography{ASKAP_PASA}

\begin{thebibliography}{}
\expandafter\ifx\csname natexlab\endcsname\relax\def\natexlab#1{#1}\fi

\bibitem[{{Ahumada} {et~al.}(2020){Ahumada}, {Allende Prieto}, {Almeida},
  {Anders}, {Anderson}, {Andrews}, {Anguiano}, {Arcodia}, {Armengaud},
  {Aubert}, {Avila}, {Avila-Reese}, {Badenes}, {Balland}, {Barger},
  {Barrera-Ballesteros}, {Basu}, {Bautista}, {Beaton}, {Beers}, {Benavides},
  {Bender}, {Bernardi}, {Bershady}, {Beutler}, {Bidin}, {Bird}, {Bizyaev},
  {Blanc}, {Blanton}, {Boquien}, {Borissova}, {Bovy}, {Brandt}, {Brinkmann},
  {Brownstein}, {Bundy}, {Bureau}, {Burgasser}, {Burtin}, {Cano-D{\'\i}az},
  {Capasso}, {Cappellari}, {Carrera}, {Chabanier}, {Chaplin}, {Chapman},
  {Cherinka}, {Chiappini}, {Doohyun Choi}, {Chojnowski}, {Chung}, {Clerc},
  {Coffey}, {Comerford}, {Comparat}, {da Costa}, {Cousinou}, {Covey}, {Crane},
  {Cunha}, {Ilha}, {Dai}, {Damsted}, {Darling}, {Davidson}, {Davies}, {Dawson},
  {De}, {de la Macorra}, {De Lee}, {Queiroz}, {Deconto Machado}, {de la Torre},
  {Dell'Agli}, {du Mas des Bourboux}, {Diamond-Stanic}, {Dillon}, {Donor},
  {Drory}, {Duckworth}, {Dwelly}, {Ebelke}, {Eftekharzadeh}, {Davis Eigenbrot},
  {Elsworth}, {Eracleous}, {Erfanianfar}, {Escoffier}, {Fan}, {Farr},
  {Fern{\'a}ndez-Trincado}, {Feuillet}, {Finoguenov}, {Fofie},
  {Fraser-McKelvie}, {Frinchaboy}, {Fromenteau}, {Fu}, {Galbany}, {Garcia},
  {Garc{\'\i}a-Hern{\'a}ndez}, {Garma Oehmichen}, {Ge}, {Geimba Maia},
  {Geisler}, {Gelfand}, {Goddy}, {Gonzalez-Perez}, {Grabowski}, {Green},
  {Grier}, {Guo}, {Guy}, {Harding}, {Hasselquist}, {Hawken}, {Hayes}, {Hearty},
  {Hekker}, {Hogg}, {Holtzman}, {Horta}, {Hou}, {Hsieh}, {Huber}, {Hunt}, {Ider
  Chitham}, {Imig}, {Jaber}, {Jimenez Angel}, {Johnson}, {Jones},
  {J{\"o}nsson}, {Jullo}, {Kim}, {Kinemuchi}, {Kirkpatrick}, {Kite}, {Klaene},
  {Kneib}, {Kollmeier}, {Kong}, {Kounkel}, {Krishnarao}, {Lacerna}, {Lan},
  {Lane}, {Law}, {Le Goff}, {Leung}, {Lewis}, {Li}, {Lian}, {Lin}, {Long},
  {Longa-Pe{\~n}a}, {Lundgren}, {Lyke}, {Mackereth}, {MacLeod}, {Majewski},
  {Manchado}, {Maraston}, {Martini}, {Masseron}, {Masters}, {Mathur},
  {McDermid}, {Merloni}, {Merrifield}, {M{\'e}sz{\'a}ros}, {Miglio}, {Minniti},
  {Minsley}, {Miyaji}, {Mohammad}, {Mosser}, {Mueller}, {Muna},
  {Mu{\~n}oz-Guti{\'e}rrez}, {Myers}, {Nadathur}, {Nair}, {Nandra}, {Correa do
  Nascimento}, {Nevin}, {Newman}, {Nidever}, {Nitschelm}, {Noterdaeme},
  {O'Connell}, {Olmstead}, {Oravetz}, {Oravetz}, {Osorio}, {Pace}, {Padilla},
  {Palanque-Delabrouille}, \& {Palicio}}]{ahumada20}
{Ahumada}, R., {Allende Prieto}, C., {Almeida}, A., {et~al.} 2020, \apjs, 249,
  3

\bibitem[{{Beck} {et~al.}(2021){Beck}, {Szapudi}, {Flewelling}, {Holmberg},
  {Magnier}, \& {Chambers}}]{beck21}
{Beck}, R., {Szapudi}, I., {Flewelling}, H., {et~al.} 2021, \mnras, 500, 1633

\bibitem[{{Bilicki} {et~al.}(2014){Bilicki}, {Jarrett}, {Peacock}, {Cluver}, \&
  {Steward}}]{bilicki14}
{Bilicki}, M., {Jarrett}, T.~H., {Peacock}, J.~A., {Cluver}, M.~E., \&
  {Steward}, L. 2014, \apjs, 210, 9

\bibitem[{{Bilicki} {et~al.}(2016){Bilicki}, {Peacock}, {Jarrett}, {Cluver},
  {Maddox}, {Brown}, {Taylor}, {Hambly}, {Solarz}, {Holwerda}, {Baldry},
  {Loveday}, {Moffett}, {Hopkins}, {Driver}, {Alpaslan}, \&
  {Bland-Hawthorn}}]{bilicki16}
{Bilicki}, M., {Peacock}, J.~A., {Jarrett}, T.~H., {et~al.} 2016, \apjs, 225, 5

\bibitem[{{Buta}(1986)}]{buta86}
{Buta}, R. 1986, \apjs, 61, 609

\bibitem[{{Buta}(1995)}]{buta99}
---. 1995, \apjs, 96, 39

\bibitem[{{Coil} {et~al.}(2024){Coil}, {Perrotta}, {Rupke}, {Lochhaas},
  {Tremonti}, {Diamond-Stanic}, {Fielding}, {Geach}, {Hickox}, {Moustakas},
  {Rudnick}, {Sell}, \& {Whalen}}]{coil24}
{Coil}, A.~L., {Perrotta}, S., {Rupke}, D. S.~N., {et~al.} 2024, \nat, 625, 459

\bibitem[{{Cutri} {et~al.}(2021){Cutri}, {Wright}, {Conrow}, {Fowler},
  {Eisenhardt}, {Grillmair}, {Kirkpatrick}, {Masci}, {McCallon}, {Wheelock},
  {Fajardo-Acosta}, {Yan}, {Benford}, {Harbut}, {Jarrett}, {Lake}, {Leisawitz},
  {Ressler}, {Stanford}, {Tsai}, {Liu}, {Helou}, {Mainzer}, {Gettngs},
  {Gonzalez}, {Hoffman}, {Marsh}, {Padgett}, {Skrutskie}, {Beck}, {Papin}, \&
  {Wittman}}]{cutri13}
{Cutri}, R.~M., {Wright}, E.~L., {Conrow}, T., {et~al.} 2021, VizieR Online
  Data Catalog, II/328

\bibitem[{{DeBoer} {et~al.}(2009){DeBoer}, {Gough}, {Bunton}, {Cornwell},
  {Beresford}, {Johnston}, {Feain}, {Schinckel}, {Jackson}, {Kesteven},
  {Chippendale}, {Hampson}, {O'Sullivan}, {Hay}, {Jacka}, {Sweetnam}, {Storey},
  {Ball}, \& {Boyle}}]{DeBoer09}
{DeBoer}, D.~R., {Gough}, R.~G., {Bunton}, J.~D., {et~al.} 2009, IEEE
  Proceedings, 97, 1507

\bibitem[{{Dolag} {et~al.}(2023){Dolag}, {B{\"o}ss}, {Koribalski},
  {Steinwandel}, \& {Valentini}}]{dolag23}
{Dolag}, K., {B{\"o}ss}, L.~M., {Koribalski}, B.~S., {Steinwandel}, U.~P., \&
  {Valentini}, M. 2023, \apj, 945, 74

\bibitem[{{Duncan}(2022)}]{duncan22}
{Duncan}, K.~J. 2022, \mnras, 512, 3662

\bibitem[{{Ekers}(2009)}]{ekers09}
{Ekers}, R.~D. 2009, in Proceedings of the special session ``Accelerating the
  Rate of Astronomical Discovery'' of the 27th IAU General Assembly. August
  11-14 2009. Rio de Janeiro, 7

\bibitem[{{Elmegreen}(1994)}]{elmegreen94}
{Elmegreen}, B.~G. 1994, \apjl, 425, L73

\bibitem[{{Fanaroff} \& {Riley}(1974)}]{fanaroff74}
{Fanaroff}, B.~L., \& {Riley}, J.~M. 1974, MNRAS, 167, 31P

\bibitem[{{Forbes} {et~al.}(1994){Forbes}, {Norris}, {Williger}, \&
  {Smith}}]{forbes94}
{Forbes}, D.~A., {Norris}, R.~P., {Williger}, G.~M., \& {Smith}, R.~C. 1994,
  \aj, 107, 984

\bibitem[{{Green}(2025)}]{green25}
{Green}, D.~A. 2025, Journal of Astrophysics and Astronomy, 46, 14

\bibitem[{{Gupta} {et~al.}(2024{\natexlab{a}}){Gupta}, {Hayder}, {Norris},
  {Huynh}, \& {Petersson}}]{gupta24a}
{Gupta}, N., {Hayder}, Z., {Norris}, R.~P., {Huynh}, M., \& {Petersson}, L.
  2024{\natexlab{a}}, \pasa, 41, e001

\bibitem[{{Gupta} {et~al.}(2023){Gupta}, {Hayder}, {Norris}, {Hyunh}, \&
  {Petersson}}]{gupta23b}
{Gupta}, N., {Hayder}, Z., {Norris}, R.~P., {Hyunh}, M., \& {Petersson}, L.
  2023, NeurIPS ML4PS 2023, arXiv:2312.06728

\bibitem[{{Gupta} {et~al.}(2022){Gupta}, {Huynh}, {Norris}, {Wang}, {Hopkins},
  {Andernach}, {Koribalski}, \& {Galvin}}]{gupta22}
{Gupta}, N., {Huynh}, M., {Norris}, R.~P., {et~al.} 2022, \pasa, 39, e051

\bibitem[{Gupta {et~al.}(2023)Gupta, Hayder, Norris, Huynh, Petersson, Wang,
  Andernach, Koribalski, Yew, Crawford, \& et~al.}]{gupta23a}
Gupta, N., Hayder, Z., Norris, R.~P., {et~al.} 2023, Publications of the
  Astronomical Society of Australia, 40, e044

\bibitem[{{Gupta} {et~al.}(2024{\natexlab{b}}){Gupta}, {Norris}, {Hayder},
  {Huynh}, {Petersson}, {Rosalind Wang}, {Hopkins}, {Andernach}, {Gordon},
  {Riggi}, {Yew}, {Crawford}, {Koribalski}, {Filipovi{\'c}}, {Kapi{\'n}ska},
  {Shabala}, {Vernstrom}, \& {Marvil}}]{gupta24b}
{Gupta}, N., {Norris}, R.~P., {Hayder}, Z., {et~al.} 2024{\natexlab{b}}, \pasa,
  41, e027

\bibitem[{{Gupta et al.}(in preparation)}]{gupta25prep}
{Gupta et al.} in preparation

\bibitem[{Hopkins {et~al.}(2025)Hopkins, Kapinska, Marvil, Vernstrom, Collier,
  Norris, Gordon, Duchesne, Rudnick, Gupta, \& et~al.}]{hopkins25}
Hopkins, A.~M., Kapinska, A., Marvil, J., {et~al.} 2025, Publications of the
  Astronomical Society of Australia, 1–32

\bibitem[{{Hotan} {et~al.}(2021){Hotan}, {Bunton}, {Chippendale}, {Whiting},
  {Tuthill}, {Moss}, {McConnell}, {Amy}, {Huynh}, {Allison}, {Anderson},
  {Bannister}, {Bastholm}, {Beresford}, {Bock}, {Bolton}, {Chapman}, {Chow},
  {Collier}, {Cooray}, {Cornwell}, {Diamond}, {Edwards}, {Feain}, {Franzen},
  {George}, {Gupta}, {Hampson}, {Harvey-Smith}, {Hayman}, {Heywood}, {Jacka},
  {Jackson}, {Jackson}, {Jeganathan}, {Johnston}, {Kesteven}, {Kleiner},
  {Koribalski}, {Lee-Waddell}, {Lenc}, {Lensson}, {Mackay}, {Mahony},
  {McClure-Griffiths}, {McConigley}, {Mirtschin}, {Ng}, {Norris}, {Pearce},
  {Phillips}, {Pilawa}, {Raja}, {Reynolds}, {Roberts}, {Roxby}, {Sadler},
  {Shields}, {Schinckel}, {Serra}, {Shaw}, {Sweetnam}, {Troup}, {Tzioumis},
  {Voronkov}, \& {Westmeier}}]{hotan21}
{Hotan}, A.~W., {Bunton}, J.~D., {Chippendale}, A.~P., {et~al.} 2021, PASA, 38,
  e009

\bibitem[{{Johnston} {et~al.}(2007){Johnston}, {Bailes}, {Bartel}, {Baugh},
  {Bietenholz}, {Blake}, {Braun}, {Brown}, {Chatterjee}, {Darling}, {Deller},
  {Dodson}, {Edwards}, {Ekers}, {Ellingsen}, {Feain}, {Gaensler}, {Haverkorn},
  {Hobbs}, {Hopkins}, {Jackson}, {James}, {Joncas}, {Kaspi}, {Kilborn},
  {Koribalski}, {Kothes}, {Landecker}, {Lenc}, {Lovell}, {Macquart},
  {Manchester}, {Matthews}, {McClure-Griffiths}, {Norris}, {Pen}, {Phillips},
  {Power}, {Protheroe}, {Sadler}, {Schmidt}, {Stairs}, {Staveley-Smith},
  {Stil}, {Taylor}, {Tingay}, {Tzioumis}, {Walker}, {Wall}, \&
  {Wolleben}}]{johnston07ASKAP}
{Johnston}, S., {Bailes}, M., {Bartel}, N., {et~al.} 2007, \pasa, 24, 174

\bibitem[{{Jonas} \& {MeerKAT Team}(2016)}]{jonas16}
{Jonas}, J., \& {MeerKAT Team}. 2016, in MeerKAT Science: On the Pathway to the
  SKA, 1

\bibitem[{{Jones} {et~al.}(2009){Jones}, {Read}, {Saunders}, {Colless},
  {Jarrett}, {Parker}, {Fairall}, {Mauch}, {Sadler}, {Watson}, {Burton},
  {Campbell}, {Cass}, {Croom}, {Dawe}, {Fiegert}, {Frankcombe}, {Hartley},
  {Huchra}, {James}, {Kirby}, {Lahav}, {Lucey}, {Mamon}, {Moore}, {Peterson},
  {Prior}, {Proust}, {Russell}, {Safouris}, {Wakamatsu}, {Westra}, \&
  {Williams}}]{jones09}
{Jones}, D.~H., {Read}, M.~A., {Saunders}, W., {et~al.} 2009, \mnras, 399, 683

\bibitem[{{Koribalski} {et~al.}(2021){Koribalski}, {Norris}, {Andernach},
  {Rudnick}, {Shabala}, {Filipovi{\'c}}, \& {Lenc}}]{koribalski21}
{Koribalski}, B.~S., {Norris}, R.~P., {Andernach}, H., {et~al.} 2021, \mnras,
  505, L11

\bibitem[{{Koribalski} {et~al.}(2024{\natexlab{a}}){Koribalski}, {Veronica},
  {Dolag}, {Reiprich}, {Br{\"u}ggen}, {Heywood}, {Andernach}, {Dettmar},
  {Hoeft}, {Zhang}, {Bulbul}, {Garrel}, {J{\'o}zsa}, \&
  {English}}]{koribalski24a}
{Koribalski}, B.~S., {Veronica}, A., {Dolag}, K., {et~al.} 2024{\natexlab{a}},
  \mnras, 531, 3357

\bibitem[{{Koribalski} {et~al.}(2024{\natexlab{b}}){Koribalski}, {Khabibullin},
  {Dolag}, {Churazov}, {Norris}, {Carretti}, {Hopkins}, {Vernstrom}, {Shabala},
  \& {Gupta}}]{koribalski24b}
{Koribalski}, B.~S., {Khabibullin}, I., {Dolag}, K., {et~al.}
  2024{\natexlab{b}}, \mnras, 532, 3682

\bibitem[{{Kumari} \& {Pal}(2024{\natexlab{a}})}]{kumari24b}
{Kumari}, S., \& {Pal}, S. 2024{\natexlab{a}}, \aap, 683, A175

\bibitem[{{Kumari} \& {Pal}(2024{\natexlab{b}})}]{kumari24a}
---. 2024{\natexlab{b}}, \mnras, 527, 11233

\bibitem[{Lang {et~al.}(2016)Lang, Hogg, \& Mykytyn}]{lang2016tractor}
Lang, D., Hogg, D.~W., \& Mykytyn, D. 2016, Astrophysics Source Code Library,
  ascl

\bibitem[{{Lasker} {et~al.}(1990){Lasker}, {Sturch}, {McLean}, {Russell},
  {Jenkner}, \& {Shara}}]{lasker90}
{Lasker}, B.~M., {Sturch}, C.~R., {McLean}, B.~J., {et~al.} 1990, \aj, 99, 2019

\bibitem[{{Lastufka} {et~al.}(2024){Lastufka}, {Bait}, {Taran}, {Drozdova},
  {Kinakh}, {Piras}, {Audard}, {Dessauges-Zavadsky}, {Holotyak}, {Schaerer}, \&
  {Voloshynovskiy}}]{Lastufka24}
{Lastufka}, E., {Bait}, O., {Taran}, O., {et~al.} 2024, \aap, 690, A310

\bibitem[{{Lochner} \& {Rudnick}(2024)}]{lochner24}
{Lochner}, M., \& {Rudnick}, L. 2024, arXiv e-prints, arXiv:2411.04188

\bibitem[{{Lochner} {et~al.}(2023){Lochner}, {Rudnick}, {Heywood}, {Knowles},
  \& {Shabala}}]{lochner23}
{Lochner}, M., {Rudnick}, L., {Heywood}, I., {Knowles}, K., \& {Shabala}, S.~S.
  2023, \mnras, 520, 1439

\bibitem[{{Masters} {et~al.}(2011){Masters}, {Maraston}, {Nichol}, {Thomas},
  {Beifiori}, {Bundy}, {Edmondson}, {Higgs}, {Leauthaud}, {Mandelbaum},
  {Pforr}, {Ross}, {Ross}, {Schneider}, {Skibba}, {Tinker}, {Tojeiro}, {Wake},
  {Brinkmann}, \& {Weaver}}]{masters11}
{Masters}, K.~L., {Maraston}, C., {Nichol}, R.~C., {et~al.} 2011, \mnras, 418,
  1055

\bibitem[{{Mohale} \& {Lochner}(2024)}]{Mohale24}
{Mohale}, K., \& {Lochner}, M. 2024, \mnras, 530, 1274

\bibitem[{{Mostert} {et~al.}(2021){Mostert}, {Duncan}, {R{\"o}ttgering},
  {Polsterer}, {Best}, {Brienza}, {Br{\"u}ggen}, {Hardcastle}, {Jurlin},
  {Mingo}, {Morganti}, {Shimwell}, {Smith}, \& {Williams}}]{mostert21}
{Mostert}, R. I.~J., {Duncan}, K.~J., {R{\"o}ttgering}, H. J.~A., {et~al.}
  2021, \aap, 645, A89

\bibitem[{{Murphy} {et~al.}(2011){Murphy}, {Condon}, {Schinnerer}, {Kennicutt},
  {Calzetti}, {Armus}, {Helou}, {Turner}, {Aniano}, {Beir{\~a}o}, {Bolatto},
  {Brandl}, {Croxall}, {Dale}, {Donovan Meyer}, {Draine}, {Engelbracht},
  {Hunt}, {Hao}, {Koda}, {Roussel}, {Skibba}, \& {Smith}}]{murphy11}
{Murphy}, E.~J., {Condon}, J.~J., {Schinnerer}, E., {et~al.} 2011, \apj, 737,
  67

\bibitem[{Nims {et~al.}(2015)Nims, Quataert, \& Faucher-Giguère}]{nims15}
Nims, J., Quataert, E., \& Faucher-Giguère, C.-A. 2015, Monthly Notices of the
  Royal Astronomical Society, 447, 3612

\bibitem[{{Norris} {et~al.}(2015){Norris}, {Basu}, {Brown}, {Carretti},
  {Kapinska}, {Prandoni}, {Rudnick}, \& {Seymour}}]{norris15}
{Norris}, R., {Basu}, K., {Brown}, M., {et~al.} 2015, in Advancing Astrophysics
  with the Square Kilometre Array (AASKA14), 86

\bibitem[{{Norris}(2011)}]{norris11}
{Norris}, R.~P. 2011, Journal of Astrophysics and Astronomy, 32, 599

\bibitem[{{Norris} {et~al.}(2025){Norris}, {Koribalski}, {Hale}, {Jarvis},
  {Macgregor}, \& {Taylor}}]{norris24}
{Norris}, R.~P., {Koribalski}, B.~S., {Hale}, C.~L., {et~al.} 2025, \mnras,
  537, L42

\bibitem[{{Norris} {et~al.}(2021{\natexlab{a}}){Norris}, {Marvil}, {Collier},
  {Kapi{\'n}ska}, {O'Brien}, {Rudnick}, {Andernach}, {Asorey}, {Brown},
  {Br{\"u}ggen}, {Crawford}, {English}, {Rahman}, {Filipovi{\'c}}, {Gordon},
  {G{\"u}rkan}, {Hale}, {Hopkins}, {Huynh}, {HyeongHan}, {James Jee},
  {Koribalski}, {Lenc}, {Luken}, {Parkinson}, {Prandoni}, {Raja}, {Reiprich},
  {Riseley}, {Shabala}, {Sheil}, {Vernstrom}, {Whiting}, {Allison}, {Anderson},
  {Ball}, {Bell}, {Bunton}, {Galvin}, {Gupta}, {Hotan}, {Jacka}, {Macgregor},
  {Mahony}, {Maio}, {Moss}, {Pandey-Pommier}, \& {Voronkov}}]{norris21}
{Norris}, R.~P., {Marvil}, J., {Collier}, J.~D., {et~al.} 2021{\natexlab{a}},
  \pasa, 38, e046

\bibitem[{{Norris} {et~al.}(2021{\natexlab{b}}){Norris}, {Intema},
  {Kapi{\'n}ska}, {Koribalski}, {Lenc}, {Rudnick}, {Alsaberi}, {Anderson},
  {Anderson}, {Crawford}, {Crocker}, {English}, {Filipovi{\'c}}, {Galvin},
  {Hopkins}, {Hurley-Walker}, {Inoue}, {Luken}, {Macgregor}, {Manojlovi{\'c}},
  {Marvil}, {O'Brien}, {Park}, {Raja}, {Shobhana}, {Venturi}, {Collier},
  {Hale}, {Hotan}, {Moss}, \& {Whiting}}]{norris21b}
{Norris}, R.~P., {Intema}, H.~T., {Kapi{\'n}ska}, A.~D., {et~al.}
  2021{\natexlab{b}}, \pasa, 38, e003

\bibitem[{{Norris} {et~al.}(2022){Norris}, {Collier}, {Crocker}, {Heywood},
  {Macgregor}, {Rudnick}, {Shabala}, {Andernach}, {da Cunha}, {English},
  {Filipovi{\'c}}, {Koribalski}, {Luken}, {Robotham}, {Sekhar}, {Thorne}, \&
  {Vernstrom}}]{norris22}
{Norris}, R.~P., {Collier}, J.~D., {Crocker}, R.~M., {et~al.} 2022, \mnras,
  513, 1300

\bibitem[{{Omar}(2022)}]{omar22}
{Omar}, A. 2022, Research Notes of the American Astronomical Society, 6, 100

\bibitem[{{Parker} {et~al.}(2017){Parker}, {Boji{\v{c}}i{\'c}}, \&
  {Frew}}]{parker17}
{Parker}, Q.~A., {Boji{\v{c}}i{\'c}}, I., \& {Frew}, D.~J. 2017, in Planetary
  Nebulae: Multi-Wavelength Probes of Stellar and Galactic Evolution, ed.
  X.~{Liu}, L.~{Stanghellini}, \& A.~{Karakas}, Vol. 323, 36--39

\bibitem[{{Perley} {et~al.}(2011){Perley}, {Chandler}, {Butler}, \&
  {Wrobel}}]{perley11}
{Perley}, R.~A., {Chandler}, C.~J., {Butler}, B.~J., \& {Wrobel}, J.~M. 2011,
  APJ, 739, L1

\bibitem[{{Planck Collaboration} {et~al.}(2020){Planck Collaboration},
  {Aghanim}, {Akrami}, {Ashdown}, {Aumont}, {Baccigalupi}, {Ballardini},
  {Banday}, {Barreiro}, {Bartolo}, {Basak}, {Battye}, {Benabed}, {Bernard},
  {Bersanelli}, {Bielewicz}, {Bock}, {Bond}, {Borrill}, {Bouchet}, {Boulanger},
  {Bucher}, {Burigana}, {Butler}, {Calabrese}, {Cardoso}, {Carron},
  {Challinor}, {Chiang}, {Chluba}, {Colombo}, {Combet}, {Contreras}, {Crill},
  {Cuttaia}, {de Bernardis}, {de Zotti}, {Delabrouille}, {Delouis}, {Di
  Valentino}, {Diego}, {Dor{\'e}}, {Douspis}, {Ducout}, {Dupac}, {Dusini},
  {Efstathiou}, {Elsner}, {En{\ss}lin}, {Eriksen}, {Fantaye}, {Farhang},
  {Fergusson}, {Fernandez-Cobos}, {Finelli}, {Forastieri}, {Frailis},
  {Fraisse}, {Franceschi}, {Frolov}, {Galeotta}, {Galli}, {Ganga},
  {G{\'e}nova-Santos}, {Gerbino}, {Ghosh}, {Gonz{\'a}lez-Nuevo}, {G{\'o}rski},
  {Gratton}, {Gruppuso}, {Gudmundsson}, {Hamann}, {Handley}, {Hansen},
  {Herranz}, {Hildebrandt}, {Hivon}, {Huang}, {Jaffe}, {Jones}, {Karakci},
  {Keih{\"a}nen}, {Keskitalo}, {Kiiveri}, {Kim}, {Kisner}, {Knox},
  {Krachmalnicoff}, {Kunz}, {Kurki-Suonio}, {Lagache}, {Lamarre}, {Lasenby},
  {Lattanzi}, {Lawrence}, {Le Jeune}, {Lemos}, {Lesgourgues}, {Levrier},
  {Lewis}, {Liguori}, {Lilje}, {Lilley}, {Lindholm}, {L{\'o}pez-Caniego},
  {Lubin}, {Ma}, {Mac{\'\i}as-P{\'e}rez}, {Maggio}, {Maino}, {Mandolesi},
  {Mangilli}, {Marcos-Caballero}, {Maris}, {Martin}, {Martinelli},
  {Mart{\'\i}nez-Gonz{\'a}lez}, {Matarrese}, {Mauri}, {McEwen}, {Meinhold},
  {Melchiorri}, {Mennella}, {Migliaccio}, {Millea}, {Mitra},
  {Miville-Desch{\^e}nes}, {Molinari}, {Montier}, {Morgante}, {Moss}, {Natoli},
  {N{\o}rgaard-Nielsen}, {Pagano}, {Paoletti}, {Partridge}, {Patanchon},
  {Peiris}, {Perrotta}, {Pettorino}, {Piacentini}, {Polastri}, {Polenta},
  {Puget}, {Rachen}, {Reinecke}, {Remazeilles}, {Renzi}, {Rocha}, {Rosset},
  {Roudier}, {Rubi{\~n}o-Mart{\'\i}n}, {Ruiz-Granados}, {Salvati}, {Sandri},
  {Savelainen}, {Scott}, {Shellard}, {Sirignano}, {Sirri}, {Spencer},
  {Sunyaev}, {Suur-Uski}, {Tauber}, {Tavagnacco}, {Tenti}, {Toffolatti},
  {Tomasi}, {Trombetti}, {Valenziano}, {Valiviita}, {Van Tent}, {Vibert},
  {Vielva}, {Villa}, {Vittorio}, {Wandelt}, {Wehus}, {White}, {White},
  {Zacchei}, \& {Zonca}}]{planck18-1}
{Planck Collaboration}, {Aghanim}, N., {Akrami}, Y., {et~al.} 2020, \aap, 641,
  A6

\bibitem[{{Riggi} {et~al.}(2024){Riggi}, {Cecconello}, {Palazzo}, {Hopkins},
  {Gupta}, {Bordiu}, {Ingallinera}, {Buemi}, {Bufano}, {Cavallaro},
  {Filipovi{\'c}}, {Leto}, {Loru}, {Ruggeri}, {Trigilio}, {Umana}, \&
  {Vitello}}]{riggi24}
{Riggi}, S., {Cecconello}, T., {Palazzo}, S., {et~al.} 2024, \pasa, 41, e085

\bibitem[{{Rupke} {et~al.}(2024){Rupke}, {Coil}, {Whalen}, {Moustakas},
  {Tremonti}, \& {Perrotta}}]{rupke24}
{Rupke}, D. S.~N., {Coil}, A.~L., {Whalen}, K.~E., {et~al.} 2024, \apj, 967, 51

\bibitem[{{Saydjari} {et~al.}(2023){Saydjari}, {Schlafly}, {Lang}, {Meisner},
  {Green}, {Zucker}, {Zelko}, {Speagle}, {Daylan}, {Lee}, {Valdes}, {Schlegel},
  \& {Finkbeiner}}]{saydjari23}
{Saydjari}, A.~K., {Schlafly}, E.~F., {Lang}, D., {et~al.} 2023, \apjs, 264, 28

\bibitem[{{Schlafly} \& {Finkbeiner}(2011)}]{schlafly11}
{Schlafly}, E.~F., \& {Finkbeiner}, D.~P. 2011, \apj, 737, 103

\bibitem[{{Schlegel} {et~al.}(2021){Schlegel}, {Dey}, {Herrera}, {Juneau},
  {Landriau}, {Lang}, {Meisner}, {Moustakas}, {Myers}, {Schlafly}, {Valdes},
  {Weaver}, {Zhang}, {Zhou}, \& {DESI Legacy Imaging Surveys
  Team}}]{schlegel21}
{Schlegel}, D., {Dey}, A., {Herrera}, D., {et~al.} 2021, in American
  Astronomical Society Meeting Abstracts, Vol.~53, American Astronomical
  Society Meeting Abstracts, 235.03

\bibitem[{{Schlegel} {et~al.}(1998){Schlegel}, {Finkbeiner}, \&
  {Davis}}]{schlegel98}
{Schlegel}, D.~J., {Finkbeiner}, D.~P., \& {Davis}, M. 1998, \apj, 500, 525

\bibitem[{{Segal} {et~al.}(2022){Segal}, {Parkinson}, {Norris}, {Hopkins},
  {Andernach}, {Alexander}, {Carretti}, {Koribalski}, {Legodi}, {Leslie},
  {Luo}, {Pierce}, {Tang}, {Vardoulaki}, \& {Vernstrom}}]{segal22}
{Segal}, G., {Parkinson}, D., {Norris}, R., {et~al.} 2022, arXiv e-prints,
  arXiv:2206.14677

\bibitem[{{Shabala} {et~al.}(2024){Shabala}, {Yates-Jones}, {Jerrim}, {Turner},
  {Krause}, {Norris}, {Koribalski}, {Filipovi{\'c}}, {Rudnick}, {Power}, \&
  {Crocker}}]{shabala24}
{Shabala}, S.~S., {Yates-Jones}, P.~M., {Jerrim}, L.~A., {et~al.} 2024, \pasa,
  41, e024

\bibitem[{{Shimwell} {et~al.}(2022){Shimwell}, {Hardcastle}, {Tasse}, {Best},
  {R{\"o}ttgering}, {Williams}, {Botteon}, {Drabent}, {Mechev}, {Shulevski},
  {van Weeren}, {Bester}, {Br{\"u}ggen}, {Brunetti}, {Callingham}, {Chy{\.z}y},
  {Conway}, {Dijkema}, {Duncan}, {de Gasperin}, {Hale}, {Haverkorn}, {Hugo},
  {Jackson}, {Mevius}, {Miley}, {Morabito}, {Morganti}, {Offringa}, {Oonk},
  {Rafferty}, {Sabater}, {Smith}, {Schwarz}, {Smirnov}, {O'Sullivan},
  {Vedantham}, {White}, {Albert}, {Alegre}, {Asabere}, {Bacon}, {Bonafede},
  {Bonnassieux}, {Brienza}, {Bilicki}, {Bonato}, {Calistro Rivera}, {Cassano},
  {Cochrane}, {Croston}, {Cuciti}, {Dallacasa}, {Danezi}, {Dettmar}, {Di
  Gennaro}, {Edler}, {En{\ss}lin}, {Emig}, {Franzen}, {Garc{\'\i}a-Vergara},
  {Grange}, {G{\"u}rkan}, {Hajduk}, {Heald}, {Heesen}, {Hoang}, {Hoeft},
  {Horellou}, {Iacobelli}, {Jamrozy}, {Jeli{\'c}}, {Kondapally}, {Kukreti},
  {Kunert-Bajraszewska}, {Magliocchetti}, {Mahatma}, {Ma{\l}ek}, {Mandal},
  {Massaro}, {Meyer-Zhao}, {Mingo}, {Mostert}, {Nair}, {Nakoneczny},
  {Nikiel-Wroczy{\'n}ski}, {Orr{\'u}}, {Pajdosz-{\'S}mierciak}, {Pasini},
  {Prandoni}, {van Piggelen}, {Rajpurohit}, {Retana-Montenegro}, {Riseley},
  {Rowlinson}, {Saxena}, {Schrijvers}, {Sweijen}, {Siewert}, {Timmerman},
  {Vaccari}, {Vink}, {West}, {Wo{\l}owska}, {Zhang}, \& {Zheng}}]{shimwell22}
{Shimwell}, T.~W., {Hardcastle}, M.~J., {Tasse}, C., {et~al.} 2022, \aap, 659,
  A1

\bibitem[{{Slijepcevic} {et~al.}(2024){Slijepcevic}, {Scaife}, {Walmsley},
  {Bowles}, {Wong}, {Shabala}, \& {White}}]{slijepcevic23}
{Slijepcevic}, I.~V., {Scaife}, A. M.~M., {Walmsley}, M., {et~al.} 2024, RAS
  Techniques and Instruments, 3, 19

\bibitem[{Sormani {et~al.}(2024)Sormani, Sobacchi, \& Sanders}]{sormani24}
Sormani, M.~C., Sobacchi, E., \& Sanders, J.~L. 2024, Monthly Notices of the
  Royal Astronomical Society, 528, 5742

\bibitem[{{Tonry} {et~al.}(2018){Tonry}, {Denneau}, {Flewelling}, {Heinze},
  {Onken}, {Smartt}, {Stalder}, {Weiland}, \& {Wolf}}]{tonry18}
{Tonry}, J.~L., {Denneau}, L., {Flewelling}, H., {et~al.} 2018, \apj, 867, 105

\bibitem[{{van Haarlem} {et~al.}(2013){van Haarlem}, {Wise}, {Gunst}, {Heald},
  {McKean}, {Hessels}, {de Bruyn}, {Nijboer}, {Swinbank}, {Fallows},
  {Brentjens}, {Nelles}, {Beck}, {Falcke}, {Fender}, {H{\"o}randel},
  {Koopmans}, {Mann}, {Miley}, {R{\"o}ttgering}, {Stappers}, {Wijers},
  {Zaroubi}, {van den Akker}, {Alexov}, {Anderson}, {Anderson}, {van Ardenne},
  {Arts}, {Asgekar}, {Avruch}, {Batejat}, {B{\"a}hren}, {Bell}, {Bell}, {van
  Bemmel}, {Bennema}, {Bentum}, {Bernardi}, {Best}, {B{\^\i}rzan}, {Bonafede},
  {Boonstra}, {Braun}, {Bregman}, {Breitling}, {van de Brink}, {Broderick},
  {Broekema}, {Brouw}, {Br{\"u}ggen}, {Butcher}, {van Cappellen}, {Ciardi},
  {Coenen}, {Conway}, {Coolen}, {Corstanje}, {Damstra}, {Davies}, {Deller},
  {Dettmar}, {van Diepen}, {Dijkstra}, {Donker}, {Doorduin}, {Dromer}, {Drost},
  {van Duin}, {Eisl{\"o}ffel}, {van Enst}, {Ferrari}, {Frieswijk}, {Gankema},
  {Garrett}, {de Gasperin}, {Gerbers}, {de Geus}, {Grie{\ss}meier}, {Grit},
  {Gruppen}, {Hamaker}, {Hassall}, {Hoeft}, {Holties}, {Horneffer}, {van der
  Horst}, {van Houwelingen}, {Huijgen}, {Iacobelli}, {Intema}, {Jackson},
  {Jelic}, {de Jong}, {Juette}, {Kant}, {Karastergiou}, {Koers}, {Kollen},
  {Kondratiev}, {Kooistra}, {Koopman}, {Koster}, {Kuniyoshi}, {Kramer},
  {Kuper}, {Lambropoulos}, {Law}, {van Leeuwen}, {Lemaitre}, {Loose}, {Maat},
  {Macario}, {Markoff}, {Masters}, {McFadden}, {McKay-Bukowski}, {Meijering},
  {Meulman}, {Mevius}, {Middelberg}, {Millenaar}, {Miller-Jones}, {Mohan},
  {Mol}, {Morawietz}, {Morganti}, {Mulcahy}, {Mulder}, {Munk}, {Nieuwenhuis},
  {van Nieuwpoort}, {Noordam}, {Norden}, {Noutsos}, {Offringa}, {Olofsson},
  {Omar}, {Orr{\'u}}, {Overeem}, {Paas}, {Pandey-Pommier}, {Pandey}, {Pizzo},
  {Polatidis}, {Rafferty}, {Rawlings}, {Reich}, {de Reijer}, {Reitsma},
  {Renting}, {Riemers}, {Rol}, {Romein}, {Roosjen}, {Ruiter}, {Scaife}, {van
  der Schaaf}, {Scheers}, {Schellart}, {Schoenmakers}, {Schoonderbeek},
  {Serylak}, {Shulevski}, {Sluman}, {Smirnov}, {Sobey}, {Spreeuw}, {Steinmetz},
  {Sterks}, {Stiepel}, {Stuurwold}, {Tagger}, {Tang}, {Tasse}, {Thomas},
  {Thoudam}, {Toribio}, {van der Tol}, {Usov}, {van Veelen}, {van der Veen},
  {ter Veen}, {Verbiest}, {Vermeulen}, {Vermaas}, {Vocks}, {Vogt}, {de Vos},
  {van der Wal}, {van Weeren}, {Weggemans}, {Weltevrede}, {White}, {Wijnholds},
  {Wilhelmsson}, {Wucknitz}, {Yatawatta}, {Zarka}, {Zensus}, \& {van
  Zwieten}}]{vanharleem13}
{van Haarlem}, M.~P., {Wise}, M.~W., {Gunst}, A.~W., {et~al.} 2013, \aap, 556,
  A2

\bibitem[{{Walmsley} {et~al.}(2022){Walmsley}, {Scaife}, {Lintott}, {Lochner},
  {Etsebeth}, {G{\'e}ron}, {Dickinson}, {Fortson}, {Kruk}, {Masters}, {Mantha},
  \& {Simmons}}]{walmsley22}
{Walmsley}, M., {Scaife}, A. M.~M., {Lintott}, C., {et~al.} 2022, \mnras, 513,
  1581

\bibitem[{{Wayth} {et~al.}(2018){Wayth}, {Tingay}, {Trott}, {Emrich},
  {Johnston-Hollitt}, {McKinley}, {Gaensler}, {Beardsley}, {Booler}, {Crosse},
  {Franzen}, {Horsley}, {Kaplan}, {Kenney}, {Morales}, {Pallot}, {Sleap},
  {Steele}, {Walker}, {Williams}, {Wu}, {Cairns}, {Filipovic}, {Johnston},
  {Murphy}, {Quinn}, {Staveley-Smith}, {Webster}, \& {Wyithe}}]{wayth18}
{Wayth}, R.~B., {Tingay}, S.~J., {Trott}, C.~M., {et~al.} 2018, \pasa, 35, e033

\bibitem[{{Wen} \& {Han}(2024)}]{wen24}
{Wen}, Z.~L., \& {Han}, J.~L. 2024, \apjs, 272, 39

\bibitem[{{Whiting} \& {Humphreys}(2012)}]{whiting12}
{Whiting}, M., \& {Humphreys}, B. 2012, \pasa, 29, 371

\bibitem[{{Whiting} {et~al.}(2017){Whiting}, {Voronkov}, {Mitchell}, \& {Askap
  Team}}]{whiting17}
{Whiting}, M., {Voronkov}, M., {Mitchell}, D., \& {Askap Team}. 2017, in
  Astronomical Society of the Pacific Conference Series, Vol. 512, Astronomical
  Data Analysis Software and Systems XXV, ed. N.~P.~F. {Lorente},
  K.~{Shortridge}, \& R.~{Wayth}, 431

\bibitem[{{Wright} {et~al.}(2010){Wright}, {Eisenhardt}, {Mainzer}, {Ressler},
  {Cutri}, {Jarrett}, {Kirkpatrick}, {Padgett}, {McMillan}, {Skrutskie},
  {Stanford}, {Cohen}, {Walker}, {Mather}, {Leisawitz}, {Gautier}, {McLean},
  {Benford}, {Lonsdale}, {Blain}, {Mendez}, {Irace}, {Duval}, {Liu}, {Royer},
  {Heinrichsen}, {Howard}, {Shannon}, {Kendall}, {Walsh}, {Larsen}, {Cardon},
  {Schick}, {Schwalm}, {Abid}, {Fabinsky}, {Naes}, \& {Tsai}}]{wright10}
{Wright}, E.~L., {Eisenhardt}, P.~R.~M., {Mainzer}, A.~K., {et~al.} 2010, \aj,
  140, 1868

\bibitem[{{Zaw} {et~al.}(2019){Zaw}, {Chen}, \& {Farrar}}]{zaw19}
{Zaw}, I., {Chen}, Y.-P., \& {Farrar}, G.~R. 2019, \apj, 872, 134

\bibitem[{{Zhou} {et~al.}(2021){Zhou}, {Newman}, {Mao}, {Meisner}, {Moustakas},
  {Myers}, {Prakash}, {Zentner}, {Brooks}, {Duan}, {Landriau}, {Levi}, {Prada},
  \& {Tarle}}]{zhou21}
{Zhou}, R., {Newman}, J.~A., {Mao}, Y.-Y., {et~al.} 2021, \mnras, 501, 3309

\bibitem[{{Zhou} {et~al.}(2025){Zhou}, {Li}, {Zou}, {Gong}, {Deng}, {Chen},
  {Yu}, {He}, \& {Ding}}]{zhou25}
{Zhou}, X., {Li}, N., {Zou}, H., {et~al.} 2025, \mnras, 536, 2260

\bibitem[{Zou {et~al.}(2019)Zou, Gao, Zhou, \& Kong}]{zou19}
Zou, H., Gao, J., Zhou, X., \& Kong, X. 2019, The Astrophysical Journal
  Supplement Series, 242, 8

\bibitem[{{Zou} {et~al.}(2022){Zou}, {Sui}, {Xue}, {Zhou}, {Ma}, {Zhou}, {Nie},
  {Zhang}, {Feng}, {Shen}, \& {Wang}}]{zou22}
{Zou}, H., {Sui}, J., {Xue}, S., {et~al.} 2022, Research in Astronomy and
  Astrophysics, 22, 065001

\end{thebibliography}

\label{lastpage}
\end{document}